\documentclass[journal]{IEEEtran}
\usepackage{amsmath,epsfig,amssymb,verbatim,amsopn,subfigure,color}
\usepackage{cite,xspace}
\usepackage{array,algorithm,algorithmic}
\usepackage{bbm}
\usepackage{array}
\usepackage{multirow}
\usepackage{multicol}
\usepackage{rotating}
\usepackage{dsfont,float}
\usepackage{stfloats}
\usepackage{enumerate,makecell}
\usepackage{url}
\usepackage{tikz,graphicx}
\usepackage{booktabs}
\usepackage[font=footnotesize]{caption}
\normalsize

\ifCLASSOPTIONpeerreview

\fi

\ifCLASSOPTIONonecolumn
    
\else
    
\fi

\makeatletter
\let\l@ENGLISH\l@english
\makeatother

\makeatletter
\renewcommand*{\@opargbegintheorem}[3]{\trivlist
  \item[\hskip \labelsep{\itshape #1\ #2}] {\itshape (#3):} {\normalfont}}
\makeatother
\usepackage{relsize}

\newcommand{\Pmax}{P_\mathrm{M}}
\newcommand{\Pt}{P_\mathrm{TBS}}
\newcommand{\Pa}{P_\mathrm{ABS}}
\newcommand{\Pc}{\mathbb{P}_\mathrm{cov}}
\newcommand{\alphab}{\alpha_\mathrm{T}}
\newcommand{\alphaL}{\alpha_\mathrm{L}}
\newcommand{\alphaN}{\alpha_\mathrm{N}}
\newcommand{\alphak}{\alpha_\mathrm{k}}
\newcommand{\rhob}{\rho_\mathrm{T}}
\newcommand{\rhod}{\rho_\mathrm{A}}
\newcommand{\gammab}{\gamma_\mathrm{T}}
\newcommand{\gammad}{\gamma_\mathrm{A}}
\newcommand{\Ib}{I_\mathrm{T}}
\newcommand{\Id}{I_\mathrm{A}}
\newcommand{\Zd}{\mathcal{Z}_A}
\newcommand{\Zc}{\mathcal{Z}_T}
\newcommand{\mL}{m_\mathrm{L}}
\newcommand{\mN}{m_\mathrm{N}}
\newcommand{\pLOS}{p_\mathrm{L}}
\newcommand{\pNLOS}{p_\mathrm{N}}
\newcommand{\etak}{\eta_\mathrm{k}}
\newcommand{\etaL}{\eta_\mathrm{L}}
\newcommand{\etaN}{\eta_\mathrm{N}}

\newtheorem{remark}{Remark}
\newtheorem{theorem}{Theorem}

\newtheorem{lemma}{Lemma}
\graphicspath{{figs/},{Figures/}}

\newcommand{\AuthorOne}{Xiaohui~Zhou, {\em{Student Member, IEEE}}}
\newcommand{\AuthorTwo}{Salman~Durrani, {\em{Senior Member, IEEE}}}
\newcommand{\AuthorThree}{Jing~Guo, {\em{Member, IEEE}}}
\newcommand{\AuthorFour}{Halim~Yanikomeroglu, {\em{Fellow, IEEE}}}

\newcommand{\ThankOne}{X. Zhou, S. Durrani and J. Guo are with the Research School of Engineering, College of Engineering and Computer Science, The Australian National University, Canberra, ACT 2601, Australia (Emails: \{xiaohui.zhou, salman.durrani, jing.guo\}@anu.edu.au). H. Yanikomeroglu is with Department of Systems and Computer Engineering, Carleton University, Ottawa, ON K1S 5B6, Canada (Email: halim@sce.carleton.ca). A preliminary version of this work has been presented at IEEE ICC Workshop, Kansas City, USA, 2018~\cite{Zhou-2018}. (Corresponding author: Jing Guo.)

This work was supported in part by the Australian Research Council's Discovery Project Funding Scheme (Project number DP170100939) and Huawei Canada Research Centre.}

\begin{document}
\title{Underlay Drone Cell for Temporary Events: Impact of Drone Height and Aerial Channel Environments}
\author{{\AuthorOne,~\AuthorTwo,~\AuthorThree,~and~\AuthorFour\thanks{\ThankOne}}}
\maketitle

\begin{abstract}
Providing seamless connection to a large number of devices is one of the biggest challenges for the Internet of Things (IoT) networks. Using a drone as an aerial base station (ABS) to provide coverage to devices or users on ground is envisaged as a promising solution for IoT networks. In this paper, we consider a communication network with an underlay ABS to provide coverage for a temporary event, such as a sporting event or a concert in a stadium. Using stochastic geometry, we propose a general analytical framework to compute the uplink and downlink coverage probabilities for both the aerial and the terrestrial cellular system. Our framework is valid for any aerial channel model for which the probabilistic functions of line-of-sight (LOS) and non-line-of-sight (NLOS) links are specified. The accuracy of the analytical results is verified by Monte Carlo simulations considering two commonly adopted aerial channel models. Our results show the non-trivial impact of the different aerial channel environments (i.e., suburban, urban, dense urban and high-rise urban) on the uplink and downlink coverage probabilities and provide design guidelines for best ABS deployment height.
\end{abstract}

\begin{IEEEkeywords}
Internet of Things (IoT), underlay drone cell, aerial channel model, uplink, downlink, stochastic geometry.
\end{IEEEkeywords}

\ifCLASSOPTIONonecolumn
    \newpage
\else
    \vspace{-0.2 in}
\fi

\section{Introduction}\label{sec:Intro}
The Internet of Things (IoT) is at present rapidly evolving and will connect a massive number of devices, both machine and human operated, in the near future. The applications of IoT will improve all aspects of our life, ranging from smart homes, such as self-managed household appliances, personal wearables and healthcare, to smart cities, like manufacturing management, infrastructure monitoring and vehicle communications. This requires effective and reliable wireless connectivity, high data rates and ultra low latency. It is not possible to rely solely on the conventional wireless networks (e.g., cellular system) to support IoT. The high data rates and the numerous number of connected network nodes would quickly and constantly overload base stations and put pressure on network resources. Moreover, in locations which experience poor coverage by base stations, IoT devices may not be able to connect to the base station. Furthermore, wireless network coverage may not reach all locations where smaller IoT devices are placed, such as rural areas, forest and sea. In this regard, the use of drones as aerial base stations (ABSs), to form drone cells and to provide flexible and agile coverage is a potential network solution for IoT applications~\cite{Yaliniz-2016,Chandrasekharan-2016,Mozaffari-2018,Zeng-2018}.

\subsection{Motivation}
Recently in \cite{Yaliniz-2016}, eight scenarios have been identified for drone cell deployment, with drones providing service to: (i) rural areas, (ii) urban areas, (iii) users with high mobility, (iv) congested urban areas, (v) congested backhaul, (vi) temporary events, (vii) temporary blind spots and (viii) sensor networks. The literature to date~\cite{Mozaffari-2016c,Chetlur-2017,Azari-2017a,Galkin-2017,Azari-2017b,Sekander-2018} has focused on the downlink, generally for Scenarios (i)--(iii). In this work, we focus on the use of drone cell for Scenario (vi), i.e., using ABS to provide additional coverage for temporary events, such as concerts or sports events. Such temporary events are happening more frequently in cities all over the world with very high number of users gathering. A large number of IoT devices are expected to be deployed in a stadium to help the event operators to provide superior experience to event participants efficiently and effectively. For example, security cameras being used to monitor all corners of the venue and keep the crowd safe. Sensors being used to provide up-to-date information on queues for merchandise stalls. Temperature sensors being used to monitor and control the air conditioning system. These devices all need to be connected to the core network. Therefore, assessing the usefulness of ABSs to provide coverage for temporary events is an important open problem in the literature which has great practical importance as well. Note that network users inside a stadium are not limited to smartphones, but can also be security cameras, noise and temperature sensors inside the stadium and performance monitoring devices on the sports players.

\subsection{Related Work}
The investigation of drone cells has drawn attention in the literature from different perspectives, such as drone channel modeling~\cite{Khawaja-2018,Amorim-2017a,Amorim-2017b,Hourani-2014b,Hourani-2014a}, drone deployment and optimization of trajectories~\cite{Fotouhi-2017,Zeng-2016a,Alzenad-2017a,He-2018,Zhang-2017,Lagum-2017,Wu-2017,Hourani-2014b} and performance analysis of drone enable cellular systems~\cite{Mozaffari-2016c,Chetlur-2017,Azari-2017a,Galkin-2017,Azari-2017b,Azari-2018,Alzenad-2018}.

The work in \cite{Khawaja-2018} presented a comprehensive overview of existing research related to aerial (also known as air-to-ground (A2G)) channel measurements, including large and small-scale fading channel models. Models for path-loss exponents and shadowing for the aerial radio channel between drones and cellular networks were presented in~\cite{Amorim-2017a,Amorim-2017b} based on field measurements. Using the general geometrical statistics of various environments provided by the International Telecommunication Union (ITU-R), the authors proposed location and environment dependent path-loss model for low altitude platforms in~\cite{Hourani-2014b,Hourani-2014a}.

Optimal drone trajectory was designed for a drone base station in cellular network in~\cite{Fotouhi-2017} and for a drone enabled relaying network in~\cite{Zeng-2016a}. An analytical model for finding optimal placement of a drone was provided in~\cite{Alzenad-2017a} to maximize the number of covered users and optimal height and antenna beamwidth was found in~\cite{He-2018} for throughput optimization. Optimal density of the underlay drones was investigated in~\cite{Zhang-2017} and optimal placement for multiple drones in a large-scale network was studied in~\cite{Lagum-2017}. The work in~\cite{Wu-2017} studied the joint user scheduling and drone trajectory optimization for maximizing the minimum average rate of ground users. In~\cite{Hourani-2014b}, the authors derived the optimal altitude enabling a single drone to achieve a maximum coverage radius on the ground.

Some previous works investigated the performance of drone network using stochastic geometry. Note that stochastic geometry is a powerful mathematical tool, which can be used to capture the randomness of the nodes' locations and fading channels. Specifically, \cite{Azari-2018} investigated the uplink performance of a drone cell in the presence of a Poisson field of ground interferers. The downlink performance of a single static drone and a single mobile drone with underlay device-to-device users was studied in \cite{Mozaffari-2016c}, while \cite{Chetlur-2017} analyzed the downlink coverage of a finite network formed by multiple drones. The downlink coverage probability of a network with multiple directional beamforming drones was investigated in \cite{Azari-2017a}. The downlink coverage performance of multiple drones in an urban environment was studied in\cite{Galkin-2017}. In \cite{Azari-2017b}, the authors studied the downlink in a cellular network with multiple ground base stations and a drone user equipment. The downlink coverage and rate in a Poisson field of drone base stations was investigated in \cite{Alzenad-2018}.

We focus on the performance analysis of drone in a temporary event scenario using stochastic geometry. Recently, there are some works investigating the application of drone in temporary events from different perspectives. Specifically, the Aerial Base Stations with Opportunistic Links for Unexpected and Temporary Events (ABSOLUTE) project in Europe aims to provide reliable network coverage through a combination of aerial, terrestrial and satellite links for unexpected and temporary events \cite{Absolute-2018}. A heuristically accelerated reinforcement learning based framework was proposed in \cite{Morozs-2015} for dynamic spectrum sharing in a temporary event scenario. In \cite{Koulali-2016}, the authors constructed a non-cooperative game and studied the equilibrium beaconing durations in terms of energy efficiency of the competing drones for temporary events. A proactive drone cell deployment scheme was investigated in \cite{Yang-2017} to cope with flash crowd traffic in different scenarios, including stadiums. A limited feedback scheme for non-orthogonal multiple access was designed for millimeter wave drone to provide coverage over a stadium in \cite{Rupasinghe-2018}.

\subsection{Contributions}
In this work, we consider a drone system coexisting with a single-cell cellular network, where an ABS is designated to provide service to IoT devices (namely the ABS-supported devices (AsDs), such as smartphones, security cameras, noise and temperature sensors and performance monitoring devices) inside a stadium for a temporary event (e.g., a concert or a sporting event). Since the ABS shares the same spectrum resources (i.e., in an underlay fashion) with the terrestrial base station (TBS), the concurrent transmission of both systems can cause interference to each other and impact the network performance. To the best of our knowledge, this scenario and its study have not yet been presented in the literature to date. In our preliminary work~\cite{Zhou-2018}, we considered a simplified aerial channel model and used stochastic geometry to assess the uplink coverage performance for an underlay drone cell. In this work, we consider a general aerial channel model and both uplink and downlink network performance. The novel contributions of this paper are summarized as follows:
\begin{itemize}
\item Leveraging tools from stochastic geometry, we develop a general analytical framework to analyze the uplink coverage probability of the TBS and the ABS and the downlink coverage probability of the TBS-supported user equipment (TsUE) and the AsD. The proposed framework is able to accommodate any aerial channel model.
\item Our proposed framework depends on the Laplace transforms of the interference power distribution at the TBS, the ABS, the TsUE and the AsD. We derive the key factors that determining the Laplace transforms of the interference power distribution, the distribution function of the 3-D distance between the ABS and an independently and uniformly distributed (i.u.d.) AsD and the distribution function of the 3-D distance between the ABS and an i.u.d. TsUE. Note that such distance distributions take into account the hole effects (i.e., the TsUE are prohibited from the ABS serving region and the AsD are contained in the ABS serving region).
\item Our results show that for urban environment and dense urban environment the ABS is best deployed at a low height (e.g., 200 m or lower), regardless of the distance between the center of the stadium and the TBS. However, for suburban environment and high-rise urban environment the best ABS deployment height depends on the distance between the center of the stadium and the TBS and the task of the system (i.e., prioritize the terrestrial link or the aerial link, prioritize the uplink or the downlink communication).
\end{itemize}

\ifCLASSOPTIONpeerreview
\begin{table}
\centering
\caption{Summary of Main Symbols Used in the Paper.}
\label{tab:notation}
\begin{tabular}{|c|l|c|l|}
\hline
Symbol & Definition & Symbol & Definition\\ \hline
$R_1$ & Radius of the network region of TBS & $\rhob$ & Uplink receiver sensitivity of TBS\\ \hline
$R_2$ & Radius of the stadium & $\rhod$ & Uplink receiver sensitivity of ABS\\ \hline
$d$ & Distance between the center of the stadium and the TBS & $\Pmax$ & AsD maximum transmit power\\ \hline
$\alphab$ & Path-loss exponent of terrestrial link & $\Pt$ & TBS transmit power\\ \hline
$\alphaL$ & Path-loss exponent of LOS aerial link & $\Pa$ & ABS transmit power\\ \hline
$\alphaN$ & Path-loss exponent of NLOS aerial link  & $\gammab^u$ & Uplink SINR threshold of TBS\\ \hline
$\etaL$ & Additional attenuation factor for LOS aerial link  & $\gammad^u$ & Uplink SINR threshold of ABS\\ \hline
$\etaN$ & Additional attenuation factor for NLOS aerial link & $\gammab^d$ & Downlink SINR threshold of TsUE\\ \hline
$\mL$ & Nakagami-$m$ fading parameter for LOS aerial link & $\gammad^d$ & Downlink SINR threshold of AsD\\ \hline
$\mN$ & Nakagami-$m$ fading parameter for NLOS aerial link & $\sigma^2$ & Noise power\\ \hline
\end{tabular}
\end{table}
\else
\begin{table}
\centering
\caption{Summary of Main Symbols Used in the Paper.}
\label{tab:notation}
\begin{tabular}{|c|l|}
\hline
Symbol & Definition\\ \hline
$R_1$ & Radius of the network region of TBS\\ \hline
$R_2$ & Radius of the stadium\\ \hline
$d$ & Distance between the center of the stadium and the TBS \\ \hline
$\alphab$ & Path-loss exponent of terrestrial link\\ \hline
$\alphaL$ & Path-loss exponent of LOS aerial link\\ \hline
$\alphaN$ & Path-loss exponent of NLOS aerial link\\ \hline
$\etaL$ & Additional attenuation factor for LOS aerial link\\ \hline
$\etaN$ & Additional attenuation factor for NLOS aerial link\\ \hline
$\mL$ & Nakagami-$m$ fading parameter for LOS aerial link\\ \hline
$\mN$ & Nakagami-$m$ fading parameter for NLOS aerial link\\ \hline
$\rhob$ & Uplink receiver sensitivity of TBS\\ \hline
$\rhod$ & Uplink receiver sensitivity of ABS\\ \hline
$\Pmax$ & AsD maximum transmit power\\ \hline
$\Pt$ & TBS transmit power\\ \hline
$\Pa$ & ABS transmit power\\ \hline
$\gammab^u$ & Uplink SINR threshold of TBS\\ \hline
$\gammad^u$ & Uplink SINR threshold of ABS\\ \hline
$\gammab^d$ & Downlink SINR threshold of TsUE\\ \hline
$\gammad^d$ & Downlink SINR threshold of AsD\\ \hline
$\sigma^2$ & Noise power\\ \hline
\end{tabular}
\end{table}
\fi

\subsection{Notation and Paper Organization}
The following notation is used in this paper. $\Pr(\cdot)$ indicates the probability measure and $\mathbb{E}[\cdot]$ denotes the expectation operator. $f_X(x)$ denotes the probability density function (pdf) of a random variable $X$. $\mathcal{L}_X(s)=\mathbb{E}[\exp(-sX)]$ denotes the Laplace transform of a random variable $X$. A list of the main mathematical symbols employed in this paper is given in Table~\ref{tab:notation}.

The rest of the paper is organized as follows: Section~\ref{sec:systemmodel} describes the system model and assumptions. Section~\ref{sec:uplink} focuses on the uplink coverage probability at the TBS and the ABS. Section~\ref{sec:downlink} details the analysis of the downlink coverage probability at the TsUE and the AsD. Section~\ref{sec:result} presents the results and the effect of the system parameters on the network performance. Finally, Section~\ref{sec:Conc} concludes the paper.

\section{System model}\label{sec:systemmodel}
A two-cell communication network comprised of a TBS and an ABS is considered in this paper, where the network region $\mathcal{S}_1$ is a disk with radius $R_1$, i.e., $|\mathcal{S}_1|=\pi R_1^2$ and a TBS is located at the center. We assume that there is a temporary event held inside a stadium within the network region and a large number of IoT devices are active inside the stadium. The stadium's building area $\mathcal{S}_2$ is modeled as a disk with radius $R_2$ and its center is at a distance $d$ from the TBS. A drone is placed as an ABS\footnote{Current drone regulations prohibit a drone from flying over stadiums or sports events. This is expected to change in the future.} to provide additional resources for the event. The ABS is assumed to be deployed at a height of $h$ above the center of the stadium, as shown in Fig.~\ref{fig:system}. The TsUEs served by the TBS are uniformly distributed over the network region excluding the stadium, i.e., $\mathcal{S}_1\setminus\mathcal{S}_2$. At the same time, the AsDs\footnote{The analysis in this work focuses on single channel use, so an assumption on the number of TsUEs or AsDs is not required.} are independently uniformly deployed on the ground inside the stadium $\mathcal{S}_2$. For tractability, we assume that all AsDs are connected to the ABS only. In this work, we focus our analysis on one terrestrial cell and its underlay drone cell without inter terrestrial cell interference. This is based on the assumption that the terrestrial adjacent cells use different frequencies, while the TBS and the underlay ABS share the same spectrum resource. Hence, the interference from far away cells becomes negligible and can be ignored \cite{Yang-2017,Lyu-2018}. Note that the single terrestrial cell model is commonly used in literature for underlay network analysis \cite{Song-2014,Zhou-2017}.

\ifCLASSOPTIONonecolumn
\begin{figure}[t]
\centering
\subfigure[3D view.]{\includegraphics[width=0.35\textwidth]{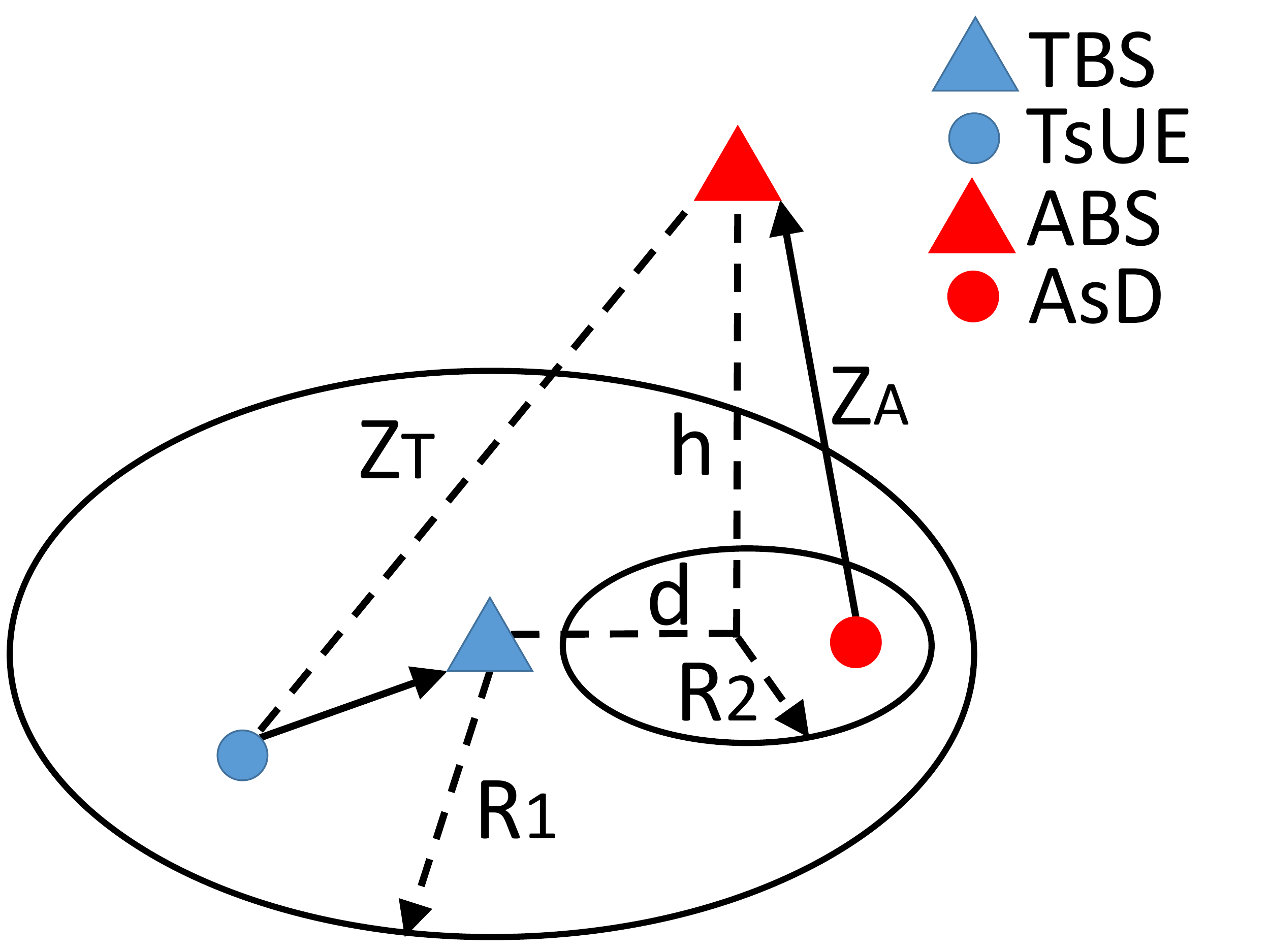}}
\subfigure[Projection on the ground.]{\includegraphics[width=0.35\textwidth]{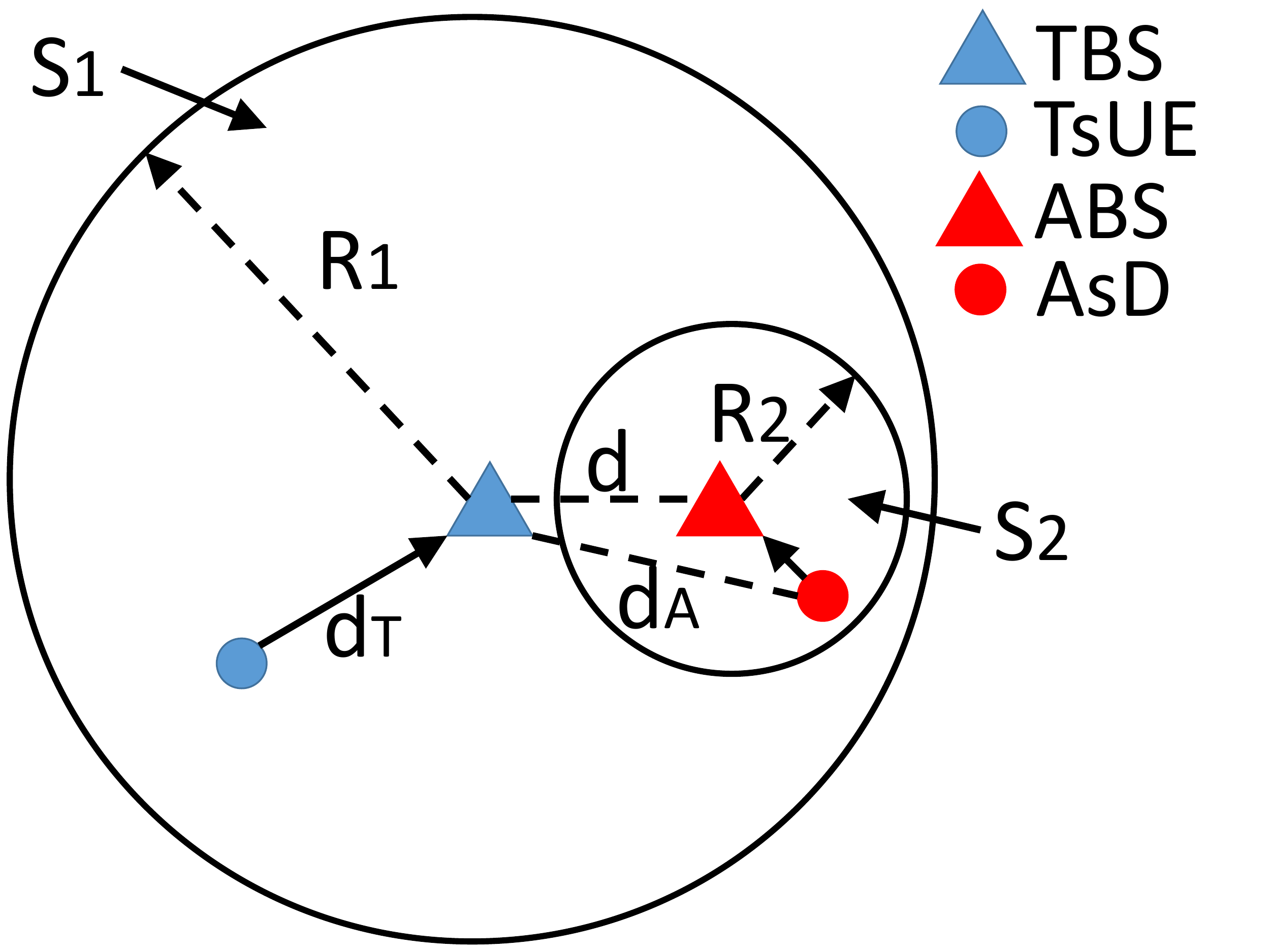}}
\vspace{-3mm}
\caption{Illustration of the system model.}
\label{fig:system}
\vspace{-6mm}
\end{figure}
\else
\begin{figure}[t]
\centering
\subfigure[3D view.]{\includegraphics[width=0.24\textwidth]{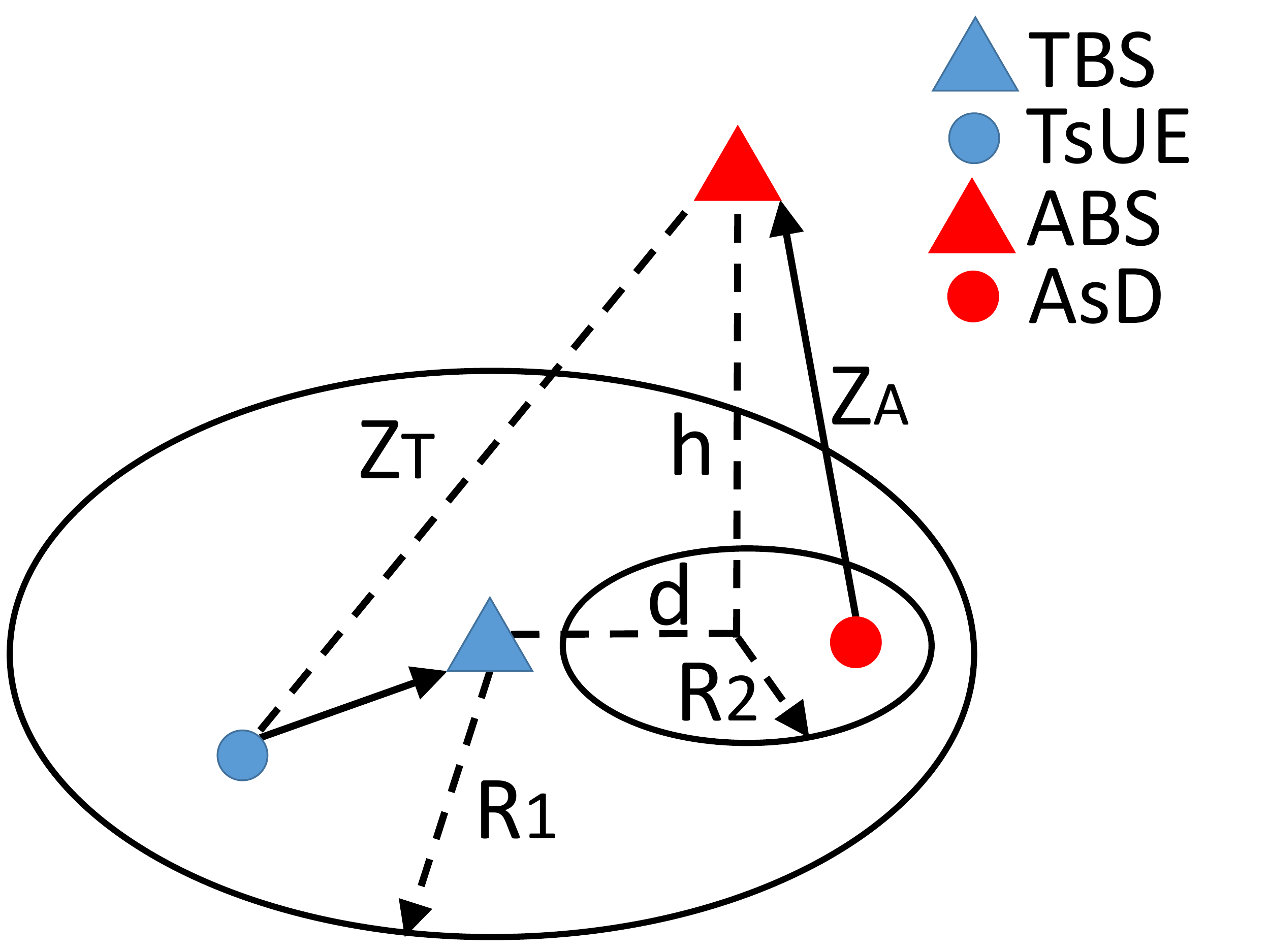}}
\subfigure[Projection on the ground.]{\includegraphics[width=0.24\textwidth]{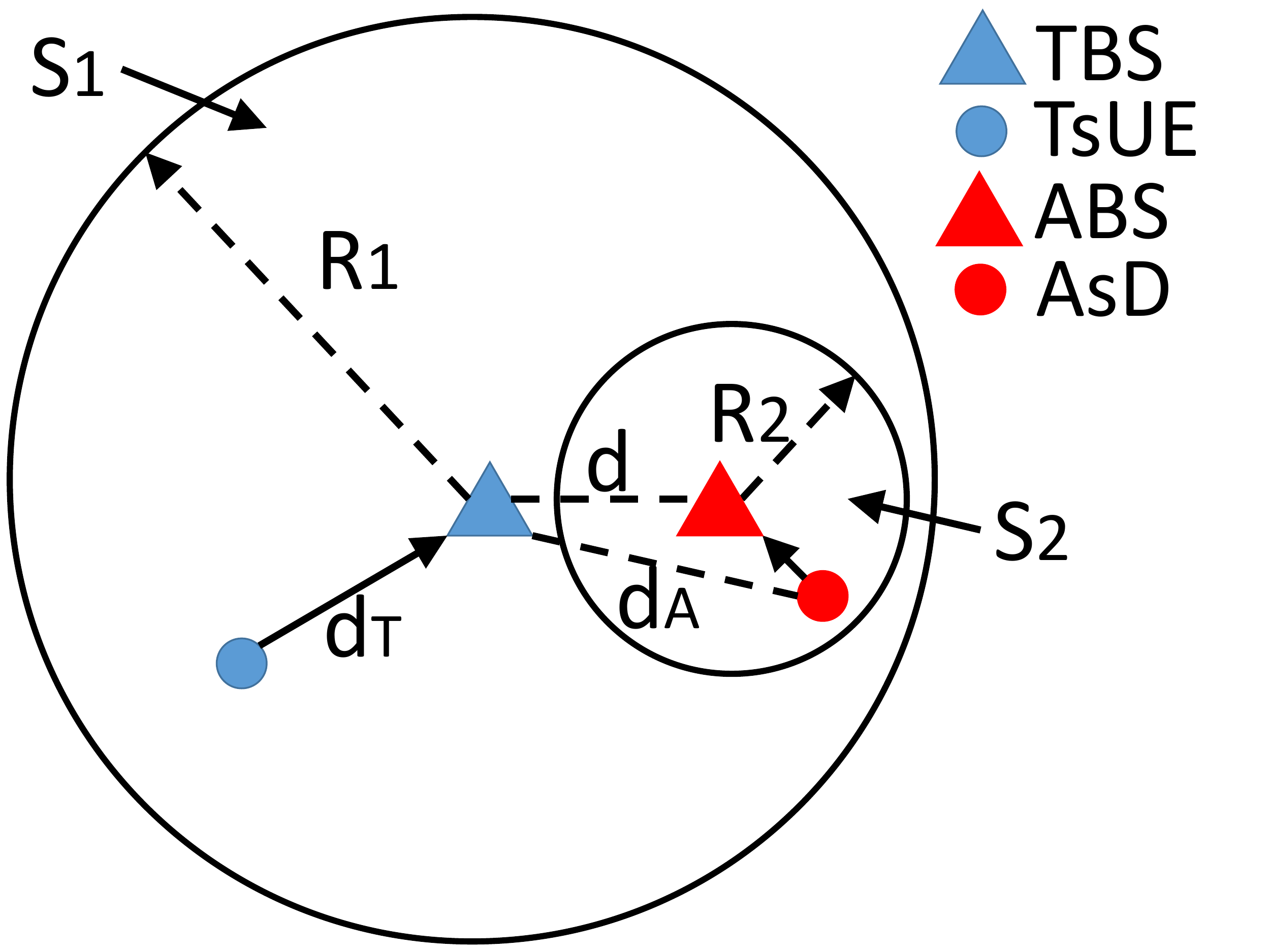}}
\vspace{-3mm}
\caption{Illustration of the system model.}
\label{fig:system}
\vspace{-6mm}
\end{figure}
\fi

\subsection{Channel Model}
There are two types of communication links in the considered system model: aerial links and terrestrial links. The link between the TsUE and the TBS and the link between the AsD and the TBS are terrestrial links. The link between the TsUE and the ABS and the link between the AsD and the ABS are aerial links.

\textit{\underline{Terrestrial links:}}
A general power-law path-loss model is considered for terrestrial links, in which the signal power decays at a rate $\ell^{-\alphab}$ with the propagation distance $\ell$ and $\alphab$ is the path-loss exponent. Furthermore, we assume the terrestrial links experience small-scale Rayleigh fading and additive white Gaussian noise (AWGN) with variance $\sigma^2$.

\textit{\underline{Aerial links:}}
The channel characteristics of the aerial links (or known as A2G links) are significantly different from the terrestrial links. Depending on altitude and type of the drone, elevation angle and type of propagation environment, the aerial links can be either line-of-sight (LOS) or non-line-of-sight (NLOS) with different probabilities of occurrence $\pLOS$ and $\pNLOS$ \cite{Hourani-2014a}.

The path-loss of the NLOS link is higher than LOS one, because of the shadowing effect and the reflection of signals from obstacles. Following \cite{Mozaffari-2016c}, the path-loss of the aerial link is modeled as
\begin{align}
PL_a(z)=\begin{cases}
\etaL z^{-\alphaL},& \mathrm{LOS}\\
\etaN z^{-\alphaN},& \mathrm{NLOS}
\end{cases},
\end{align}
where $z$ is the 3-D propagation distance between the TsUE and the ABS and between the AsD and the ABS, $\alphaL$ and $\alphaN$ is the path-loss exponent of LOS aerial link and NLOS aerial link respectively, $\etaL$ and $\etaN$ is the additional attenuation factor for LOS aerial link and NLOS aerial link respectively and $\etaL>\etaN$.

Most previous works using aerial channel model ignore the impact of small-scale fading~\cite{Mozaffari-2016c,Alzenad-2017a}. However, small-scale fading characteristics are measured and reported in the literature recently for various aerial propagation environments~\cite{Khawaja-2018}. In this paper, the small-scale fading of the aerial link is modeled as Nakagami-$m$ fading, which is a flexible model that mimics various fading environments. For example, Nakagami-$m$ fading is equivalent to Rayleigh fading when $m$ equals to 1 and Nakagami-$m$ fading can also closely approximate Rician fading by matching the $m$ values. The fading parameter for the LOS aerial link and NLOS aerial link is denoted by $\mL$ and $\mN$ respectively. The difference between including and ignoring the small-scale fading will be discussed in the result section. The aerial links also experience AWGN with variance $\sigma^2$.

\setcounter{equation}{1}
\begin{figure*}[t]
\begin{align}\label{eq:Pu}
P_\mathrm{AsD}=\begin{cases}
                \left.\begin{aligned} \Pmax,\;\;\;\;\;\;\;\;\;\;& \textup{cond.} 1\mathrm{L}\;\;\;||\;\;\;\left(\textup{cond.} 2\mathrm{L}\;\;\; \&\;\;\; \Zd>(\frac{\Pmax\etaL}{\rhod})^{\frac{1}{\alphaL}}\right)\\
                \frac{\rhod}{\etaL} \Zd^{\alphaL},\;\;\;\;\;\;& \textup{cond.} 3\mathrm{L}\;\;\;||\;\;\;\left(\textup{cond.} 2\mathrm{L}\;\;\; \&\;\;\;\Zd<(\frac{\Pmax\etaL}{\rhod})^{\frac{1}{\alphaL}}\right)\end{aligned}\right\}\mathrm{LOS}\\
                \left.\begin{aligned} \Pmax,\;\;\;\;\;\;\;\;\;\;& \textup{cond.} 1\mathrm{N}\;\;\;||\;\;\;\left(\textup{cond.} 2\mathrm{N}\;\;\; \&\;\;\; \Zd>(\frac{\Pmax\etaN}{\rhod})^{\frac{1}{\alphaN}}\right)\\
                \frac{\rhod}{\etaN} \Zd^{\alphaN},\;\;\;\;\;\;& \textup{cond.} 3\mathrm{N}\;\;\;||\;\;\;\left(\textup{cond.} 2\mathrm{N}\;\;\; \&\;\;\;\Zd<(\frac{\Pmax\etaN}{\rhod})^{\frac{1}{\alphaN}}\right)\end{aligned}\right\}\mathrm{NLOS}
		\end{cases}.
\end{align}
\rule{18.2cm}{0.5pt}
\vspace{-10mm}
\end{figure*}

\subsection{Uplink Network Model}
For uplink, we assume that orthogonal multiple access technique is employed \cite{Novlan-2013}. Hence, there is no interference among TsUEs (or AsDs). However, both the TBS and the ABS share the same spectrum resource, i.e., in an underlay fashion. We assume that the number of the TsUEs and the AsDs are sufficiently high. That is to say, there will always be one TsUE and one AsD to be served per each channel at the same time. Therefore, interference exists between TsUEs and AsDs. In this work, we focus our analysis on one uplink channel since other channels share the same interference statistics.

For reliable and successful uplink communication, the TsUE controls its transmit power such that the average signal received at the TBS is equal to the receiver sensitivity $\rhob$. Power control is deployed at the AsD as well. Perfect channel state information (CSI) knowledge is assumed at TsUE and AsD. We also set a maximum transmit power constraint at the AsD, to avoid the transmit power for AsD going to very large when the ABS is placed at a high altitude. In other words, the AsD compensates for the path-loss to keep the average signal power at the ABS equal to the receiver sensitivity $\rhod$ if the transmit power required for the path-loss inversion is less than $\Pmax$. Otherwise, the AsD tries to establish an uplink connection with the ABS by transmitting with a power of $\Pmax$. Therefore, the instantaneous transmit power for the AsD, $P_\mathrm{AsD}$, depends on the propagation distance between the AsD and the ABS and can be shown as \eqref{eq:Pu} at the top of this page, where $\Zd$ is the Euclidean distance between the AsD and the ABS and the conditions are:
\setcounter{equation}{2}
\begin{align}
&\textup{cond.} 1\mathrm{k}: h\geqslant\left(\frac{\Pmax\etak}{\rhod}\right)^{\frac{1}{\alphak}},\label{cd:1}\\
&\textup{cond.} 2\mathrm{k}: \sqrt{\left(\frac{\Pmax\etak}{\rhod}\right)^{\frac{2}{\alphak}}\!-\!R_2^2}<h<\left(\frac{\Pmax\etak}{\rhod}\right)^{\frac{1}{\alphak}},\label{cd:2}\\
&\textup{cond.} 3\mathrm{k}: h\leqslant\sqrt{\left(\frac{\Pmax\etak}{\rhod}\right)^{\frac{2}{\alphak}}-R_2^2},\label{cd:3}
\end{align}
where $\mathrm{k}$ (in $\etak$ and $\alphak$) is $\mathrm{L}$ for LOS case or $\mathrm{N}$ for NLOS case. These conditions come from the fact that AsD will transmit with its maximum power regardless of where it is located inside the stadium, if the ABS is placed at a high enough altitude. On the other hand, the AsD will always be under full channel inversion if the altitude of the ABS is low.

\textit{\underline{SINR:}}
For the considered setup, the instantaneous uplink signal-to-interference-plus-noise ratio (SINR) at the TBS is given as

\setcounter{equation}{5}
\begin{align}\label{eq:sinrBS}
\textsf{SINR}^u_\mathrm{T}\!=\!\frac{P_\mathrm{TsUE} H^u_T d_T^{-\alphab}}{P_\mathrm{AsD} H^u_A d_A^{-\alphab}\!+\!\sigma^2}\!=\!\frac{\rhob H^u_T}{P_\mathrm{AsD} H^u_A d_A^{-\alphab}\!+\!\sigma^2},
\end{align}
where $P_\mathrm{TsUE}=\rhob d_T^{\alphab}$ is the TsUE transmit power. $H^u_T$ and $H^u_A$ are the uplink fading power gain between the TsUE and the TBS and between the AsD and the TBS, respectively, which follow exponential distribution. $d_T$ and $d_A$ are the Euclidean distance between the TsUE and the TBS and between the AsD and the TBS, respectively. The transmit power of the AsD $P_\mathrm{AsD}$ is given in \eqref{eq:Pu}.

The instantaneous uplink SINR at the ABS is given as
\begin{align}\label{eq:sinrUAV}
\textsf{SINR}^u_\mathrm{A}=\frac{P_\mathrm{AsD} G^u_A PL_a(\Zd)}{P_\mathrm{TsUE} G^u_T PL_a(\Zc)+\sigma^2},
\end{align}
where $G^u_A$ and $G^u_T$ are the uplink fading power gain between the AsD and the ABS and between the TsUE and the ABS, respectively, which follow Gamma distribution. $\Zd$ and $\Zc$ are the Euclidean distance between the AsD and the ABS and between the TsUE and the ABS, respectively.

\subsection{Downlink Network Model}
Different from uplink transmission, the TBS and ABS are assumed to transmit at a constant power $\Pt$ and $\Pa$ respectively.

\textit{\underline{SINR:}}
For the considered setup, the instantaneous downlink SINR at the TsUE is given as
\begin{align}\label{eq:sinrTUE}
\textsf{SINR}^d_\mathrm{T}\!=\!\frac{\Pt H^d_T d_T^{-\alphab}}{\Pa G^d_T PL_a(\Zc)+\sigma^2},
\end{align}
where $H^d_T$ is the downlink fading power gain between the TsUE and the TBS, which follow exponential distribution and $G^d_T$ is the downlink fading power gain between the TsUE and the ABS, which follow Gamma distribution.

The instantaneous downlink SINR at the AsD is given as
\begin{align}\label{eq:sinrAUE}
\textsf{SINR}^d_\mathrm{A}=\frac{\Pa G^d_A PL_a(\Zd)}{\Pt H^d_A d_A^{-\alphab}+\sigma^2},
\end{align}
where $G^d_A$ is the downlink fading power gain between the AsD and the ABS, which follow Gamma distribution and $H^d_A$ is the downlink fading power gain between the AsD and the TBS, which follow exponential distribution.

\ifCLASSOPTIONonecolumn
\setcounter{equation}{10}
\begin{figure*}[t]
{\small
\begin{align}\label{eq:LIB}
\mathcal{L}_{\Ib^u}(s)\!=\!\begin{cases}
                \int_h^{\sqrt{h^2+R_2^2}}\int_{0}^{2\pi}\mathcal{L}_{\Ib^u}(s,\Pmax|\theta,z)\frac{1}{2\pi}f_{\Zd}(z)\textup{d}\theta \textup{d}z, \quad\quad\quad\quad\quad\quad\quad\quad\quad\quad\quad\quad\quad \textup{cond.} 1\mathrm{L}\\
                \int_h^{(\frac{\Pmax\etaL}{\rhod})^\frac{1}{\alphaL}}\int_{0}^{2\pi}\pLOS\mathcal{L}_{\Ib^u}(s,\frac{\rhod}{\etaL} z^{\alphaL}|\theta,z)\frac{1}{2\pi}f_{\Zd}(z)\textup{d}\theta \textup{d}z+\int_{(\frac{\Pmax\etaL}{\rhod})^\frac{1}{\alphaL}}^{\sqrt{h^2+R_2^2}}\int_{0}^{2\pi}\pLOS\mathcal{L}_{\Ib^u}(s,\Pmax|\theta,z)\frac{1}{2\pi}f_{\Zd}(z)\textup{d}\theta \textup{d}z\\
                +\int_h^{\sqrt{h^2+R_2^2}}\int_{0}^{2\pi}\pNLOS\mathcal{L}_{\Ib^u}(s,\Pmax|\theta,z)\frac{1}{2\pi}f_{\Zd}(z)\textup{d}\theta \textup{d}z, \quad\quad\quad\quad\quad\quad\quad\quad\quad\quad\quad\textup{cond.} 4\;||\;\textup{cond.} 8\\
                \int_h^{(\frac{\Pmax\etaL}{\rhod})^\frac{1}{\alphaL}}\int_{0}^{2\pi}\pLOS\mathcal{L}_{\Ib^u}(s,\frac{\rhod}{\etaL} z^{\alphaL}|\theta,z)\frac{1}{2\pi}f_{\Zd}(z)\textup{d}\theta \textup{d}z+\int_{(\frac{\Pmax\etaL}{\rhod})^\frac{1}{\alphaL}}^{\sqrt{h^2+R_2^2}}\int_{0}^{2\pi}\pLOS\mathcal{L}_{\Ib^u}(s,\Pmax|\theta,z)\frac{1}{2\pi}f_{\Zd}(z)\textup{d}\theta \textup{d}z\\
                +\!\!\int_h^{(\frac{\Pmax\etaN}{\rhod})^\frac{1}{\alphaN}}\!\!\!\int_{0}^{2\pi}\!\pNLOS\mathcal{L}_{\Ib^u}(s,\frac{\rhod}{\etaN} z^{\alphaN}|\theta,z)\frac{1}{2\pi}f_{\Zd}(z)\textup{d}\theta \textup{d}z\!+\!\int_{(\frac{\Pmax\etaN}{\rhod})^\frac{1}{\alphaN}}^{\sqrt{h^2+R_2^2}}\!\!\!\int_{0}^{2\pi}\!\pNLOS\mathcal{L}_{\Ib^u}(s,\Pmax|\theta,z)\frac{1}{2\pi}f_{\Zd}(z)\textup{d}\theta \textup{d}z,\\
\quad\quad\quad\quad\quad\quad\quad\quad\quad\quad\quad\quad\quad\quad\quad\quad\quad\quad\quad\quad\quad\quad\quad\quad\quad\quad\quad\quad\quad\quad\quad\quad\quad\textup{cond.} 5\\
                \int_h^{\sqrt{h^2+R_2^2}}\int_{0}^{2\pi}\pLOS\mathcal{L}_{\Ib^u}(s,\frac{\rhod}{\etaL} z^{\alphaL}|\theta,z)\frac{1}{2\pi}f_{\Zd}(z)\textup{d}\theta \textup{d}z+\int_h^{\sqrt{h^2+R_2^2}}\int_{0}^{2\pi}\pNLOS\mathcal{L}_{\Ib^u}(s,\Pmax|\theta,z)\frac{1}{2\pi}f_{\Zd}(z)\textup{d}\theta \textup{d}z,\\
\quad\quad\quad\quad\quad\quad\quad\quad\quad\quad\quad\quad\quad\quad\quad\quad\quad\quad\quad\quad\quad\quad\quad\quad\quad\quad\quad\quad\quad\quad\quad\quad\quad\textup{cond.} 6\\
                \int_h^{\sqrt{h^2+R_2^2}}\int_{0}^{2\pi}\pLOS\mathcal{L}_{\Ib^u}(s,\frac{\rhod}{\etaL} z^{\alphaL}|\theta,z)\frac{1}{2\pi}f_{\Zd}(z)\textup{d}\theta \textup{d}z+\int_h^{(\frac{\Pmax\etaN}{\rhod})^\frac{1}{\alphaN}}\int_{0}^{2\pi}\pNLOS\mathcal{L}_{\Ib^u}(s,\frac{\rhod}{\etaN} z^{\alphaN}|\theta,z)\frac{1}{2\pi}f_{\Zd}(z)\textup{d}\theta \textup{d}z\\
                +\int_{(\frac{\Pmax\etaN}{\rhod})^\frac{1}{\alphaN}}^{\sqrt{h^2+R_2^2}}\int_{0}^{2\pi}\pNLOS\mathcal{L}_{\Ib^u}(s,\Pmax|\theta,z)\frac{1}{2\pi}f_{\Zd}(z)\textup{d}\theta \textup{d}z, \quad\quad\quad\quad\quad\quad\quad\quad\quad\quad\textup{cond.} 7\;||\;\textup{cond.} 9\\
                \int_h^{\sqrt{h^2+R_2^2}}\int_{0}^{2\pi}\pLOS\mathcal{L}_{\Ib^u}(s,\frac{\rhod}{\etaL} z^{\alphaL}|\theta,z)\frac{1}{2\pi}f_{\Zd}(z)\textup{d}\theta \textup{d}z+\int_h^{\sqrt{h^2+R_2^2}}\int_{0}^{2\pi}\pNLOS\mathcal{L}_{\Ib^u}(s,\frac{\rhod}{\etaN} z^{\alphaN}|\theta,z)\frac{1}{2\pi}f_{\Zd}(z)\textup{d}\theta \textup{d}z,\\
\quad\quad\quad\quad\quad\quad\quad\quad\quad\quad\quad\quad\quad\quad\quad\quad\quad\quad\quad\quad\quad\quad\quad\quad\quad\quad\quad\quad\quad\quad\quad\quad\quad\textup{cond.} 3\mathrm{N}
		\end{cases}.
\end{align}
}
\rule{18.2cm}{0.5pt}
\vspace{-10mm}
\end{figure*}
\else
\setcounter{equation}{10}
\begin{figure*}[t]
{\small
\begin{align}\label{eq:LIB}
\mathcal{L}_{\Ib^u}(s)\!=\!\begin{cases}
                \int_h^{\sqrt{h^2+R_2^2}}\int_{0}^{2\pi}\mathcal{L}_{\Ib^u}(s,\Pmax|\theta,z)\frac{1}{2\pi}f_{\Zd}(z)\textup{d}\theta \textup{d}z, \quad\quad\quad\quad\quad\quad\quad\quad\quad\quad\quad\quad\quad\quad\quad \textup{cond.} 1\mathrm{L}\\
                \int_h^{(\frac{\Pmax\etaL}{\rhod})^\frac{1}{\alphaL}}\int_{0}^{2\pi}\pLOS\mathcal{L}_{\Ib^u}(s,\frac{\rhod}{\etaL} z^{\alphaL}|\theta,z)\frac{1}{2\pi}f_{\Zd}(z)\textup{d}\theta \textup{d}z+\int_{(\frac{\Pmax\etaL}{\rhod})^\frac{1}{\alphaL}}^{\sqrt{h^2+R_2^2}}\int_{0}^{2\pi}\pLOS\mathcal{L}_{\Ib^u}(s,\Pmax|\theta,z)\frac{1}{2\pi}f_{\Zd}(z)\textup{d}\theta \textup{d}z\\
                +\int_h^{\sqrt{h^2+R_2^2}}\int_{0}^{2\pi}\pNLOS\mathcal{L}_{\Ib^u}(s,\Pmax|\theta,z)\frac{1}{2\pi}f_{\Zd}(z)\textup{d}\theta \textup{d}z, \quad\quad\quad\quad\quad\quad\quad\quad\quad\quad\quad\quad\quad\textup{cond.} 4\;||\;\textup{cond.} 8\\
                \int_h^{(\frac{\Pmax\etaL}{\rhod})^\frac{1}{\alphaL}}\int_{0}^{2\pi}\pLOS\mathcal{L}_{\Ib^u}(s,\frac{\rhod}{\etaL} z^{\alphaL}|\theta,z)\frac{1}{2\pi}f_{\Zd}(z)\textup{d}\theta \textup{d}z+\int_{(\frac{\Pmax\etaL}{\rhod})^\frac{1}{\alphaL}}^{\sqrt{h^2+R_2^2}}\int_{0}^{2\pi}\pLOS\mathcal{L}_{\Ib^u}(s,\Pmax|\theta,z)\frac{1}{2\pi}f_{\Zd}(z)\textup{d}\theta \textup{d}z\\
                +\!\!\int_h^{(\frac{\Pmax\etaN}{\rhod})^\frac{1}{\alphaN}}\!\!\!\int_{0}^{2\pi}\!\pNLOS\mathcal{L}_{\Ib^u}(s,\frac{\rhod}{\etaN} z^{\alphaN}|\theta,z)\frac{1}{2\pi}f_{\Zd}(z)\textup{d}\theta \textup{d}z\!+\!\int_{(\frac{\Pmax\etaN}{\rhod})^\frac{1}{\alphaN}}^{\sqrt{h^2+R_2^2}}\!\!\!\int_{0}^{2\pi}\!\pNLOS\mathcal{L}_{\Ib^u}(s,\Pmax|\theta,z)\frac{1}{2\pi}f_{\Zd}(z)\textup{d}\theta \textup{d}z,\\
\quad\quad\quad\quad\quad\quad\quad\quad\quad\quad\quad\quad\quad\quad\quad\quad\quad\quad\quad\quad\quad\quad\quad\quad\quad\quad\quad\quad\quad\quad\quad\quad\quad\quad\quad\textup{cond.} 5\\
                \int_h^{\sqrt{h^2+R_2^2}}\int_{0}^{2\pi}\pLOS\mathcal{L}_{\Ib^u}(s,\frac{\rhod}{\etaL} z^{\alphaL}|\theta,z)\frac{1}{2\pi}f_{\Zd}(z)\textup{d}\theta \textup{d}z+\int_h^{\sqrt{h^2+R_2^2}}\int_{0}^{2\pi}\pNLOS\mathcal{L}_{\Ib^u}(s,\Pmax|\theta,z)\frac{1}{2\pi}f_{\Zd}(z)\textup{d}\theta \textup{d}z,\\
\quad\quad\quad\quad\quad\quad\quad\quad\quad\quad\quad\quad\quad\quad\quad\quad\quad\quad\quad\quad\quad\quad\quad\quad\quad\quad\quad\quad\quad\quad\quad\quad\quad\quad\quad\textup{cond.} 6\\
                \int_h^{\sqrt{h^2+R_2^2}}\int_{0}^{2\pi}\pLOS\mathcal{L}_{\Ib^u}(s,\frac{\rhod}{\etaL} z^{\alphaL}|\theta,z)\frac{1}{2\pi}f_{\Zd}(z)\textup{d}\theta \textup{d}z+\int_h^{(\frac{\Pmax\etaN}{\rhod})^\frac{1}{\alphaN}}\int_{0}^{2\pi}\pNLOS\mathcal{L}_{\Ib^u}(s,\frac{\rhod}{\etaN} z^{\alphaN}|\theta,z)\frac{1}{2\pi}f_{\Zd}(z)\textup{d}\theta \textup{d}z\\
                +\int_{(\frac{\Pmax\etaN}{\rhod})^\frac{1}{\alphaN}}^{\sqrt{h^2+R_2^2}}\int_{0}^{2\pi}\pNLOS\mathcal{L}_{\Ib^u}(s,\Pmax|\theta,z)\frac{1}{2\pi}f_{\Zd}(z)\textup{d}\theta \textup{d}z, \quad\quad\quad\quad\quad\quad\quad\quad\quad\quad\quad\quad\textup{cond.} 7\;||\;\textup{cond.} 9\\
                \int_h^{\sqrt{h^2+R_2^2}}\int_{0}^{2\pi}\pLOS\mathcal{L}_{\Ib^u}(s,\frac{\rhod}{\etaL} z^{\alphaL}|\theta,z)\frac{1}{2\pi}f_{\Zd}(z)\textup{d}\theta \textup{d}z+\int_h^{\sqrt{h^2+R_2^2}}\int_{0}^{2\pi}\pNLOS\mathcal{L}_{\Ib^u}(s,\frac{\rhod}{\etaN} z^{\alphaN}|\theta,z)\frac{1}{2\pi}f_{\Zd}(z)\textup{d}\theta \textup{d}z,\\
\quad\quad\quad\quad\quad\quad\quad\quad\quad\quad\quad\quad\quad\quad\quad\quad\quad\quad\quad\quad\quad\quad\quad\quad\quad\quad\quad\quad\quad\quad\quad\quad\quad\quad\quad\textup{cond.} 3\mathrm{N}
		\end{cases}.
\end{align}
}
\rule{18.2cm}{0.5pt}
\vspace{-10mm}
\end{figure*}
\fi

\section{Uplink Coverage Probability}\label{sec:uplink}
In this section, we propose the analytical framework to compute the uplink performance by adopting coverage probability as the performance metric. The uplink coverage probability is formally defined as
\setcounter{equation}{9}
\begin{align}\label{eq:covUP}
\Pc^{u,b}\triangleq\Pr(\textsf{SINR}^u_b>\gamma^u_b),
\end{align}
where superscript $b$ is $\mathrm{T}$ for TBS and $\mathrm{A}$ for ABS, and $\gamma^u_b$ is the uplink SINR threshold. $\textsf{SINR}^u_\mathrm{T}$ and $\textsf{SINR}^u_\mathrm{A}$ can be found in \eqref{eq:sinrBS} and \eqref{eq:sinrUAV}, respectively. The results for the uplink coverage probability of the TBS and the ABS are presented in the next two subsections.

\subsection{TBS Uplink Coverage Probability}
First we present two lemmas, which are used in deriving the coverage probability of the TBS in Theorem~\ref{th:covBS}.

\begin{lemma}\label{le:LIB}
The Laplace transform of the interference power distribution at the TBS is given as \eqref{eq:LIB} at the top of this page,
where
\setcounter{equation}{11}
\begin{align}
\mathcal{L}_{\Ib^u}(s,p|\theta,z)\!=\!\frac{1}{1\!+\!\frac{sp}{\left(z^2\!-\!h^2\!+\!d^2\!-\!2\sqrt{z^2\!-\!h^2}d\cos(\theta)\right)^{\frac{\alphab}{2}}}},
\end{align}
\end{lemma}
$\textup{conds.} 1\mathrm{k}$ and $3\mathrm{k}$ is given in~\eqref{cd:1} and \eqref{cd:3} respectively and the other conditions are given below:
\begin{align}
&\textup{cond.} 4: \sqrt{\!\left(\!\frac{\Pmax\etaL}{\rhod}\!\right)^{\!\frac{2}{\alphaL}}\!\!\!\!\!-\!\!R_2^2}\!<\!\!\left(\!\frac{\Pmax\etaN}{\rhod}\!\right)^{\!\frac{1}{\alphaN}}\!\!\!\!<\!h\!<\!\!\left(\!\frac{\Pmax\etaL}{\rhod}\!\right)^{\!\frac{1}{\alphaL}},\label{cd:4}\\
&\textup{cond.} 5: \sqrt{\left(\frac{\Pmax\etaL}{\rhod}\right)^{\frac{2}{\alphaL}}-R_2^2}<h\leqslant\left(\frac{\Pmax\etaN}{\rhod}\right)^{\frac{1}{\alphaN}},\label{cd:5}\\
&\textup{cond.} 6: \left(\frac{\Pmax\etaN}{\rhod}\right)^{\frac{1}{\alphaN}}\leqslant h<\sqrt{\left(\frac{\Pmax\etaL}{\rhod}\right)^{\frac{2}{\alphaL}}-R_2^2},\label{cd:6}
\end{align}
\begin{align}
&\textup{cond.} 7: \!\!\sqrt{\!\!\left(\!\frac{\Pmax\etaN}{\rhod}\!\right)^{\!\!\frac{2}{\alphaN}}\!\!\!\!\!\!-\!\!R_2^2}\!<\!h\!\leqslant\!\!\sqrt{\!\!\left(\!\frac{\Pmax\etaL}{\rhod}\!\right)^{\!\!\frac{2}{\alphaL}}\!\!\!\!\!\!-\!\!R_2^2}\!<\!\!\left(\!\frac{\Pmax\etaN}{\rhod}\!\right)^{\!\!\frac{1}{\alphaN}}\!\!\!\!\!,\label{cd:7}\\
&\textup{cond.} 8: \left(\!\frac{\Pmax\etaN}{\rhod}\!\right)^{\!\frac{1}{\alphaN}}\!\!<\!\!\sqrt{\left(\!\frac{\Pmax\etaL}{\rhod}\!\right)^{\!\frac{2}{\alphaL}}-R_2^2}\!\leqslant\!h\!<\!\left(\!\frac{\Pmax\!\etaL}{\rhod}\!\right)^{\!\frac{1}{\alphaL}},\label{cd:8}\\
&\textup{cond.} 9: \sqrt{\!\left(\!\frac{\Pmax\etaN}{\rhod}\!\right)^{\!\frac{2}{\alphaN}}\!\!\!\!\!-\!\!R_2^2}\!\!<\!h\!<\!\!\left(\!\frac{\Pmax\etaN}{\rhod}\!\right)^{\!\frac{1}{\alphaN}}\!\!\!\!\!<\!\!\sqrt{\!\left(\!\frac{\Pmax\etaL}{\rhod}\!\right)^{\!\frac{2}{\alphaL}}\!\!\!\!\!\!-\!R_2^2}\label{cd:9}.
\end{align}
These conditions come from the fact that as the height of the ABS increases, the AsD is first under full channel inversion and then reaches its maximum power constraint. Depending on whether $\left(\frac{\Pmax\etaN}{\rhod}\right)^{\frac{1}{\alphaN}}$ is greater or smaller than $\sqrt{\left(\frac{\Pmax\etaL}{\rhod}\right)^{\frac{2}{\alphaL}}-R_2^2}$, we can further specify conds. 1N, 2L, 2N and 3L as the conditions above.
\begin{IEEEproof}
See Appendix~\ref{ap:LIB}.
\end{IEEEproof}

From Lemma~\ref{le:LIB}, we can see that the transmit power of the AsD $P_\textrm{AsD}$ and the distance between the AsD and the TBS $d_A$ are related to the distance between the AsD and the ABS $\Zd$. This important distance distribution is presented in the following lemma.
\begin{lemma}\label{le:disUU}
The pdf of the distance $\Zd$ between the ABS at height $h$ above the center of $\mathcal{S}_2$ and an i.u.d. AsD inside $\mathcal{S}_2$ is
\begin{align}\label{eq:fZd}
f_{\Zd}(z)=\frac{2z}{R_2^2}, \;\;h\leqslant z\leqslant\sqrt{R_2^2+h^2}.
\end{align}
\end{lemma}
\begin{IEEEproof}
See Appendix~\ref{ap:disUU}.
\end{IEEEproof}

\setcounter{equation}{25}
\begin{figure*}[t]
{\small
\begin{align}\label{eq:covD}
\Pc^{u,\mathrm{A}}\!\!=\!\!\begin{cases}
                \int_h^{\sqrt{h^2\!+R_2^2}}\!\Pc^{u,\mathrm{L}}(\Pmax|z)f_{\Zd}(z)\textup{d}z+\int_h^{\sqrt{h^2\!+R_2^2}}\!\Pc^{u,\mathrm{N}}(\Pmax|z)f_{\Zd}(z)\textup{d}z,&\textup{cond.} 1\mathrm{L}\\
                \int_h^{(\frac{\Pmax\etaL}{\rhod})^{\frac{1}{\alphaL}}}\!\Pc^{u,\mathrm{L}}(\frac{\rhod}{\etaL} z^{\alphaL}|z)f_{\Zd}(z)\textup{d}z+\int_{(\frac{\Pmax}{\rhod})^{\frac{1}{\alphaL}}}^{\sqrt{h^2\!+R_2^2}}\!\Pc^{u,\mathrm{L}}(\Pmax|z)f_{\Zd}(z)\textup{d}z\\
                +\int_h^{\sqrt{h^2\!+R_2^2}}\!\Pc^{u,\mathrm{N}}(\Pmax|z)f_{\Zd}(z)\textup{d}z,&\textup{cond.} 4\;||\;\textup{cond.} 8\\
                \int_h^{(\frac{\Pmax\etaL}{\rhod})^{\frac{1}{\alphaL}}}\!\Pc^{u,\mathrm{L}}(\frac{\rhod}{\etaL} z^{\alphaL}|z)f_{\Zd}(z)\textup{d}z+\int_{(\frac{\Pmax}{\rhod})^{\frac{1}{\alphaL}}}^{\sqrt{h^2\!+R_2^2}}\!\Pc^{u,\mathrm{L}}(\Pmax|z)f_{\Zd}(z)\textup{d}z\\
                +\int_h^{(\frac{\Pmax\etaN}{\rhod})^{\frac{1}{\alphaN}}}\!\Pc^{u,\mathrm{N}}(\frac{\rhod}{\etaN} z^{\alphaN}|z)f_{\Zd}(z)\textup{d}z+\int_{(\frac{\Pmax}{\rhod})^{\frac{1}{\alphaN}}}^{\sqrt{h^2\!+R_2^2}}\!\Pc^{u,\mathrm{N}}(\Pmax|z)f_{\Zd}(z)\textup{d}z,&\textup{cond.} 5\\
                \int_h^{\sqrt{h^2\!+R_2^2}}\!\Pc^{u,\mathrm{L}}(\frac{\rhod}{\etaL}z^{\alphaL}|z)f_{\Zd}(z)\textup{d}z+\int_h^{\sqrt{h^2\!+R_2^2}}\!\Pc^{u,\mathrm{N}}(\Pmax|z)f_{\Zd}(z)\textup{d}z,&\textup{cond.} 6\\
                \int_h^{\sqrt{h^2\!+R_2^2}}\!\Pc^{u,\mathrm{L}}(\frac{\rhod}{\etaL}z^{\alphaL}|z)f_{\Zd}(z)\textup{d}z+\int_h^{(\frac{\Pmax\etaN}{\rhod})^{\frac{1}{\alphaN}}}\!\Pc^{u,\mathrm{N}}(\frac{\rhod}{\etaN} z^{\alphaN}|z)f_{\Zd}(z)\textup{d}z\\
                +\int_{(\frac{\Pmax}{\rhod})^{\frac{1}{\alphaN}}}^{\sqrt{h^2\!+R_2^2}}\!\Pc^{u,\mathrm{N}}(\Pmax|z)f_{\Zd}(z)\textup{d}z,&\textup{cond.} 7\;||\;\textup{cond.} 9\\
                \int_h^{\sqrt{h^2\!+R_2^2}}\!\Pc^{u,\mathrm{L}}(\frac{\rhod}{\etaL}z^{\alphaL}|z)f_{\Zd}(z)\textup{d}z+\int_h^{\sqrt{h^2\!+R_2^2}}\!\Pc^{u,\mathrm{N}}(\frac{\rhod}{\etaN}z^{\alphaN}|z)f_{\Zd}(z)\textup{d}z,&\textup{cond.} 3\textrm{N}\\
		\end{cases}.
\end{align}
}
\rule{18.2cm}{0.5pt}
\vspace{-8mm}
\end{figure*}

\begin{theorem}\label{th:covBS}
Based on the system model in Section~\ref{sec:systemmodel}, the uplink coverage probability of the TBS is
\setcounter{equation}{19}
\begin{align}\label{eq:covB}
\Pc^{u,\mathrm{T}}=\exp\left(-\frac{\gammab^u}{\rhob}\sigma^2\right)\mathcal{L}_{\Ib^u}(s),
\end{align}
where $\Ib^u=P_\text{AsD} H^u_A d_A^{-\alphab}$, $s=\frac{\gammab^u}{\rhob}$, and $\mathcal{L}_{\Ib^u}(s)$ is given by Lemma~\ref{le:LIB}.
\end{theorem}
\begin{IEEEproof}
From \eqref{eq:sinrBS} and \eqref{eq:covUP}, we can have
\begin{subequations}
\begin{align}
\Pc^{u,\mathrm{T}}=&\Pr(\textsf{SINR}^u_\mathrm{T}>\gammab^u)\!=\!\Pr\!\left(\!\frac{\rhob H^u_T}{P_\mathrm{AsD} H^u_A d_A^{-\alphab}\!+\!\sigma^2}\!>\!\gammab^u\!\right)\nonumber\\
=&\Pr\left(H^u_T>\frac{\gammab^u}{\rhob}(P_\mathrm{AsD} H^u_Ad_A^{-\alphab}+\sigma^2)\right)\nonumber\\
=&\exp\left(-\frac{\gammab^u}{\rhob}(P_\mathrm{AsD} H^u_Ad_A^{-\alphab}+\sigma^2)\right)\label{eq:covB1}\\
=&\exp\left(-\frac{\gammab^u}{\rhob}\sigma^2\right)\exp\left(-\frac{\gammab^u}{\rhob}P_\mathrm{AsD} H^u_Ad_A^{-\alphab}\right),\label{eq:covB2}
\end{align}
\end{subequations}
where \eqref{eq:covB1} follows from the fact that the link between the TsUE and the TBS experiences Rayleigh fading with a pdf of $f_{H^u_T}(h)=\exp(-h)$. Letting $\Ib^u=P_\text{AsD} H^u_A d_A^{-\alphab}$ and $s=\frac{\gammab^u}{\rhob}$ in \eqref{eq:covB2}, we can arrive at Theorem~\ref{th:covBS}.
\end{IEEEproof}

Substituting \eqref{eq:LIB} and \eqref{eq:fZd} into \eqref{eq:covB}, we can obtain the uplink coverage probability of the TBS.

\subsection{ABS Uplink Coverage Probability}
We begin by presenting three lemmas, which will then be used to compute the uplink coverage probability of the ABS in Theorem~\ref{th:covD}.

\begin{lemma}\label{le:LIU}
The Laplace transform of the interference power distribution at the ABS is
\ifCLASSOPTIONonecolumn
\setcounter{equation}{21}
\begin{align}
&\mathcal{L}_{\Id^u}\!(s)\!=\!\!\!\int_{\sqrt{R_2^2+h^2}}^{\sqrt{(R_1-d)^2+h^2}}\!\!\int_{0}^{2\pi}\!\!\!\!\mathcal{L}_{\Id^u}(s|\omega,z)f_\Omega(\omega|z)f_{\Zc}(z)\textup{d}\omega \textup{d}z\!\!+\!\!\int_{\sqrt{(R_1-d)^2+h^2}}^{\sqrt{(R_1+d)^2+h^2}}\int_{-\widehat{\omega}}^{\widehat{\omega}}\mathcal{L}_{\Id^u}(s|\omega,z)f_\Omega(\omega|z)f_{\Zc}(z)\textup{d}\omega \textup{d}z,
\end{align}
\else
\setcounter{equation}{21}
\begin{align}
&\mathcal{L}_{\Id^u}\!(s)\!=\!\!\!\int_{\sqrt{R_2^2+h^2}}^{\sqrt{(R_1-d)^2+h^2}}\!\!\int_{0}^{2\pi}\!\!\!\!\mathcal{L}_{\Id^u}(s|\omega,z)f_\Omega(\omega|z)f_{\Zc}(z)\textup{d}\omega \textup{d}z\nonumber\\
&+\!\!\int_{\sqrt{(R_1-d)^2+h^2}}^{\sqrt{(R_1+d)^2+h^2}}\!\!\!\int_{-\widehat{\omega}}^{\widehat{\omega}}\mathcal{L}_{\Id^u}(s|\omega,z)f_\Omega(\omega|z)f_{\Zc}(z)\textup{d}\omega \textup{d}z,
\end{align}
\fi
where
\ifCLASSOPTIONonecolumn
\begin{align}
&\mathcal{L}_{\Id^u}(s|\omega,z)=\pLOS\mL^{\mL}\left(\mL+s\rhob z^{-\alphaL}\etaL\left(z^2\!-\!h^2\!+\!d^2\!-\!2\sqrt{z^2\!-\!h^2}d\cos(\omega)\right)^{\frac{\alphaL}{2}}\right)^{-\mL}\nonumber\\
&+\pNLOS\mN^{\mN}\left(\mN+s\rhob z^{-\alphaN}\etaN\left(z^2\!-\!h^2\!+\!d^2\!-\!2\sqrt{z^2\!-\!h^2}d\cos(\omega)\right)^{\frac{\alphaN}{2}}\right)^{-\mN},
\end{align}
\else
\begin{align}
&\mathcal{L}_{\Id^u}(s|\omega,z)=\frac{\pLOS\mL^{\mL}}{\left(\mL+\frac{s\rhob\etaL\left(z^2\!-\!h^2\!+\!d^2\!-\!2\sqrt{z^2\!-\!h^2}d\cos(\omega)\right)^{\frac{\alphaL}{2}}}{z^{\alphaL}}\right)^{\mL}}\nonumber\\
&+\frac{\pNLOS\mN^{\mN}}{\left(\mN+\frac{s\rhob\etaN\left(z^2\!-\!h^2\!+\!d^2\!-\!2\sqrt{z^2\!-\!h^2}d\cos(\omega)\right)^{\frac{\alphaN}{2}}}{z^{\alphaN}}\right)^{\mN}},
\end{align}
\fi
and $\widehat{\omega}=\mathrm{arcsec}\left(\frac{2d\sqrt{z^2-h^2}}{d^2+z^2-h^2-R_1^2}\right)$.
\end{lemma}
\begin{IEEEproof}
See Appendix~\ref{ap:LIU}.
\end{IEEEproof}

The pdf of the distance between the TsUE and the ABS $f_{\Zc}(z)$, and the conditional pdf of the angle, $f_\Omega(\omega|z)$, between the ground projection of $\Zc$ and $d_T$ are given in Lemma~\ref{le:disCU} and Lemma~\ref{le:degCU}, respectively.

\begin{lemma}\label{le:disCU}
The pdf of the distance $\Zc$ between the ABS at height $h$ above the center of $\mathcal{S}_2$ and an i.u.d. TsUE inside $\mathcal{S}_1\setminus\mathcal{S}_2$ is
\begin{align}\label{eq:disCU}
f_{\Zc}\!(\!z\!)\!=\!\!\begin{cases}\frac{2z}{R_1^2\!-\!R_2^2}, &\sqrt{R_2^2\!+\!h^2}\!\leqslant\! z\!\leqslant\!\sqrt{(R_1\!-\!d)^2\!+\!h^2}\\
            \frac{2z\widehat{\omega}}{\pi (R_1^2\!-\!R_2^2)}, &\sqrt{(R_1\!-\!d)^2\!+\!h^2}\!<\! z\!\leqslant\!\!\sqrt{(R_1\!+\!d)^2\!+\!h^2}
\end{cases}.
\end{align}
\end{lemma}
\begin{IEEEproof}
See Appendix~\ref{ap:disCU}.
\end{IEEEproof}

\begin{lemma}\label{le:degCU}
The pdf of the angle, $f_\Omega(\omega|z)$, between the ground projection of $\Zc$ and $d_T$ conditioned on $\Zc$ is
\begin{align}\label{eq:degCU}
f_\Omega(\omega|z)\!=\!\begin{cases}\frac{1}{2\pi}, &\sqrt{R_2^2\!+\!h^2}\!\leqslant\! z\!\leqslant\!\sqrt{(R_1\!-\!d)^2\!+\!h^2}\\
            \frac{1}{2\widehat{\omega}}, &\sqrt{(R_1\!-\!d)^2\!+\!h^2}\!<\! z\!\leqslant\!\sqrt{(R_1\!+\!d)^2\!+\!h^2}
\end{cases}.
\end{align}
\end{lemma}
\begin{IEEEproof}
This lemma can be proved by using cosine rule and simple trigonometry.
\end{IEEEproof}

\begin{theorem}\label{th:covD}
Based on the system model in Section~\ref{sec:systemmodel}, the uplink coverage probability of the ABS is given as \eqref{eq:covD} at the top of this page, where

\ifCLASSOPTIONonecolumn
\setcounter{equation}{26}
\begin{align}
\Pc^{u,\mathrm{L}}(p|z)=&\sum_{n=0}^{\mL-1}\frac{(-s_1)^n}{n!}\exp(-s_1\sigma^2)\sum_{k=0}^n\binom{n}{k}(-\sigma^2)^{n-k}\frac{\textup{d}^k}{\textup{d}s_1^k}\mathcal{L}_{\Id^u}(s_1)\pLOS,
\end{align}
\else
\setcounter{equation}{26}
\begin{align}
\Pc^{u,\mathrm{L}}(p|z)=&\sum_{n=0}^{\mL-1}\frac{(-s_1)^n}{n!}\exp(-s_1\sigma^2)\nonumber\\
&\times\sum_{k=0}^n\binom{n}{k}(-\sigma^2)^{n-k}\frac{\textup{d}^k}{\textup{d}s_1^k}\mathcal{L}_{\Id^u}(s_1)\pLOS,
\end{align}
\fi
$s_1=\frac{\mL\gammad^u z^{\alphaL}}{\etaL p}$ and $\mathcal{L}_{\Id^u}(s_1)$ is given by Lemma~\ref{le:LIU} and
\ifCLASSOPTIONonecolumn
\begin{align}
\Pc^{u,\mathrm{N}}(p|z)=&\sum_{n=0}^{\mN-1}\frac{(-s_2)^n}{n!}\exp(-s_2\sigma^2)\sum_{k=0}^n\binom{n}{k}(-\sigma^2)^{n-k}\frac{\textup{d}^k}{\textup{d}s_2^k}\mathcal{L}_{\Id^u}(s_2)\pNLOS,
\end{align}
\else
\begin{align}
\Pc^{u,\mathrm{N}}(p|z)=&\sum_{n=0}^{\mN-1}\frac{(-s_2)^n}{n!}\exp(-s_2\sigma^2)\nonumber\\
&\times\sum_{k=0}^n\binom{n}{k}(-\sigma^2)^{n-k}\frac{\textup{d}^k}{\textup{d}s_2^k}\mathcal{L}_{\Id^u}(s_2)\pNLOS,
\end{align}
\fi
$s_2=\frac{\mN\gammad^u z^{\alphaN}}{\etaN p}$ and $\mathcal{L}_{\Id^u}(s_2)$ is given by Lemma~\ref{le:LIU}. The pdf of the distance between the AsD and the ABS $f_{\Zd}(z)$ is provided in Lemma~\ref{le:disUU}. $\textup{conds.} 1\textrm{k}$, $3\textrm{k}$, 4--9 are given in \eqref{cd:1}, \eqref{cd:3} and \eqref{cd:4}--\eqref{cd:9}. These conditions come from the fact that the AsD is first under full channel inversion and then transmits with its maximum power as the height of the ABS increases. Note that we can further specify conds. 1N, 2L, 2N and 3L as conds. 4--9 depending on whether $\left(\frac{\Pmax\etaN}{\rhod}\right)^{\frac{1}{\alphaN}}$ is larger or smaller than $\sqrt{\left(\frac{\Pmax\etaL}{\rhod}\right)^{\frac{2}{\alphaL}}-R_2^2}$.
\end{theorem}
\begin{IEEEproof}
See Appendix~\ref{ap:covD}.
\end{IEEEproof}

Combining Lemma~\ref{le:LIU},~\ref{le:disCU}, and~\ref{le:degCU} with Theorem~\ref{th:covD}, we can calculate the uplink coverage probability of the ABS.

\section{Downlink Coverage Probability}\label{sec:downlink}
In this section, we present the analytical framework to analyze the performance metric, the downlink coverage probability, which is formally defined as
\begin{align}
\Pc^{d,b}\triangleq\Pr(\textsf{SINR}^d_b>\gamma^d_b),
\end{align}
where superscript $b$ is $\mathrm{T}$ for TsUE and $\mathrm{A}$ for AsD, and $\gamma^d_b$ is the downlink SINR threshold. $\textsf{SINR}^d_\mathrm{T}$ and $\textsf{SINR}^d_\mathrm{A}$ can be found in \eqref{eq:sinrTUE} and \eqref{eq:sinrAUE}, respectively. The next two subsections investigate the downlink coverage probability of the TsUE and the AsD.

\subsection{TsUE Downlink Coverage Probability}
First we present a lemma, which is used in deriving the coverage probability of the TsUE in Theorem~\ref{th:covTUE}
\begin{lemma}\label{le:LITUE}
The conditional Laplace transform of the interference power distribution at the TsUE is
\ifCLASSOPTIONonecolumn
\begin{align}
&\mathcal{L}_{\Ib^d}(s|z)=\pLOS\mL^{\mL}\left(\mL+s\Pa\etaL z^{-\alphaL}\right)^{-\mL}+\pNLOS\mN^{\mN}\left(\mN+s\Pa\etaN z^{-\alphaN}\right)^{-\mN}.
\end{align}
\else
\begin{align}
\mathcal{L}_{\Ib^d}(s|z)=&\pLOS\mL^{\mL}\left(\mL+s\Pa\etaL z^{-\alphaL}\right)^{-\mL}\nonumber\\
&+\pNLOS\mN^{\mN}\left(\mN+s\Pa\etaN z^{-\alphaN}\right)^{-\mN}.
\end{align}
\fi
\end{lemma}
\begin{IEEEproof}
The proof follows the same lines as Lemma~\ref{le:LIU} and is skipped for the sake of brevity.
\end{IEEEproof}

\begin{theorem}\label{th:covTUE}
Based on the system model in Section~\ref{sec:systemmodel}, the downlink coverage probability of the TsUE is given as
\begin{align}
&\Pc^{d,\mathrm{T}}\!\!=\!\!\!\int_{\sqrt{R_2^2+h^2}}^{\sqrt{(R_1\!-\!d)^2\!+\!h^2}}\!\!\!\!\int_{0}^{2\pi}\!\!\!\!\mathcal{L}_{\Ib^d}\!(s|z)\exp(\!-s\sigma^2)f_\Omega(\omega|z)f_{\Zc}\!(z)\textup{d}\omega \textup{d}z\nonumber\\
&+\!\!\!\int_{\sqrt{(R_1\!-\!d)^2\!+\!h^2}}^{\sqrt{(R_1\!+\!d)^2\!+\!h^2}}\!\!\!\int_{-\widehat{\omega}}^{\widehat{\omega}}\mathcal{L}_{\Ib^d}(s|z)\exp(\!-s\sigma^2)f_\Omega(\omega|z)f_{\Zc}(z)\textup{d}\omega \textup{d}z,
\end{align}
where $s=\frac{\gammab^d}{\Pt}\left(z^2\!-\!h^2\!+\!d^2\!-\!2\sqrt{z^2\!-\!h^2}d\cos(\omega)\right)^{\frac{\alphab}{2}}$, $\mathcal{L}_{\Ib^d}(s|z)$ is given by Lemma~\ref{le:LITUE}, $f_\Omega(\omega|z)$ is given by Lemma~\ref{le:degCU} and $f_{\Zc}(z)$ is given by Lemma~\ref{le:disCU}.
\end{theorem}
\begin{IEEEproof}
Using the fact that the downlink fading power gain between the TBS and the TsUE $H^d_T$ follows exponential distribution with unit mean and cosine rule, we can derive the TsUE downlink coverage probability.
\end{IEEEproof}

\subsection{AsD Downlink Coverage Probability}
First we present a lemma, which is used in deriving the coverage probability of the AsD in Theorem~\ref{th:covAUE}
\begin{lemma}\label{le:LIAUE}
The conditional Laplace transform of the interference power distribution at the AsD is
\begin{align}
&\mathcal{L}_{\Id^d}(s|z,\theta)=\frac{1}{1+\frac{s\Pt}{\left(z^2\!-\!h^2\!+\!d^2\!-\!2\sqrt{z^2\!-\!h^2}d\cos(\theta)\right)^{\frac{\alphab}{2}}}}.
\end{align}
\end{lemma}
\begin{IEEEproof}
The proof follows the same lines as Lemma~\ref{le:LIB} and is skipped for the sake of brevity.
\end{IEEEproof}

\begin{theorem}\label{th:covAUE}
Based on the system model in Section~\ref{sec:systemmodel}, the downlink coverage probability of the AsD is given as
\ifCLASSOPTIONonecolumn
\begin{align}
\Pc^{d,\mathrm{A}}&=\!\int^{\sqrt{R_2^2+h^2}}_h\!\!\int_{0}^{2\pi}\sum_{n=0}^{\mL-1}\frac{(-s_1)^n}{n!}\exp(-s_1\sigma^2)\sum_{k=0}^n\binom{n}{k}(-\sigma^2)^{n-k}
\frac{\textup{d}^k}{\textup{d}s_1^k}\mathcal{L}_{\Id^d}(s_1|z,\theta)\pLOS\frac{1}{2\pi}f_{\Zd}(z)\textup{d}\theta \textup{d}z\nonumber\\
&+\int^{\sqrt{R_2^2+h^2}}_h\!\!\int_{0}^{2\pi}\sum_{n=0}^{\mN-1}\frac{(-s_2)^n}{n!}\exp(-s_2\sigma^2)\sum_{k=0}^n\binom{n}{k}(-\sigma^2)^{n-k}
\frac{\textup{d}^k}{\textup{d}s_2^k}\mathcal{L}_{\Id^d}(s_2|z,\theta)\pNLOS\frac{1}{2\pi}f_{\Zd}(z)\textup{d}\theta \textup{d}z,
\end{align}
\else
\begin{align}
&\Pc^{d,\mathrm{A}}=\!\int^{\sqrt{R_2^2+h^2}}_h\!\!\int_{0}^{2\pi}\sum_{n=0}^{\mL-1}\frac{(-s_1)^n}{n!}\exp(-s_1\sigma^2)\nonumber\\
&\times\sum_{k=0}^n\binom{n}{k}(-\sigma^2)^{n-k}\frac{\textup{d}^k}{\textup{d}s_1^k}\mathcal{L}_{\Id^d}(s_1|z,\theta)\frac{\pLOS}{2\pi}f_{\Zd}(z)\textup{d}\theta \textup{d}z\nonumber\\
&+\int^{\sqrt{R_2^2+h^2}}_h\!\!\int_{0}^{2\pi}\sum_{n=0}^{\mN-1}\frac{(-s_2)^n}{n!}\exp(-s_2\sigma^2)\nonumber\\
&\times\!\!\sum_{k=0}^n\binom{n}{k}\!(-\sigma^2)^{n-k}\frac{\textup{d}^k}{\textup{d}s_2^k}\mathcal{L}_{\Id^d}(s_2|z,\theta)\frac{\pNLOS}{2\pi}f_{\Zd}(z)\textup{d}\theta \textup{d}z,
\end{align}
\fi
where $s_1=\frac{\mL\gammad^d z^{\alphaL}}{\Pa\etaL}$, $s_2=\frac{\mN\gammad^d z^{\alphaN}}{\Pa\etaN}$, $\mathcal{L}_{\Id^d}(s|z,\theta)$ is given by Lemma~\ref{le:LIAUE} and $f_{\Zd}(z)$ is given by Lemma~\ref{le:disUU}.
\end{theorem}
\begin{IEEEproof}
The proof follows the same lines as Theorem~\ref{th:covD} and is skipped for the sake of brevity.
\end{IEEEproof}

\section{Results}\label{sec:result}
In this section, we first validate the analytical results and then discuss the design insights of an underlay drone system for IoT devices inside a stadium. The simulation results are generated using Matlab by averaging over $10^7$ Monte Carlo simulation runs. Similar to \cite{Galkin-2017,Azari-2017b,Fotouhi-2017}, we set the path-loss exponents of LOS and NLOS aerial links as 2.5 and 4 respectively. Unless stated otherwise, the values of the parameters summarized in Table~\ref{tab:values} are used.

\ifCLASSOPTIONpeerreview
\begin{table}
\centering
\caption{Parameter Values.}
\label{tab:values}
\begin{tabular}{|c|c||c|c||c|c||c|c|}
\hline
Parameter & Value & Parameter & Value & Parameter & Value & Parameter & Value\\ \hline
$R_1$ & 500 m & $\alphab$ & 4 & $\rhob$ & -75 dBm & $\Pa$ & 20 dBm\\ \hline
$R_2$ & 100 m & $\alphaL$ & 2.5 & $\rhod$ & -50 dBm & $\Pt$ & 40 dBm\\ \hline
$d$ & 200 m & $\alphaN$ & 4 & $\Pmax$ & 20 dBm & $\gammad^d$ & 0 dB\\ \hline
$\mL$ & 5 & $\etaL$ & 0 dB & $\gammad^u$ & 0 dB & $\gammab^d$ & 0 dB\\ \hline
$\mN$ & 1 & $\etaN$ & -20 dB & $\gammab^u$ & 0 dB & $\sigma^2$ & -100 dBm\\ \hline
\end{tabular}
\end{table}
\else
\begin{table}
\centering
\caption{Parameter Values.}
\label{tab:values}
\begin{tabular}{|c|c||c|c|}
\hline
Parameter & Value & Parameter & Value\\ \hline
$R_1$ & 500 m & $\rhob$ & -75 dBm\\ \hline
$R_2$ & 100 m & $\rhod$ & -50 dBm\\ \hline
$d$ & 200 m &$\Pmax$ & 20 dBm\\ \hline
$\mL$ & 5 & $\gammad^u$ & 0 dB\\ \hline
$\mN$ & 1 & $\gammab^u$ & 0 dB\\ \hline
$\alphab$ & 4 & $\Pa$ & 20 dBm\\ \hline
$\alphaL$ & 2.5 & $\Pt$ & 40 dBm\\ \hline
$\alphaN$ & 4 & $\gammad^d$ & 0 dB\\ \hline
$\etaL$ & 0 dB & $\gammab^d$ & 0 dB\\ \hline
$\etaN$ & -20 dB & $\sigma^2$ & -100 dBm\\ \hline
\end{tabular}
\end{table}
\fi

\ifCLASSOPTIONpeerreview
\begin{table}
\centering
\caption{Aerial Channel Model Parameter Values~\cite{Hourani-2014b,Yaliniz-2016b}.}
\label{tab:pLOS}
\begin{tabular}{|c|c|c|c|c|}
\hline
Model 1 Environment & Suburban & Urban & Dense Urban & High-rise Urban\\ \hline
Parameter $(C_1,B_1)$ & (4.88, 0.43) & (9.6117, 0.1581) & (11.95, 0.136) & (27.23, 0.08)\\ \hline
\end{tabular}
\end{table}
\else
\begin{table}
\centering
\caption{Aerial Channel Model Parameter Values~\cite{Hourani-2014b,Yaliniz-2016b}.}
\label{tab:pLOS}
\begin{tabular}{|c|c|}
\hline
Model 1 Environment & Parameter $(C_1,B_1)$\\ \hline
Suburban & (4.88, 0.43)\\ \hline
Urban & (9.6117, 0.1581)\\ \hline
Dense Urban & (11.95, 0.136)\\ \hline
High-rise Urban & (27.23, 0.08)\\ \hline
\end{tabular}
\end{table}
\fi

\subsection{Aerial Channel Model Parameter Values}
Before presenting results to validate our analytical model, we first discuss about the ariel channel model. The probabilities of LOS and NLOS are functions of the environment, density and height of buildings, altitude of the drone and elevation angle between the drone and the devices on ground. There are two models commonly used in literature~\cite{Hourani-2014a,Hourani-2014b}, which are both based on the statistical parameters provided by the ITU-R.

\underline{Model 1}: The LOS probability is given by
\begin{align}
\pLOS=\frac{1}{1+C_1\exp\left(-B_1\left[\frac{180}{\pi}\sin^{-1}\left(\frac{h}{z}\right)-C_1\right]\right)}.
\end{align}
The NLOS probability is
\begin{align}
\pNLOS=1-\pLOS,
\end{align}
where $C_1$ and $B_1$ are constant values that depend on the environment (suburban, urban, dense urban, high-rise urban) and typical values are listed in Table~\ref{tab:pLOS}.

\underline{Model 2}: The LOS probability is expressed as
\begin{align}
\pLOS=C_2\left(\frac{180}{\pi}\sin^{-1}\left(\frac{h}{z}\right)-15\right)^{B_2}.
\end{align}
The NLOS probability is
\begin{align}
\pNLOS=1-\pLOS,
\end{align}
where $C_2$ and $B_2$ are environment dependent parameters. $C_2=0.6$ and $B_2=0.11$ for 2 GHz signal transmission in an urban environment~\cite{Hourani-2014a}.

\begin{remark}
Both models above are focusing on the lower stratosphere (for drone altitude between 200 m and 3000 m). A broader aerial communication model that can fit for different operational environments (e.g., hill, sea, rural and urban areas) is still an open problem. The 3rd generation partnership project (3GPP) is actively engaged in developing a 3-D aerial channel model valid for drone altitude from 10 m to 300 m~\cite{3GPP-2017}. However, such a model is still under development and is currently not available. Thus, in the figures, we consider drone altitude between 200 m and 1000 m. However, it must be noted that current drone regulations generally limit drone height to below 150 m. The analytical framework proposed in this paper is able to accommodate any aerial channel model for which the probabilistic functions of LOS and NLOS are given, such as Model 1 and 2.
\end{remark}

\subsection{Coverage Probabilities}
\begin{figure*}[t]
\centering
\subfigure[TBS uplink coverage probability with different aerial channel models and simulations.]{\includegraphics[scale=0.59]{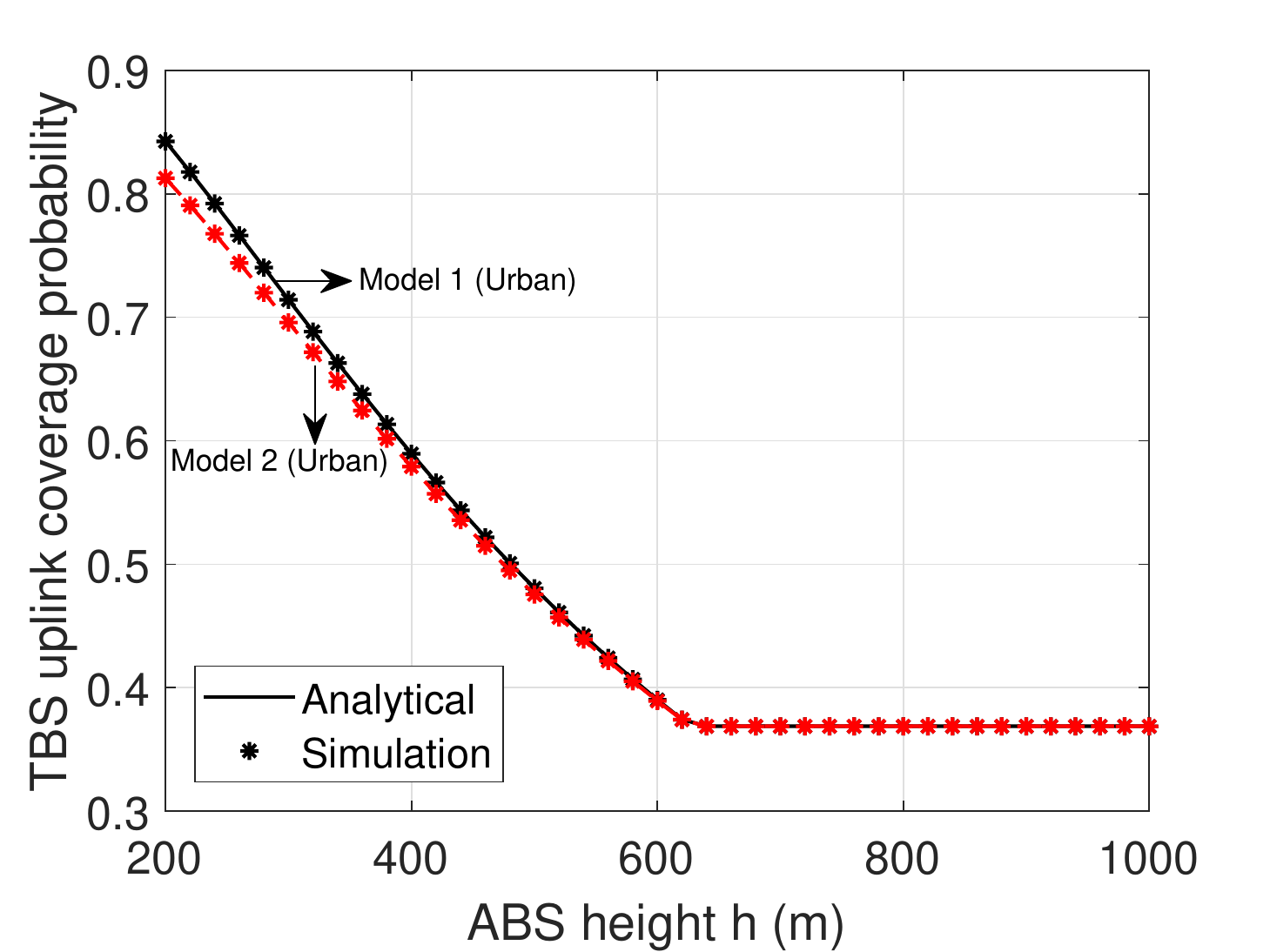}\label{fig:upT1}}\hspace{2.1mm}
\subfigure[ABS uplink coverage probability with different aerial channel models and simulations with and without Nakagami-$m$ fading.]{\includegraphics[scale=0.59]{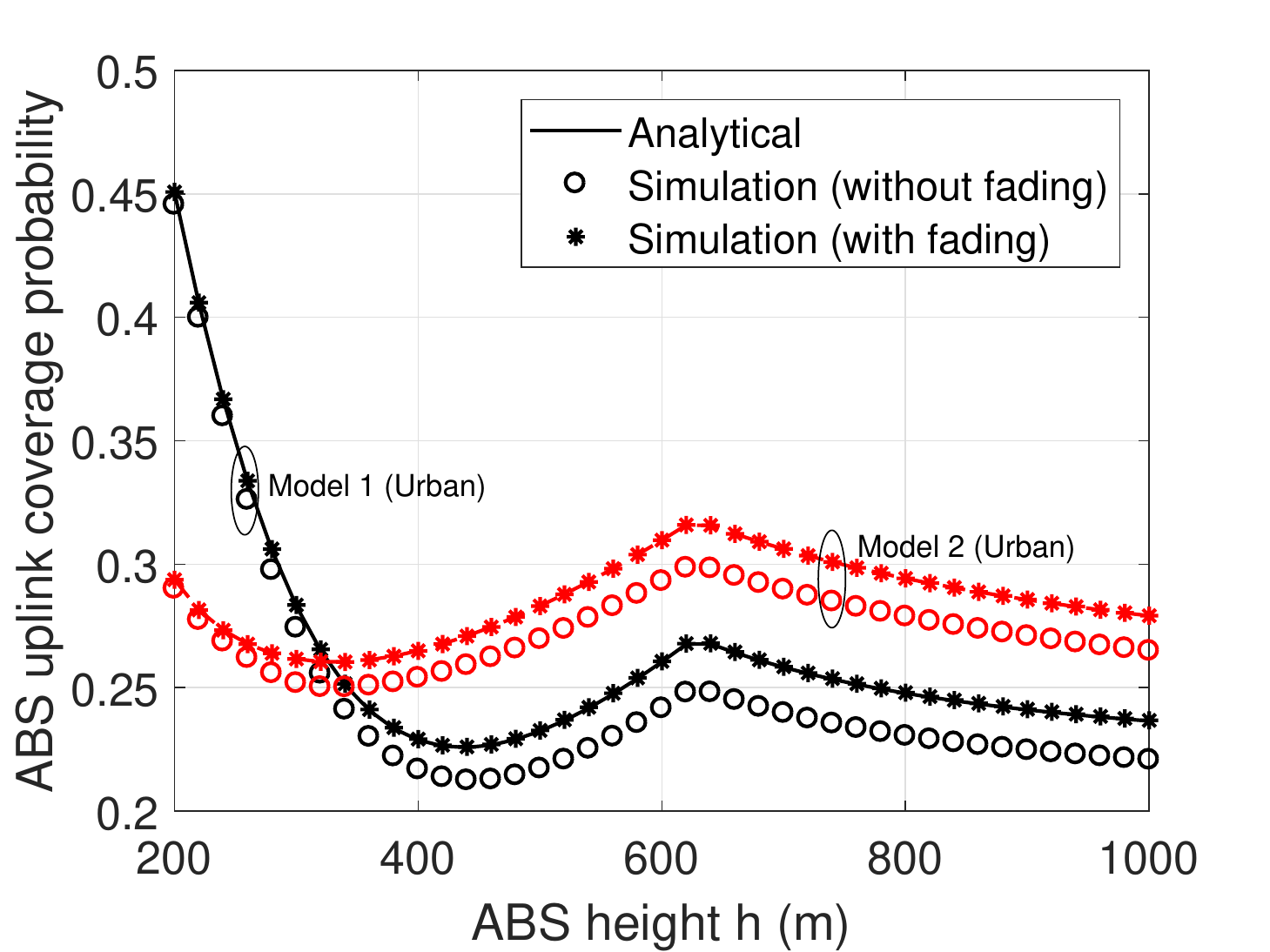}\label{fig:upA1}}
\vspace{-3mm}
\caption{Uplink coverage probabilities versus height of ABS $h$ with simulations.}
\label{fig:up1}
\end{figure*}

\begin{figure*}[t]
\centering
\subfigure[TsUE downlink coverage probability with different aerial channel models and simulations with and without Nakagami-$m$ fading.]{\includegraphics[scale=0.59]{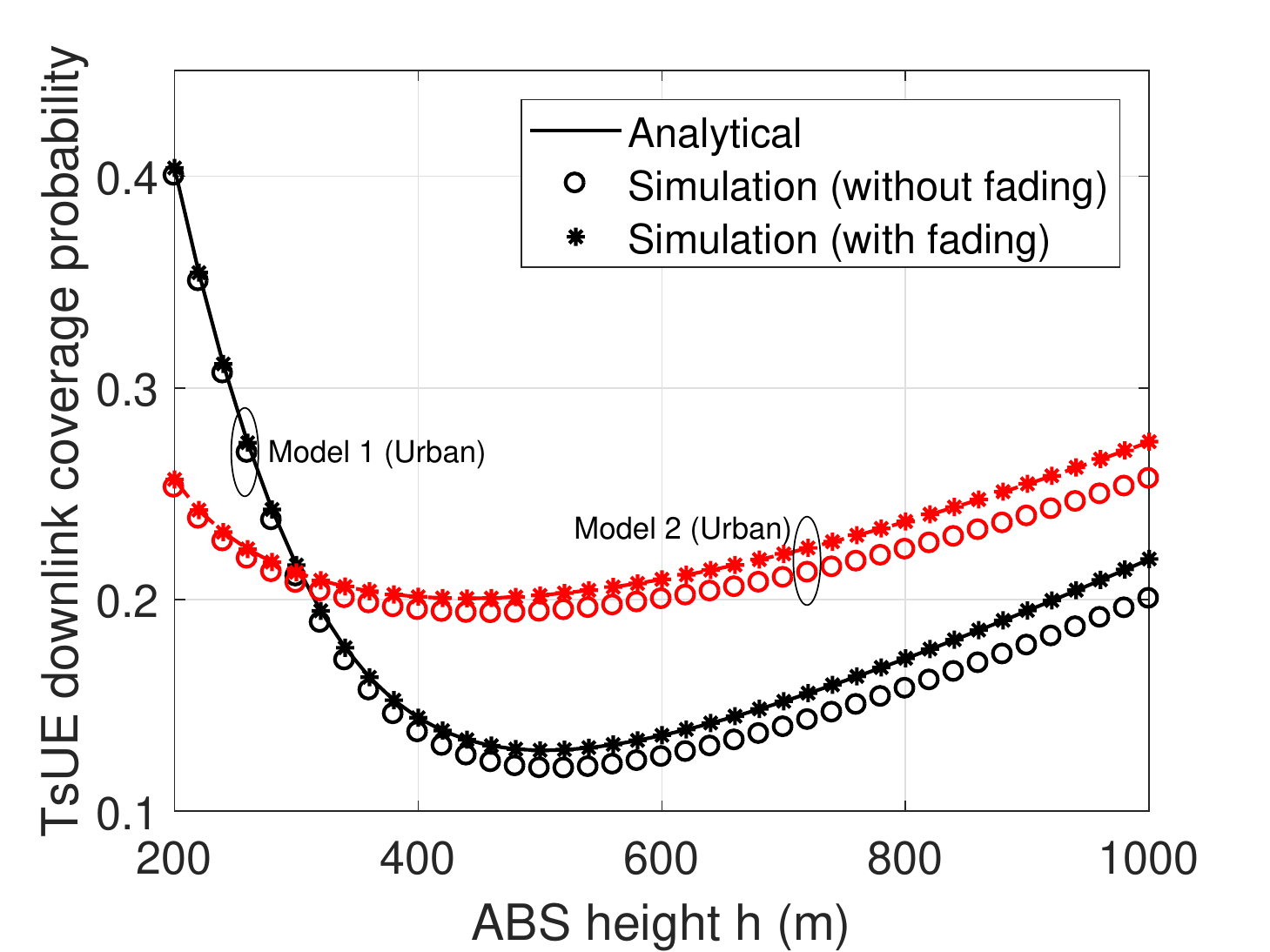}\label{fig:downT1}}\hspace{2.1mm}
\subfigure[AsD downlink coverage probability with different aerial channel models and simulations with and without Nakagami-$m$ fading.]{\includegraphics[scale=0.59]{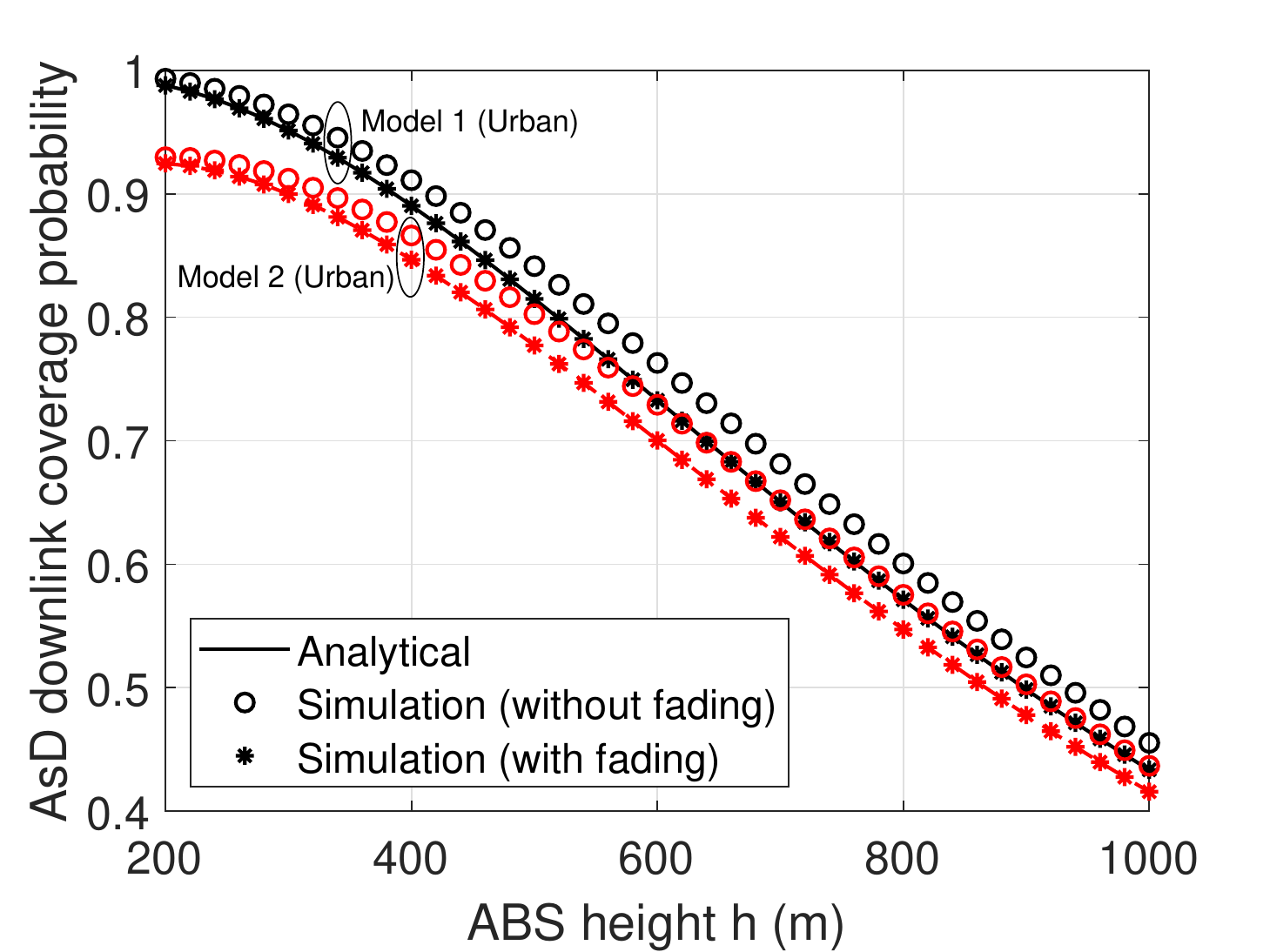}\label{fig:downA1}}
\vspace{-3mm}
\caption{Downlink coverage probabilities versus height of ABS $h$ with simulations.}
\label{fig:down1}
\end{figure*}

Fig.~\ref{fig:up1} and Fig.~\ref{fig:down1} plot the uplink coverage probability of the TBS and the ABS and the downlink coverage probability of the TsUE and the AsD against the ABS height respectively with Model 1 and Model 2 urban parameters. The analytical results are obtained using Theorem~\ref{th:covBS}, Theorem~\ref{th:covD}, Theorem~\ref{th:covTUE} and Theorem~\ref{th:covAUE}. In Fig.~\ref{fig:upA1}, Fig.~\ref{fig:downT1} and Fig.~\ref{fig:downA1}, two sets of simulation results are generated. One with Nakagami-$m$ fading for the aerial channel model and one without small-scale fading. For the TBS uplink coverage probability, the simulation results match very well with the analytical results. For the ABS uplink coverage probability, the TsUE downlink coverage probability and the AsD downlink coverage probability, the analytical results agree with the simulation results with Nakagami-$m$ fading and hold similar trends with the simulation results without small-scale fading. This validates the accuracy of our analytical framework.

Fig.~\ref{fig:up1} and Fig.~\ref{fig:down1} show almost the same trends for Model 1 and Model 2. Therefore, we focus on Model 1 for the results presented later in this paper. Also we only show the numerical results in the later subsections, since the numerical results are verified by comparison with the simulation.

\subsection{Impact of ABS Height:}
\begin{figure*}[t]
\centering
\subfigure[TBS uplink coverage probability with different aerial channel environments.]{\includegraphics[scale=0.59]{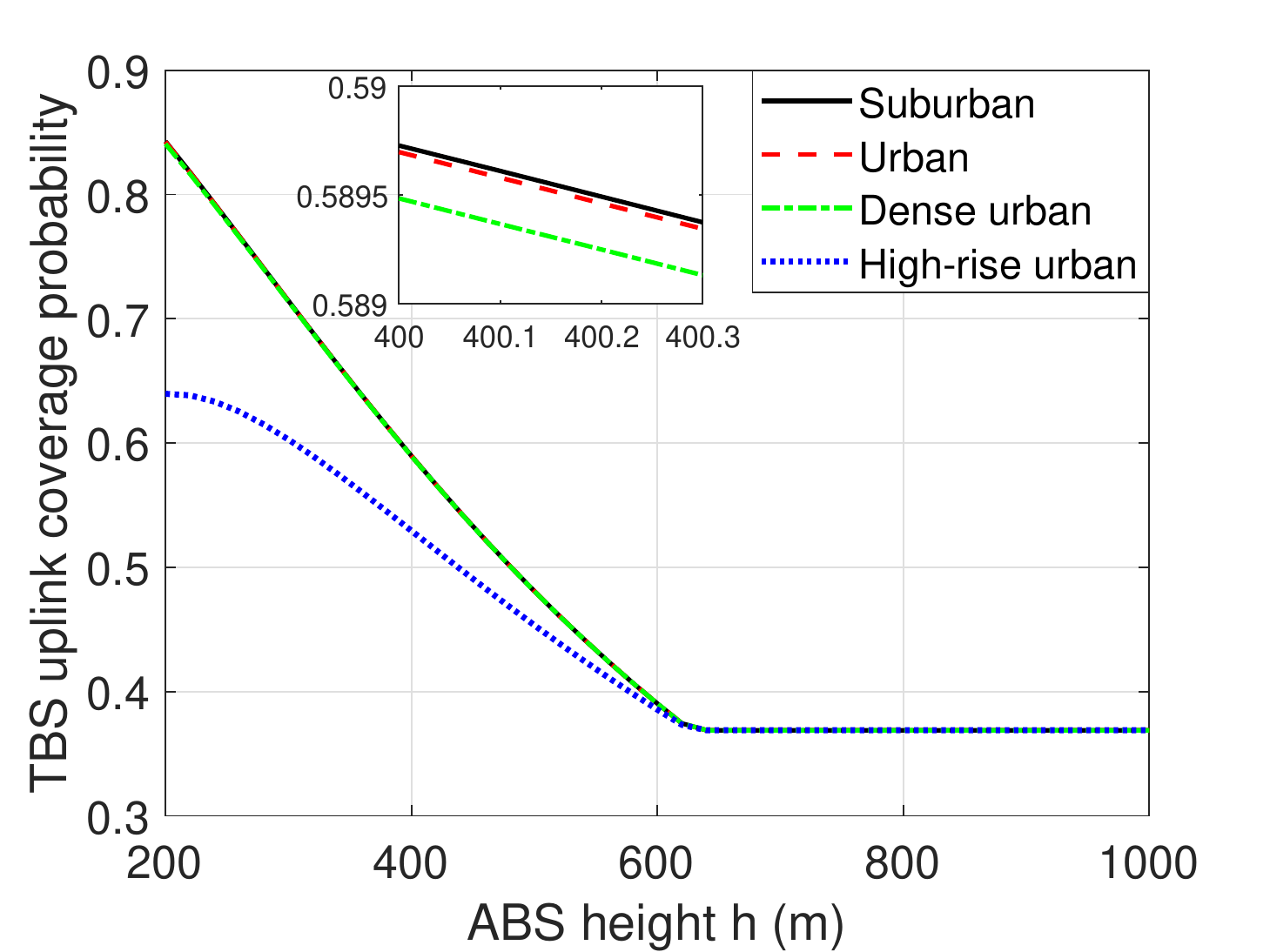}\label{fig:upT2}}\hspace{2.1mm}
\subfigure[ABS uplink coverage probability with different aerial channel environments.]{\includegraphics[scale=0.59]{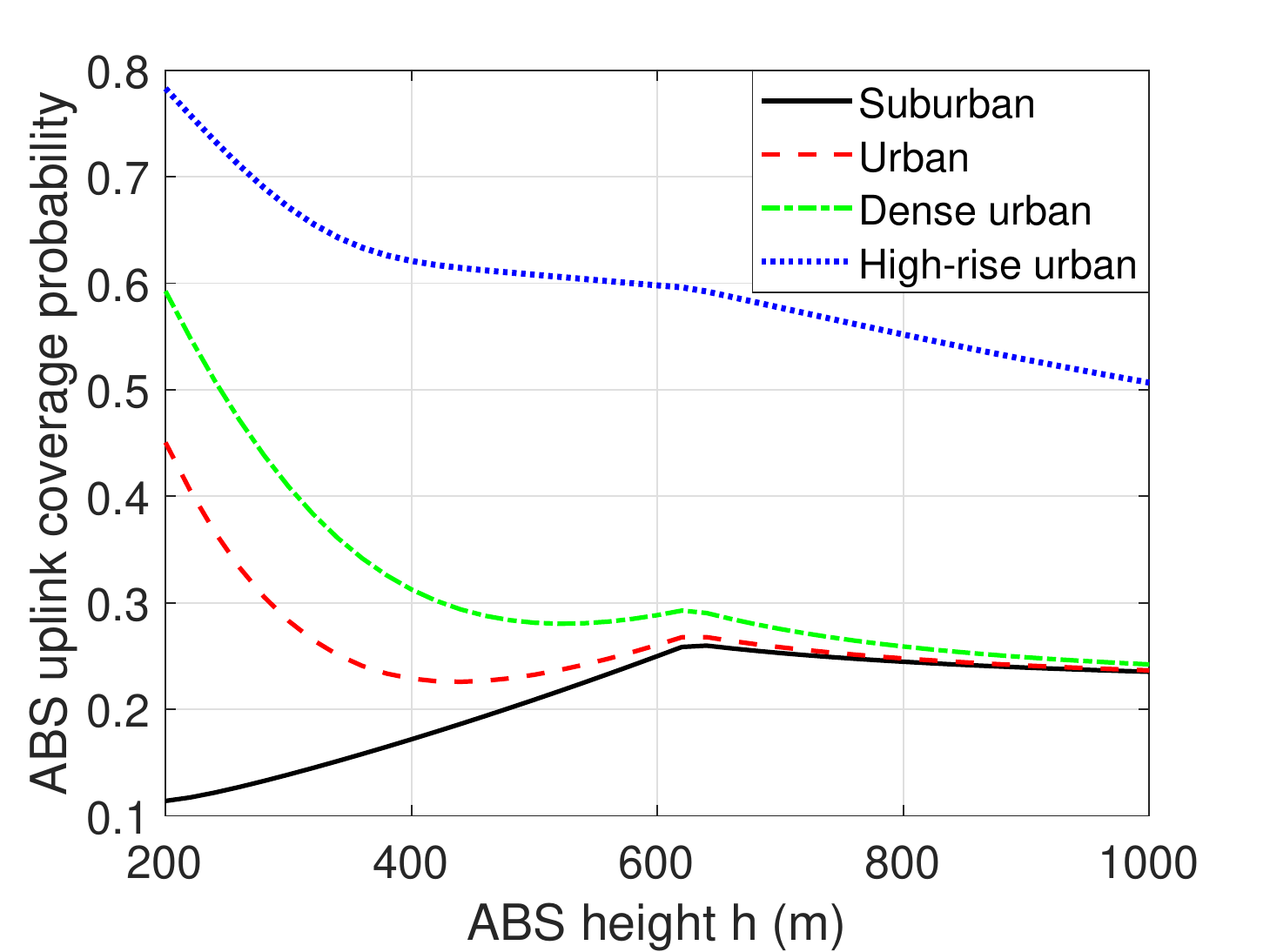}\label{fig:upA2}}
\vspace{-3mm}
\caption{Uplink coverage probabilities versus height of ABS $h$.}
\label{fig:up2}
\end{figure*}

\begin{figure*}[t]
\centering
\subfigure[TsUE downlink coverage probability with different aerial channel environments.]{\includegraphics[scale=0.59]{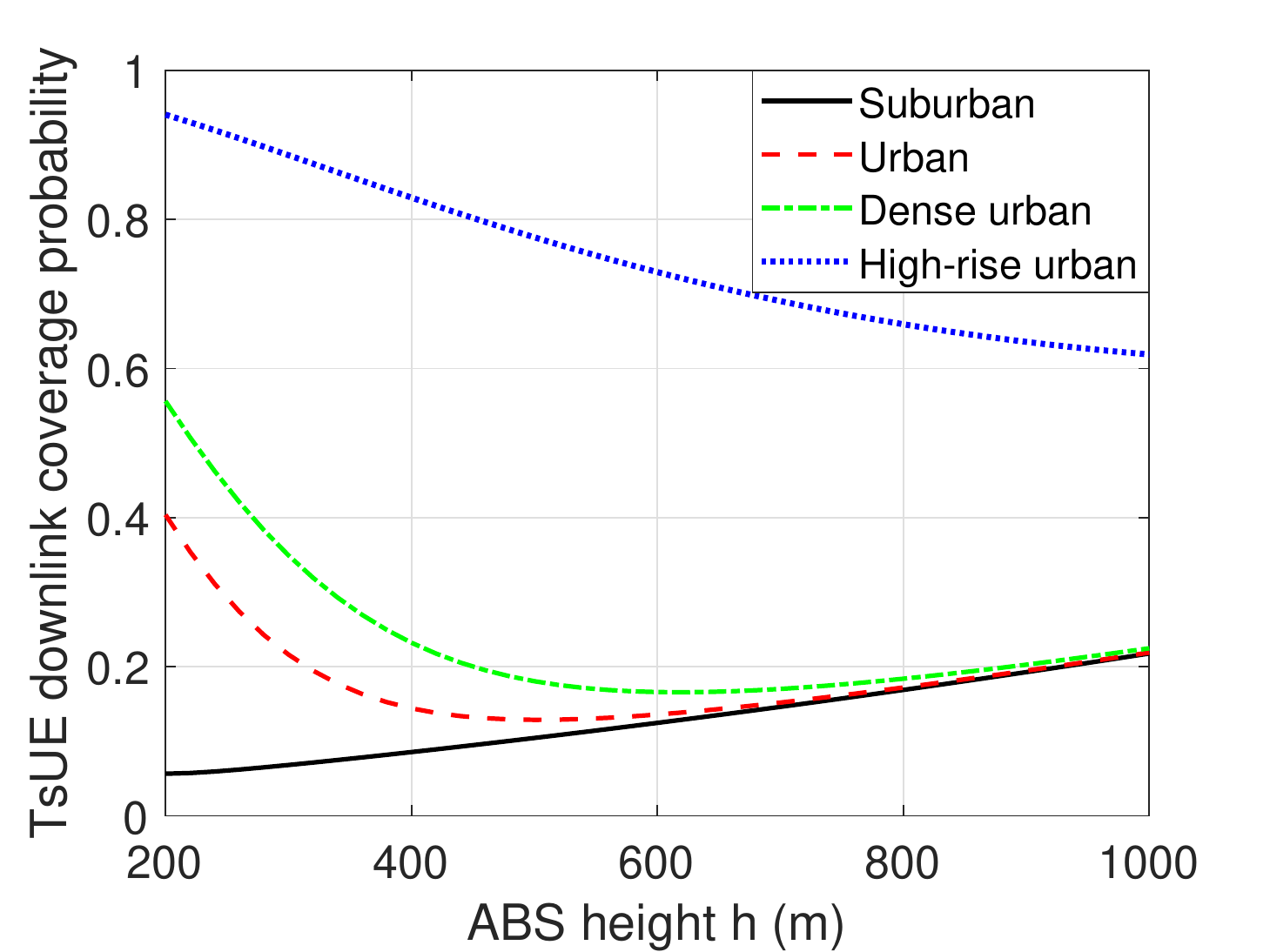}\label{fig:downT2}}\hspace{2.1mm}
\subfigure[AsD downlink coverage probability with different aerial channel environments.]{\includegraphics[scale=0.59]{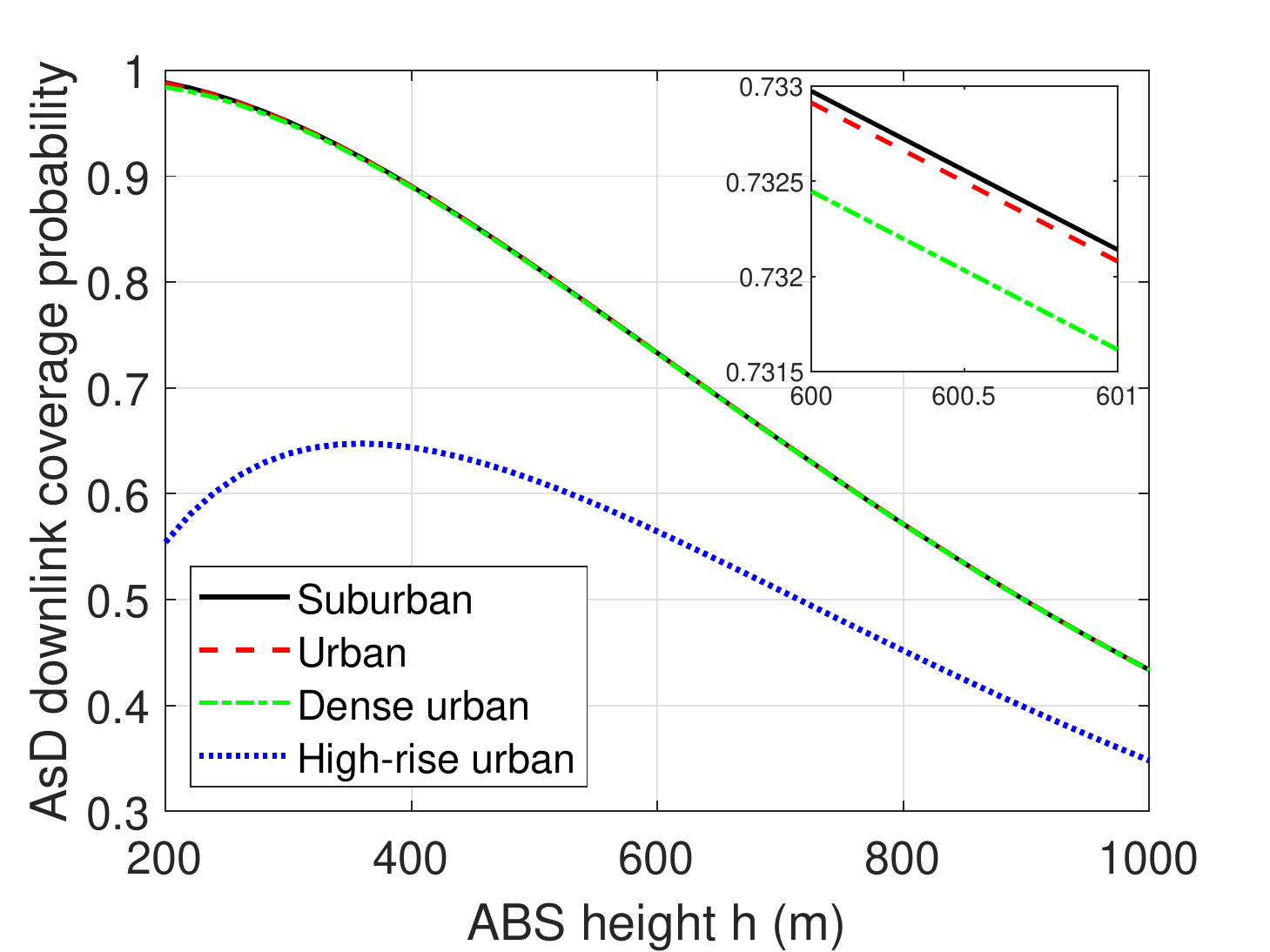}\label{fig:downA2}}
\vspace{-3mm}
\caption{Downlink coverage probabilities versus height of ABS $h$.}
\label{fig:down2}
\end{figure*}

In Fig.~\ref{fig:up2} and Fig.~\ref{fig:down2}, we investigate the effect of ABS height on the uplink coverage performance at the TBS and the ABS and the downlink coverage performance at the TsUE and the AsD under different considered environments.

\textit{Insights:} Fig.~\ref{fig:upT2} plots the uplink coverage probability of the TBS against the height of the ABS with different propagation environments (i.e., suburban, urban, dense urban and high-rise urban). From the figure, we can see that for most cases the uplink coverage probability of the TBS first decreases as the ABS height increases. This is because the transmit power of the AsD increases with the height of ABS, whereby the interference at the TBS increases. This decreases the coverage probability. After a certain ABS height, the coverage probability of the TBS stays as a constant. This is due to the fact that the AsD has reached its maximum transmit power and the interference generated at the TBS keeps the same on average. Note that the height where the coverage probability of the TBS starts to level off is independent of the considered propagation environments.

Fig.~\ref{fig:upA2} plots the uplink coverage probability of the ABS against the height of the ABS with different considered environments in Table~\ref{tab:pLOS}. For suburban environment, the ABS uplink coverage probability first increases and then decreases with the ABS height. In contrast, the ABS uplink coverage probability first decreases as the ABS height increases, then increases to its local maximum for other propagation environments. Thereafter it decreases again as the ABS height further increases. When the ABS height is low, there is higher probability that the interference link between TsUE and ABS is in LOS as the height of ABS rises. Therefore, the ABS uplink coverage probability drops. As the height increases further, the interference link is highly likely in LOS, but the interference power drops and the average desired signal power received at the ABS stays the same under full channel inversion. Hence, the coverage probability increases. When the ABS height is above the global/local optimal height (depends on considered propagation environments), the AsD transmits with its maximum power $\Pmax$ and the received power of the desired signal reduces as the height increases further. This leads to the drop of the uplink coverage probability at the ABS.

Fig.~\ref{fig:downT2} shows the downlink coverage probability of the TsUE versus the height of the ABS with different propagation environments in Table~\ref{tab:pLOS}. The TsUE downlink coverage probability increases with the ABS height for suburban environment, but decreases for high-rise urban environment. For the other environments, the TsUE downlink coverage probability decreases with the increased ABS height and then increases as the ABS height further increases. When the ABS height increases, the interference link between TsUE and ABS has a higher chance of being in LOS, but the 3-D propagation distance and the path-loss also increases. This interplay leads to the above mentioned trends.

From Fig.~\ref{fig:downA2}, we can see that the AsD downlink coverage probability first increases and then decreases as the ABS height increases in high-rise urban environment. For the other environments, the downlink coverage probability of the AsD drops with the increase in the height of ABS. Unlike uplink power control, the ABS transmits with a constant power for downlink communication. Therefore, the received power of the desired signal at the AsD reduces as the height increases and so does the AsD downlink coverage probability.

\begin{figure*}[t]
\centering
\subfigure[Maximum TBS uplink coverage probability with different aerial channel environments.]{\includegraphics[scale=0.57]{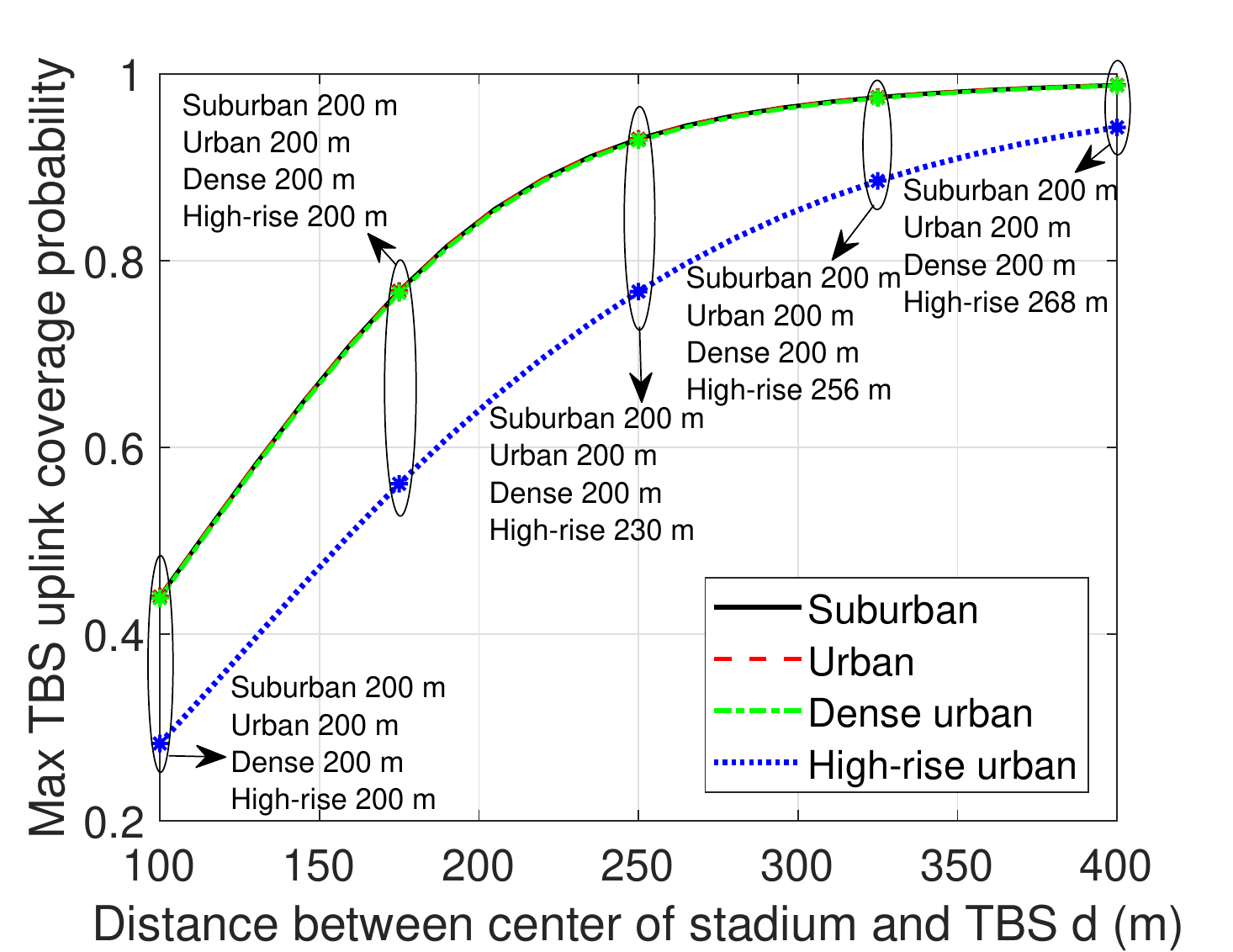}\label{fig:upT3}}\hspace{2.1mm}
\subfigure[Maximum ABS uplink coverage probability with different aerial channel environments.]{\includegraphics[scale=0.57]{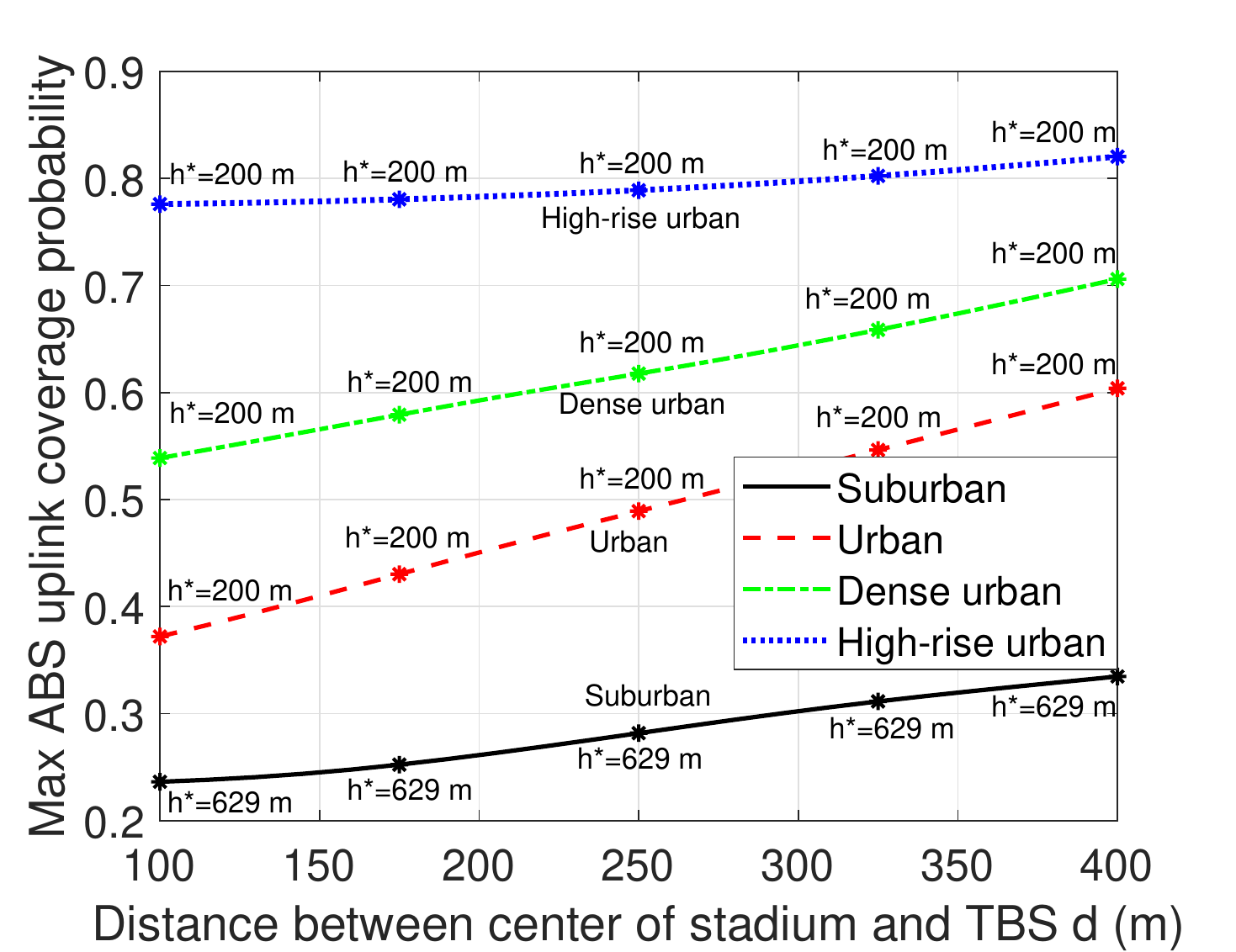}\label{fig:upA3}}
\vspace{-3mm}
\caption{Maximum uplink coverage probabilities versus distance between the center of the stadium and the TBS $d$.}
\label{fig:up3}
\end{figure*}

\begin{figure*}[t]
\centering
\subfigure[Maximum TsUE downlink coverage probability with different aerial channel environments.]{\includegraphics[scale=0.57]{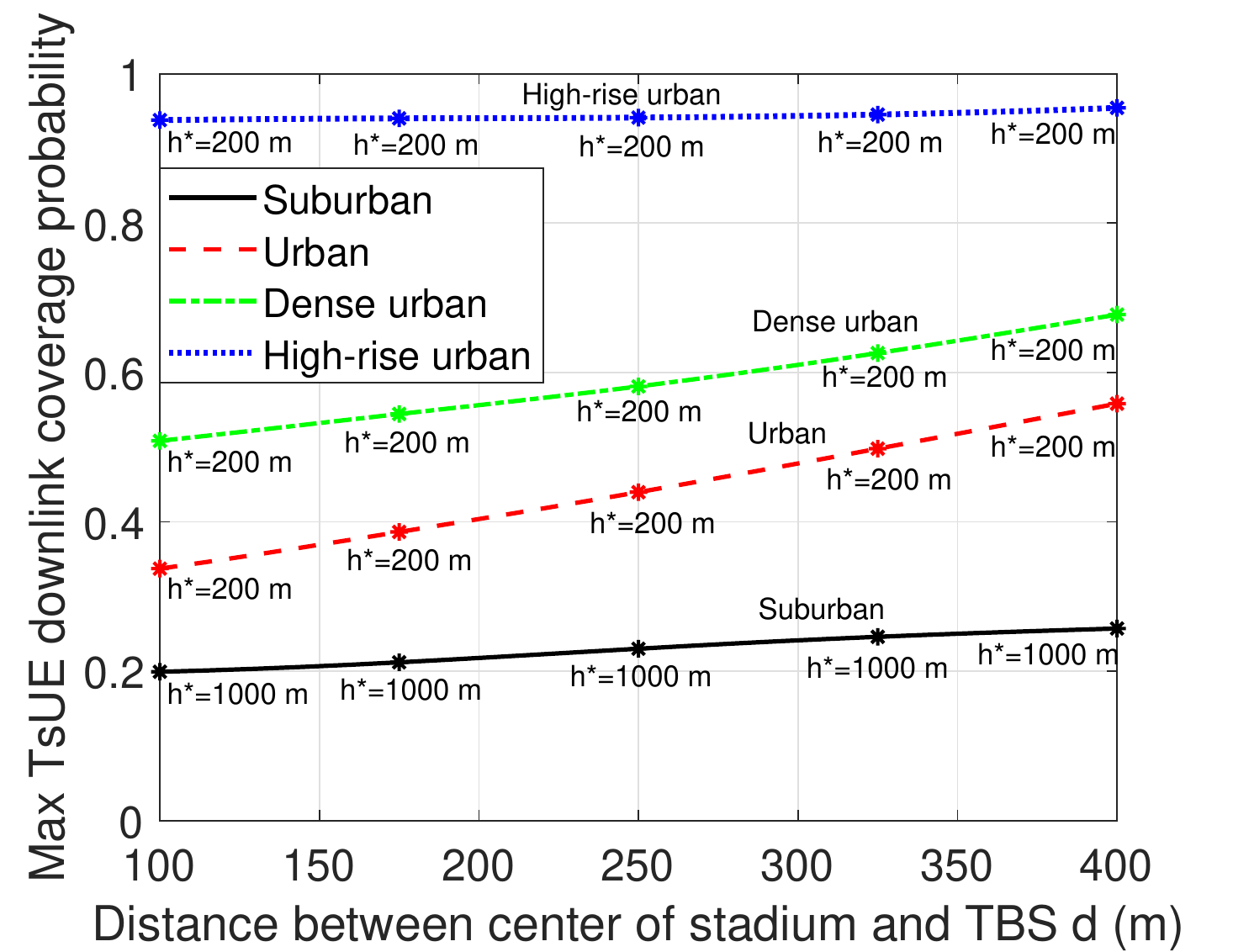}\label{fig:downT3}}\hspace{2.1mm}
\subfigure[Maximum AsD downlink coverage probability with different aerial channel environments.]{\includegraphics[scale=0.57]{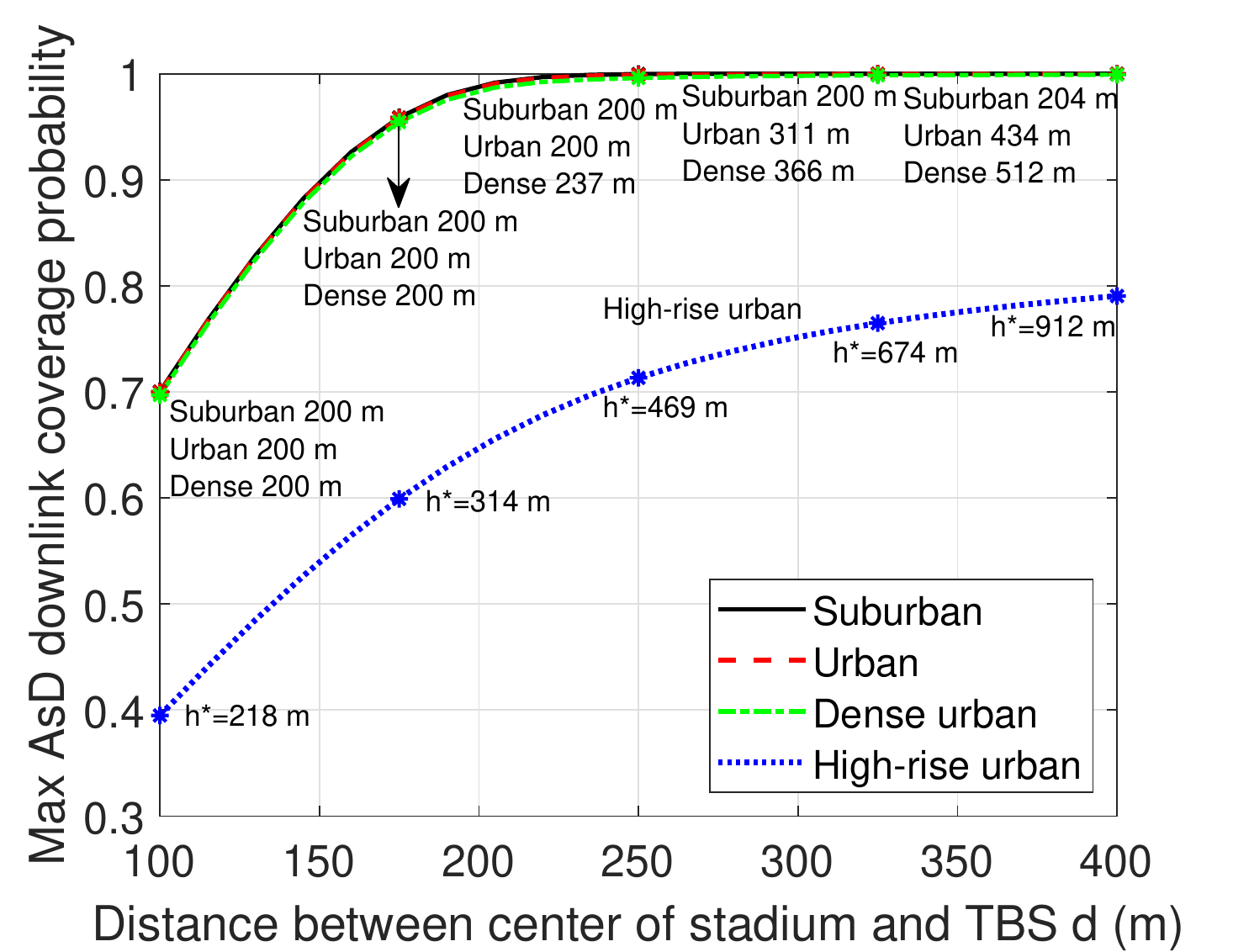}\label{fig:downA3}}
\vspace{-3mm}
\caption{Maximum downlink coverage probabilities versus distance between the center of the stadium and the TBS $d$.}
\label{fig:down3}
\end{figure*}

\subsection{Impact of Environment:}
\textit{Insights:} From Fig.~\ref{fig:up2} and Fig.~\ref{fig:down2}, we can also see that different propagation environments heavily impact the network performance. Fig.~\ref{fig:upT2} shows that the TBS uplink coverage probability is ordered from highest to lowest as follows: suburban, urban, dense urban and high-rise urban when the ABS height is low. Fig.~\ref{fig:upA2} illustrates that the ABS uplink coverage probability is ordered from highest to lowest as follows: high-rise urban, dense urban, urban and suburban. This is because building blockage is severe for dense urban and high-rise urban environment. Therefore, the path-loss between the ABS and the TsUE and between the ABS and the AsD is strong. The received interference power at the ABS from the TsUE is less. The AsD however needs to transmit with a higher power to overcome the path-loss, thus generates a higher interference power for the TBS.

The TsUE downlink coverage probability is ordered from highest to lowest as follows: high-rise urban, dense urban, urban and suburban in Fig.~\ref{fig:downT2}, but the AsD downlink coverage probability is lowest for high-rise urban environment in Fig.~\ref{fig:downA2}. The severe building blockage in the dense urban and high-rise urban environment reduces the received interfering power at the TsUE from the ABS and the received power of the desired signal at the AsD from the ABS.

These figures reveal that the height of the ABS and the propagation environment affect the network performance. In the next subsection, we are going to find the ABS heights which maximize TBS uplink coverage probability, ABS uplink coverage probability, TsUE downlink coverage probability and AsD downlink coverage probability under different propagation environments and how the optimal heights of ABS change with the distance between the center of the stadium and the TBS $d$ in the next section.

\subsection{Impact of Distance Between Center of Stadium and TBS}
Fig.~\ref{fig:upT3}, Fig.~\ref{fig:upA3}, Fig.~\ref{fig:downT3} and Fig.~\ref{fig:downA3} plot the maximum uplink coverage probability at the TBS, the maximum uplink coverage probability at the ABS, the maximum downlink coverage probability at the TsUE and the maximum downlink coverage probability at the AsD against the distance between the center of the stadium and the TBS $d$ with the corresponding ABS heights marked, respectively, for different propagation environment parameters.

\textit{Insights:} From the figures, we can see that both the uplink and downlink network performance improve when the distance between the center of the stadium and the TBS $d$ increases. This is because, the interference experienced at the drone cell and the terrestrial cell decreases, when the two cells are further apart.

From Fig.~\ref{fig:upT3}, we can see that for most cases, the uplink coverage probability at the TBS is maximized when the ABS is placed as low as possible (which is 200 m for the aerial channel model considered). The optimal ABS height which maximizes the TBS uplink coverage probability increases with the distance between the center of the stadium and the TBS $d$ for high-rise urban environment. If we refer to Fig.~\ref{fig:upT2}, we can see that the TBS uplink coverage probability for high-rise urban environment increases a little before dropping to a constant level. When $d$ increases, the drop starts at a higher ABS altitude.

Fig.~\ref{fig:upA3} shows that the uplink coverage probability at the ABS is maximized when the ABS is placed as low as possible (i.e., 200 m) for urban, dense urban and high-rise urban environments and at 629 m for suburban environment. Referring to Fig.~\ref{fig:upA2}, the global maximum of ABS uplink coverage probability exists at 629 m for suburban environment, and at a lower altitude for other environments. The optimal ABS height is independent to the distance between the center of the stadium and the TBS.

Fig.~\ref{fig:downT3} illustrates that the TsUE downlink coverage probability is maximized when the ABS is placed as low as possible (i.e., 200 m) for urban, dense urban and high-rise urban environments, but as high as possible (i.e., 1000 m) for the suburban environments. As shown in Fig.~\ref{fig:downT2}, the TsUE downlink coverage probability increases with the ABS height for the suburban environment, but decreases for high-rise urban environment. For other environments, the maximum TsUE downlink coverage probability can be found at 200 m, even though the downlink coverage probability increases when the ABS height further increases. The optimal height of ABS is independent to the distance between the center of the stadium and the TBS.

Fig.~\ref{fig:downA3} shows that the height of ABS which maximizes the AsD downlink coverage probability increases with the distance between the center of the stadium and the TBS $d$. From Fig.~\ref{fig:downA2}, we can see that the AsD downlink coverage probability first increases and then decreases as the ABS height increases. When $d$ increases, the drop starts at a higher ABS altitude.

From Fig.~\ref{fig:up3} and Fig.~\ref{fig:down3}, we can find that the uplink coverage probabilities at the TBS and the ABS and the downlink coverage probability at the TsUE are maximized when the ABS is deployed at 200 m for urban environment and dense urban environment. Although the maximum AsD downlink coverage probability is not achieved with an ABS height of 200 m and a large distance $d$, the downlink coverage probability at the AsD is much higher than the one at the TsUE for urban environment and dense urban environment. \textit{Therefore, it is best to place the ABS at a low height (e.g., 200 m or lower) for urban environment and dense urban environment regardless of the distance between the center of the stadium and the TBS $d$. In contrast, the ABS should be placed at different heights depending on the distance between the center of the stadium and the TBS $d$ and the task of the system (i.e., prioritize the terrestrial link or the aerial link, prioritize the uplink or the downlink communication) for suburban environment and high-rise urban environment.}

\vspace{-1.1mm}
\section{Conclusions}\label{sec:Conc}
In this paper, a two-cell network with a TBS and an underlay ABS for IoT device coverage in temporary events was considered. We presented a general analytical framework for uplink coverage probability of the TBS and the ABS and downlink coverage probability of the TsUE and the AsD in terms of the Laplace transforms of the interference power distribution and the distance distribution between the ABS and an i.u.d. AsD and between the ABS and an i.u.d. TsUE. The framework is able to accommodate any aerial channel model. The simulation results confirmed the accuracy of the proposed model. The results have shown that the ABS is best to be deployed at 200 m or lower for urban environment and dense urban environment regardless of the distance between the center of the stadium and the TBS. Future work can consider the impact of beamforming at the ABS, user scheduling if there are multiple UEs per channel, load balancing between the ABS and the TBS and effect of imperfect CSI.

\appendices
\section{Proof of Lemma~\ref{le:LIB}}\label{ap:LIB}
Following the definition of the Laplace transform, the Laplace transform of the interference power distribution at the TBS is expressed as
\ifCLASSOPTIONpeerreview
\begin{align}
\mathcal{L}_{\Ib^u}(s)\!=\!\mathbb{E}_{\Ib^u}[\exp(\!-s\Ib^u\!)]\!=\!\mathbb{E}_{P,h,d}[\exp(\!-sP_\mathrm{AsD} H^u_A d_A^{-\alphab}\!)]=\!\mathbb{E}_{P,d}\left[\frac{1}{1+sP_\mathrm{AsD} d_A^{-\alphab}}\right],
\end{align}
\else
\begin{align}
\mathcal{L}_{\Ib^u}(s)&\!=\!\mathbb{E}_{\Ib^u}[\exp(\!-s\Ib^u\!)]\!=\!\mathbb{E}_{P,h,d}[\exp(\!-sP_\mathrm{AsD} H^u_A d_A^{-\alphab}\!)]\nonumber\\
&=\!\mathbb{E}_{P,d}\left[\frac{1}{1+sP_\mathrm{AsD} d_A^{-\alphab}}\right],
\end{align}
\fi
where the third step comes from the fact that $H^u_A$ follows exponential distribution with unit mean. Conditioned on the value of $h$, there are six possible cases for $\mathcal{L}_{\Ib^u}(s)$. When $\sqrt{(\frac{\Pmax\etaL}{\rhod})^{\frac{2}{\alphaL}}-R_2^2}<h<(\frac{\Pmax\etaN}{\rhod})^{\frac{1}{\alphaN}}$, the Laplace transform of the interference power distribution at the TBS equals to
\ifCLASSOPTIONpeerreview
\begin{subequations}
\begin{align}
&\mathcal{L}_{\Ib^u}(s)\!=\!\int_h^{(\frac{\Pmax\etaL}{\rhod})^\frac{1}{\alphaL}}\!\!\mathbb{E}_{d}\left[\frac{\pLOS}{1\!+\!s\frac{\rhod z^{\alphaL}}{\etaL d_A^{\alphab}}}\right]f_{\Zd}(z)\textup{d}z+\int_{(\frac{\Pmax\etaL}{\rhod})^\frac{1}{\alphaL}}^{\sqrt{h^2+R_2^2}}\mathbb{E}_{d}\left[\frac{\pLOS}{1+\frac{s\Pmax}{d_A^{\alphaL}}}\right]f_{\Zd}(z)\textup{d}z\nonumber\\
&+\int_h^{(\frac{\Pmax\etaN}{\rhod})^\frac{1}{\alphaN}}\!\!\mathbb{E}_{d}\left[\frac{\pNLOS}{1\!+\!s\frac{\rhod z^{\alphaN}}{\etaN d_A^{\alphab}}}\right]f_{\Zd}(z)\textup{d}z+\int_{(\frac{\Pmax\etaN}{\rhod})^\frac{1}{\alphaN}}^{\sqrt{h^2+R_2^2}}\mathbb{E}_{d}\left[\frac{\pNLOS}{1+\frac{s\Pmax}{d_A^{\alphaN}}}\right]f_{\Zd}(z)\textup{d}z\nonumber\\
&=\!\!\!\int_h^{\left(\!\frac{\Pmax\etaL}{\rhod}\!\right)^\frac{1}{\alphaL}}\!\!\!\!\!\mathbb{E}_{\theta}\!\!\left[\frac{\pLOS}{1\!+\!\frac{s\rhod z^{\alphaL}}{\etaL\left(z^2\!-\!h^2\!+\!d^2\!-\!2\sqrt{z^2\!-\!h^2}d\cos\Theta\right)^{\frac{\alphab}{2}}}}\right]\!\!f_{\Zd}\!(z)\textup{d}z\!\!+\!\!\!\int_{\left(\!\frac{\Pmax\etaL}{\rhod}\!\right)^\frac{1}{\alphaL}}^{\sqrt{h^2\!+\!R_2^2}}\!\mathbb{E}_{\theta}\!\!\left[\frac{\pLOS}{1\!+\!\frac{s\Pmax}{\left(z^2\!-\!h^2\!+\!d^2\!-\!2\sqrt{z^2\!-\!h^2}d\cos\Theta\right)^{\frac{\alphab}{2}}}}\right]\!\!f_{\Zd}\!(z)\textup{d}z\nonumber\\
&+\!\!\!\int_h^{\left(\!\frac{\Pmax\etaN}{\rhod}\!\right)^\frac{1}{\alphaN}}\!\!\!\!\!\mathbb{E}_{\theta}\!\!\left[\frac{\pNLOS}{1\!+\!\frac{s\rhod z^{\alphaN}}{\etaN\left(z^2\!-\!h^2\!+\!d^2\!-\!2\sqrt{z^2\!-\!h^2}d\cos\Theta\right)^{\frac{\alphab}{2}}}}\right]\!\!f_{\Zd}\!(z)\textup{d}z\!\!+\!\!\!\int_{\left(\!\frac{\Pmax\etaN}{\rhod}\!\right)^\frac{1}{\alphaN}}^{\sqrt{h^2\!+\!R_2^2}}\!\mathbb{E}_{\theta}\!\!\left[\frac{\pNLOS}{1\!+\!\frac{s\Pmax}{\left(z^2\!-\!h^2\!+\!d^2\!-\!2\sqrt{z^2\!-\!h^2}d\cos\Theta\right)^{\frac{\alphab}{2}}}}\right]\!\!f_{\Zd}\!(z)\textup{d}z\label{eq:prLIB1}\\
&=\int_h^{(\frac{\Pmax\etaL}{\rhod})^\frac{1}{\alphaL}}\int_{0}^{2\pi}\mathcal{L}_{\Ib^u}(s,\frac{\rhod}{\etaL} z^{\alphaL}|\theta,z)\frac{\pLOS}{2\pi}f_{\Zd}(z)\textup{d}\theta \textup{d}z+\!\!\!\int_{(\frac{\Pmax\etaL}{\rhod})^\frac{1}{\alphaL}}^{\sqrt{h^2+R_2^2}}\!\int_{0}^{2\pi}\!\!\!\mathcal{L}_{\Ib^u}\!(s,\!\Pmax|\theta,z)\frac{\pLOS}{2\pi}f_{\Zd}\!(z)\textup{d}\theta \textup{d}z\nonumber\\
&+\int_h^{(\frac{\Pmax\etaN}{\rhod})^\frac{1}{\alphaN}}\int_{0}^{2\pi}\mathcal{L}_{\Ib^u}(s,\frac{\rhod}{\etaN} z^{\alphaN}|\theta,z)\frac{\pNLOS}{2\pi}f_{\Zd}(z)\textup{d}\theta \textup{d}z+\!\!\!\int_{(\frac{\Pmax\etaN}{\rhod})^\frac{1}{\alphaN}}^{\sqrt{h^2+R_2^2}}\!\int_{0}^{2\pi}\!\!\!\mathcal{L}_{\Ib^u}\!(s,\!\Pmax|\theta,z)\frac{\pNLOS}{2\pi}f_{\Zd}\!(z)\textup{d}\theta \textup{d}z\label{eq:prLIB2},
\end{align}
\end{subequations}
\else
\begin{subequations}
\begin{align}
&\mathcal{L}_{\Ib^u}(s)\!=\!\int_h^{(\frac{\Pmax\etaL}{\rhod})^\frac{1}{\alphaL}}\!\!\mathbb{E}_{d}\left[\frac{\pLOS}{1\!+\!s\frac{\rhod z^{\alphaL}}{\etaL d_A^{\alphab}}}\right]f_{\Zd}(z)\textup{d}z\nonumber\\
&+\int_{(\frac{\Pmax\etaL}{\rhod})^\frac{1}{\alphaL}}^{\sqrt{h^2+R_2^2}}\mathbb{E}_{d}\left[\frac{\pLOS}{1+\frac{s\Pmax}{d_A^{\alphaL}}}\right]f_{\Zd}(z)\textup{d}z\nonumber\\
&+\int_h^{(\frac{\Pmax\etaN}{\rhod})^\frac{1}{\alphaN}}\!\!\mathbb{E}_{d}\left[\frac{\pNLOS}{1\!+\!s\frac{\rhod z^{\alphaN}}{\etaN d_A^{\alphab}}}\right]f_{\Zd}(z)\textup{d}z\nonumber\\
&+\int_{(\frac{\Pmax\etaN}{\rhod})^\frac{1}{\alphaN}}^{\sqrt{h^2+R_2^2}}\mathbb{E}_{d}\left[\frac{\pNLOS}{1+\frac{s\Pmax}{d_A^{\alphaN}}}\right]f_{\Zd}(z)\textup{d}z\nonumber\\
&=\!\!\!\int_h^{\!\left(\!\frac{\Pmax\etaL}{\rhod}\!\right)^\frac{1}{\alphaL}}\!\!\!\!\!\mathbb{E}_{\theta}\!\!\left[\frac{\pLOS}{1\!\!+\!\frac{s\rhod z^{\alphaL}}{\etaL\left(z^2\!-\!h^2\!+\!d^2\!-\!2\sqrt{z^2\!-\!h^2}d\cos\Theta\right)^{\!\frac{\alphab}{2}}}}\right]\!\!f_{\Zd}\!(z)\textup{d}z\nonumber\\
&+\!\!\!\int_{\!\left(\!\frac{\Pmax\etaL}{\rhod}\!\right)^\frac{1}{\alphaL}}^{\sqrt{h^2\!+\!R_2^2}}\!\mathbb{E}_{\theta}\!\!\left[\frac{\pLOS}{1\!+\!\frac{s\Pmax}{\left(z^2\!-\!h^2\!+\!d^2\!-\!2\sqrt{z^2\!-\!h^2}d\cos\Theta\right)^{\!\frac{\alphab}{2}}}}\right]\!\!f_{\Zd}\!(z)\textup{d}z\nonumber\\
&+\!\!\!\int_h^{\!\left(\!\frac{\Pmax\etaN}{\rhod}\!\right)^\frac{1}{\alphaN}}\!\!\!\!\!\mathbb{E}_{\theta}\!\!\left[\frac{\pNLOS}{1\!+\!\frac{s\rhod z^{\alphaN}}{\etaN\left(z^2\!-\!h^2\!+\!d^2\!-\!2\sqrt{z^2\!-\!h^2}d\cos\Theta\right)^{\frac{\alphab}{2}}}}\right]\!\!f_{\Zd}\!(z)\textup{d}z\nonumber\\
&+\!\!\!\int_{\!\left(\!\frac{\Pmax\etaN}{\rhod}\!\right)^{\!\frac{1}{\alphaN}}}^{\sqrt{h^2\!+\!R_2^2}}\!\mathbb{E}_{\theta}\!\!\left[\frac{\pNLOS}{1\!+\!\frac{s\Pmax}{\left(z^2\!-\!h^2\!+\!d^2\!-\!2\sqrt{z^2\!-\!h^2}d\cos\Theta\right)^{\!\frac{\alphab}{2}}}}\right]\!\!f_{\Zd}\!(z)\textup{d}z\label{eq:prLIB1}\\
&=\int_h^{(\frac{\Pmax\etaL}{\rhod})^\frac{1}{\alphaL}}\int_{0}^{2\pi}\mathcal{L}_{\Ib^u}(s,\frac{\rhod}{\etaL} z^{\alphaL}|\theta,z)\frac{\pLOS}{2\pi}f_{\Zd}(z)\textup{d}\theta \textup{d}z\nonumber\\
&+\!\!\!\int_{(\frac{\Pmax\etaL}{\rhod})^\frac{1}{\alphaL}}^{\sqrt{h^2+R_2^2}}\!\int_{0}^{2\pi}\!\!\!\mathcal{L}_{\Ib^u}\!(s,\!\Pmax|\theta,z)\frac{\pLOS}{2\pi}f_{\Zd}\!(z)\textup{d}\theta \textup{d}z\nonumber\\
&+\int_h^{(\frac{\Pmax\etaN}{\rhod})^\frac{1}{\alphaN}}\int_{0}^{2\pi}\mathcal{L}_{\Ib^u}(s,\frac{\rhod}{\etaN} z^{\alphaN}|\theta,z)\frac{\pNLOS}{2\pi}f_{\Zd}(z)\textup{d}\theta \textup{d}z\nonumber\\
&+\!\!\!\int_{(\frac{\Pmax\etaN}{\rhod})^\frac{1}{\alphaN}}^{\sqrt{h^2+R_2^2}}\!\int_{0}^{2\pi}\!\!\!\mathcal{L}_{\Ib^u}\!(s,\!\Pmax|\theta,z)\frac{\pNLOS}{2\pi}f_{\Zd}\!(z)\textup{d}\theta \textup{d}z\label{eq:prLIB2},
\end{align}
\end{subequations}
\fi
\noindent where $d_A$ is expressed in terms of $z$, $h$, $d$, and $\theta$ by cosine rule in \eqref{eq:prLIB1} and \eqref{eq:prLIB2} is obtained by taking the expectation over $\Theta$, which has a conditional pdf as $f_\Theta(\theta|z)=\frac{1}{2\pi}$ for $h\leqslant z\leqslant\sqrt{R_2^2+h^2}$.

Following similar steps, we can work out the Laplace transform of the interference power distribution at the TBS for the other five cases.

\section{Proof of Lemma~\ref{le:disUU}}\label{ap:disUU}
The relation between the length of the AsD to ABS link $\Zd$ with its projection distance on the ground $r_A$ is $\Zd=\sqrt{r_A^2+h^2}$. The distance distribution of the projection distance on the ground is $f_{r_A}(r)=\frac{2r}{R_2^2}$. Thus, we can get the pdf of $\Zd$ as
\ifCLASSOPTIONpeerreview
\begin{align}
f_{\Zd}(z)=\frac{\textup{d}(\sqrt{z^2-h^2})}{\textup{d}z}f_{r_A}\left(\sqrt{z^2-h^2}\right)=\frac{z}{\sqrt{z^2-h^2}}\frac{2\sqrt{z^2-h^2}}{R_2^2}=\frac{2z}{R_2^2}.
\end{align}
\else
\begin{align}
f_{\Zd}(z)&=\frac{\textup{d}(\sqrt{z^2-h^2})}{\textup{d}z}f_{r_A}\left(\sqrt{z^2-h^2}\right)\nonumber\\
&=\frac{z}{\sqrt{z^2-h^2}}\frac{2\sqrt{z^2-h^2}}{R_2^2}=\frac{2z}{R_2^2}.
\end{align}
\fi

\section{Proof of Lemma~\ref{le:LIU}}\label{ap:LIU}
Following the definition of the Laplace transform, the Laplace transform of the interference power distribution at the ABS is expressed as
\ifCLASSOPTIONpeerreview
\begin{subequations}
\begin{align}
&\mathcal{L}_{\Id^u}(s)\!=\!\mathbb{E}_{\Id^u}[\exp(\!-s\Id^u\!)]\!=\!\mathbb{E}_{P,g,z}[\exp(\!-sP_\mathrm{TsUE} G^u_T PL_a(\Zc)\!)]\!=\!\mathbb{E}_{d,g,z}[\exp(\!-s\rhob d_T^{\alphab} G^u_T PL_a(\Zc)\!)]\nonumber\\
&=\!\mathbb{E}_{d,z}\left[\mL^{\mL}(\mL+s\rhob d_T^{\alphab}\Zc^{-\alphaL}\etaL)^{-\mL}\pLOS+\mN^{\mN}(\mN+s\rhob d_T^{\alphab}\Zc^{-\alphaN}\etaN)^{-\mN}\pNLOS\right]\label{eq:prLIU1}\\
&=\!\mathbb{E}_{\omega,z}\left[\mL^{\mL}\left(\mL+s\rhob\left(\Zc^2\!-\!h^2\!+\!d^2\!-\!2\sqrt{\Zc^2\!-\!h^2}d\cos\Omega\right)^{\frac{\alphab}{2}}\Zc^{-\alphaL}\etaL\right)^{-\mL}\pLOS\right.\nonumber\\
&\left.+\mN^{\mN}\left(\mN+s\rhob \left(\Zc^2\!-\!h^2\!+\!d^2\!-\!2\sqrt{\Zc^2\!-\!h^2}d\cos\Omega\right)^{\frac{\alphab}{2}}\Zc^{-\alphaN}\etaN\right)^{-\mN}\pNLOS\right]\label{eq:prLIU2}\\
&=\int_{\sqrt{R_2^2+h^2}}^{\sqrt{(R_1-d)^2+h^2}}\!\!\int_{0}^{2\pi}\!\!\!\!\mathcal{L}_{\Id^u}(s|\omega,z)f_\Omega(\omega|z)f_{\Zc}(z)\textup{d}\omega \textup{d}z\!\!+\!\!\int_{\sqrt{(R_1-d)^2+h^2}}^{\sqrt{(R_1+d)^2+h^2}}\int_{-\widehat{\omega}}^{\widehat{\omega}}\mathcal{L}_{\Id^u}(s|\omega,z)f_\Omega(\omega|z)f_{\Zc}(z)\textup{d}\omega \textup{d}z,\nonumber
\end{align}
\end{subequations}
\else
\begin{subequations}
\begin{align}
&\mathcal{L}_{\Id^u}(s)=\mathbb{E}_{\Id^u}[\exp(-s\Id^u)]\nonumber\\
&=\mathbb{E}_{P,g,z}[\exp(-sP_\mathrm{TsUE} G^u_T PL_a(\Zc))]\nonumber\\
&=\mathbb{E}_{d,g,z}[\exp(-s\rhob d_T^{\alphab} G^u_T PL_a(\Zc))]\nonumber\\
&=\!\mathbb{E}_{d,z}\!\!\left[\!\frac{\mL^{\mL}\pLOS}{\left(\mL\!\!+\!\frac{s\rhob d_T^{\alphab}\etaL}{\Zc^{\alphaL}}\right)^{\!\mL}}\!+\!\frac{\mN^{\mN}\pNLOS}{\left(\mN\!\!+\!\frac{s\rhob d_T^{\alphab}\etaN}{\Zc^{\alphaN}}\right)^{\!\mN}}\!\right]\label{eq:prLIU1}\\
&=\!\mathbb{E}_{\omega,z}\left[\frac{\mL^{\mL}\pLOS}{\left(\mL+\frac{s\rhob\etaL\left(\Zc^2\!-\!h^2\!+\!d^2\!-\!2\sqrt{\Zc^2\!-\!h^2}d\cos\Omega\right)^{\frac{\alphab}{2}}}{\Zc^{\alphaL}}\right)^{\mL}}\right.\nonumber\\
&\left.+\frac{\mN^{\mN}\pNLOS}{\left(\mN\!+\!\frac{s\rhob\etaN\left(\Zc^2\!-\!h^2\!+\!d^2\!-\!2\sqrt{\Zc^2\!-\!h^2}d\cos\Omega\right)^{\frac{\alphab}{2}}}{\Zc^{\alphaN}}\right)^{\mN}}\right]\label{eq:prLIU2}\\
&=\int_{\sqrt{R_2^2+h^2}}^{\sqrt{(R_1-d)^2+h^2}}\!\!\int_{0}^{2\pi}\!\!\!\!\mathcal{L}_{\Id^u}(s|\omega,z)f_\Omega(\omega|z)f_{\Zc}(z)\textup{d}\omega \textup{d}z\nonumber\\
&+\!\!\int_{\sqrt{(R_1-d)^2+h^2}}^{\sqrt{(R_1+d)^2+h^2}}\int_{-\widehat{\omega}}^{\widehat{\omega}}\mathcal{L}_{\Id^u}(s|\omega,z)f_\Omega(\omega|z)f_{\Zc}(z)\textup{d}\omega \textup{d}z,\nonumber
\end{align}
\end{subequations}
\fi
where \eqref{eq:prLIU1} comes from the fact that $G^u_T$ follows Gamma distribution with parameter $\mL$ and $\mN$ for LOS and NLOS aerial link respectively. In \eqref{eq:prLIU2}, $d_T$ is expressed in terms of $\Zc$, $h$, $d$, and $\Omega$ by cosine rule.

\section{Proof of Lemma~\ref{le:disCU}}\label{ap:disCU}
The relation between the length of the TsUE to ABS link $\Zc$ with its projection distance on the ground $r_T$ is $\Zc=\sqrt{r_T^2+h^2}$. In order to find the distance distribution of $\Zc$, the distance distribution of $r_T$ is needed. As shown in Fig.~\ref{fig:disCU}, the total area of the region where the TsUE is located at is $\pi R_1^2-\pi R_2^2$. When $R_2\leqslant r_T\leqslant R_1-d$, the TsUE falls onto the ring. Therefore, the distance distribution is $f_{r_T}(r)=\frac{2\pi r}{\pi R_1^2-\pi R_2^2}=\frac{2r}{R_1^2-R_2^2}$. When $R_1-d<r_T\leqslant R_1+d$, the TsUE lies in the arc. The distance distribution is $f_{r_T}(r)=\frac{2\hat{\omega} r}{\pi R_1^2-\pi R_2^2}$. Based on cosine rule, $\hat{\omega}=\mathrm{arccos}\left(\frac{d^2+r^2-R_1^2}{2dr}\right)$. Using the pdf of the auxiliary random variable $r_T$, we can derive the distance distribution of $\Zc$ in Lemma~\ref{le:disCU} as

\ifCLASSOPTIONpeerreview
\begin{align}\label{eq:zc}
f_{\Zc}\!(z)&=\frac{\textup{d}(\sqrt{z^2-h^2})}{\textup{d}z}f_{r_T}\left(\sqrt{z^2-h^2}\right)\nonumber\\
&=\!\begin{cases}\!\!\frac{2z}{R_1^2-R_2^2},&\sqrt{R_2^2+h^2}\!\leqslant \! z\!\leqslant\! \sqrt{(R_1\!-\! d)^2+h^2}\\
            \!\!\frac{2z}{\pi R_1^2-\pi R_2^2}\mathrm{arcsec}\left(\frac{2d\sqrt{z^2-h^2}}{d^2+z^2-h^2-R_1^2}\right),&\sqrt{(R_1\!-\! d)^2+h^2}\!<\! z\!\leqslant\! \sqrt{(R_1\!+\! d)^2+h^2}
\end{cases}.
\end{align}
\else
\begin{align}\label{eq:zc}
&f_{\Zc}\!(z)=\frac{\textup{d}(\sqrt{z^2-h^2})}{\textup{d}z}f_{r_T}\left(\sqrt{z^2-h^2}\right)\nonumber\\
&=\!\begin{cases}\!\!\frac{2z}{R_1^2-R_2^2},\\
\quad\quad\quad\sqrt{R_2^2+h^2}\!\leqslant \! z\!\leqslant\! \sqrt{(R_1\!-\! d)^2+h^2}\\
            \!\!\frac{2z}{\pi R_1^2-\pi R_2^2}\mathrm{arcsec}\left(\frac{2d\sqrt{z^2-h^2}}{d^2+z^2-h^2-R_1^2}\right),\\
\quad\quad\quad\sqrt{(R_1\!-\! d)^2\!+\!h^2}\!<\! z\!\leqslant\! \sqrt{(R_1\!+\! d)^2\!+\!h^2}
\end{cases}\!\!\!\!\!\!.
\end{align}
\fi

\ifCLASSOPTIONpeerreview
\begin{figure}[t]
\centering
\includegraphics[width=0.3 \textwidth]{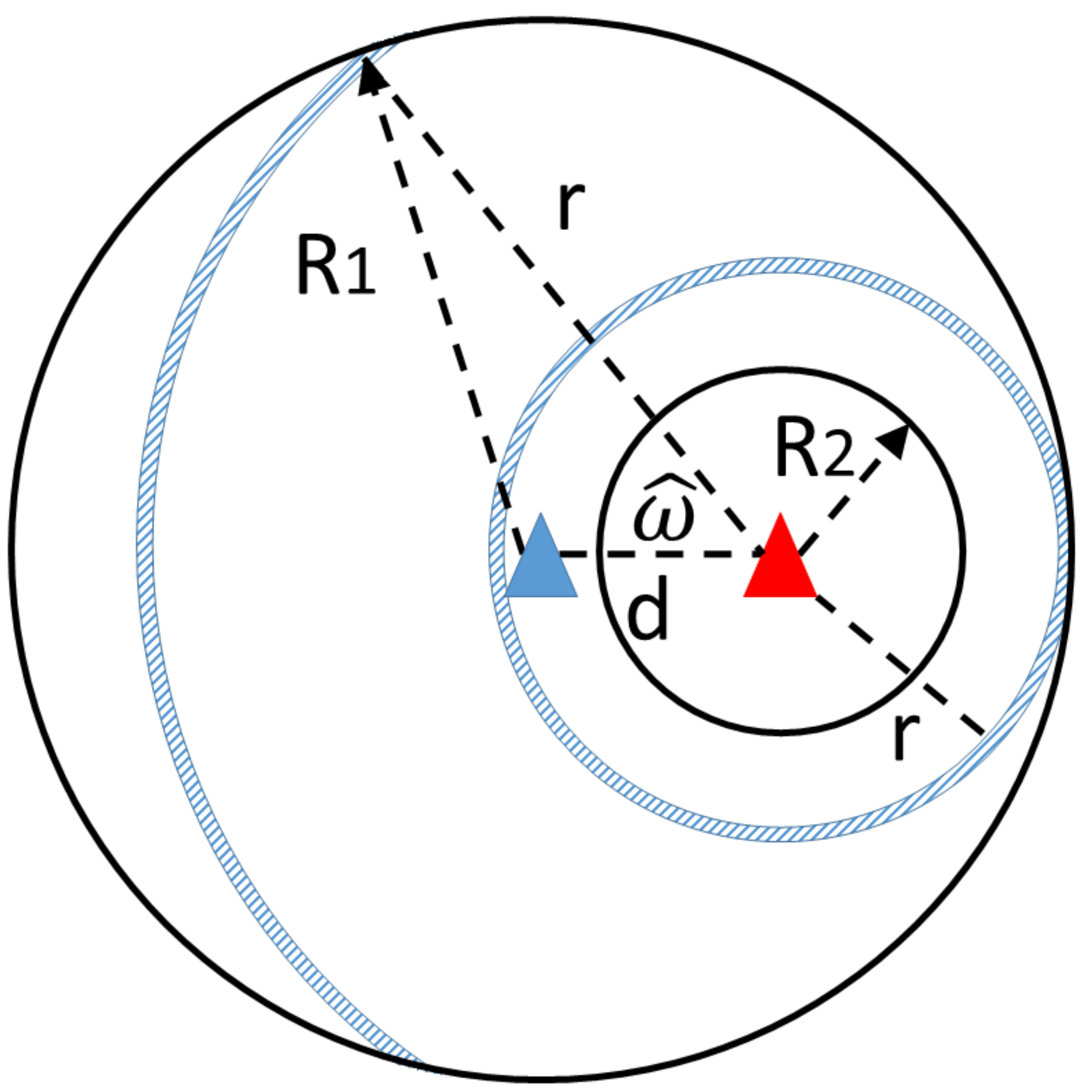}
\centering
\vspace{-1mm}
\caption{Illustration of a disk region of radius $R_1$ with a circular hole with radius $R_2$. Their centers are $d$ apart.}
\label{fig:disCU}
\centering
\end{figure}
\else
\begin{figure}[t]
\centering
\includegraphics[width=0.25 \textwidth]{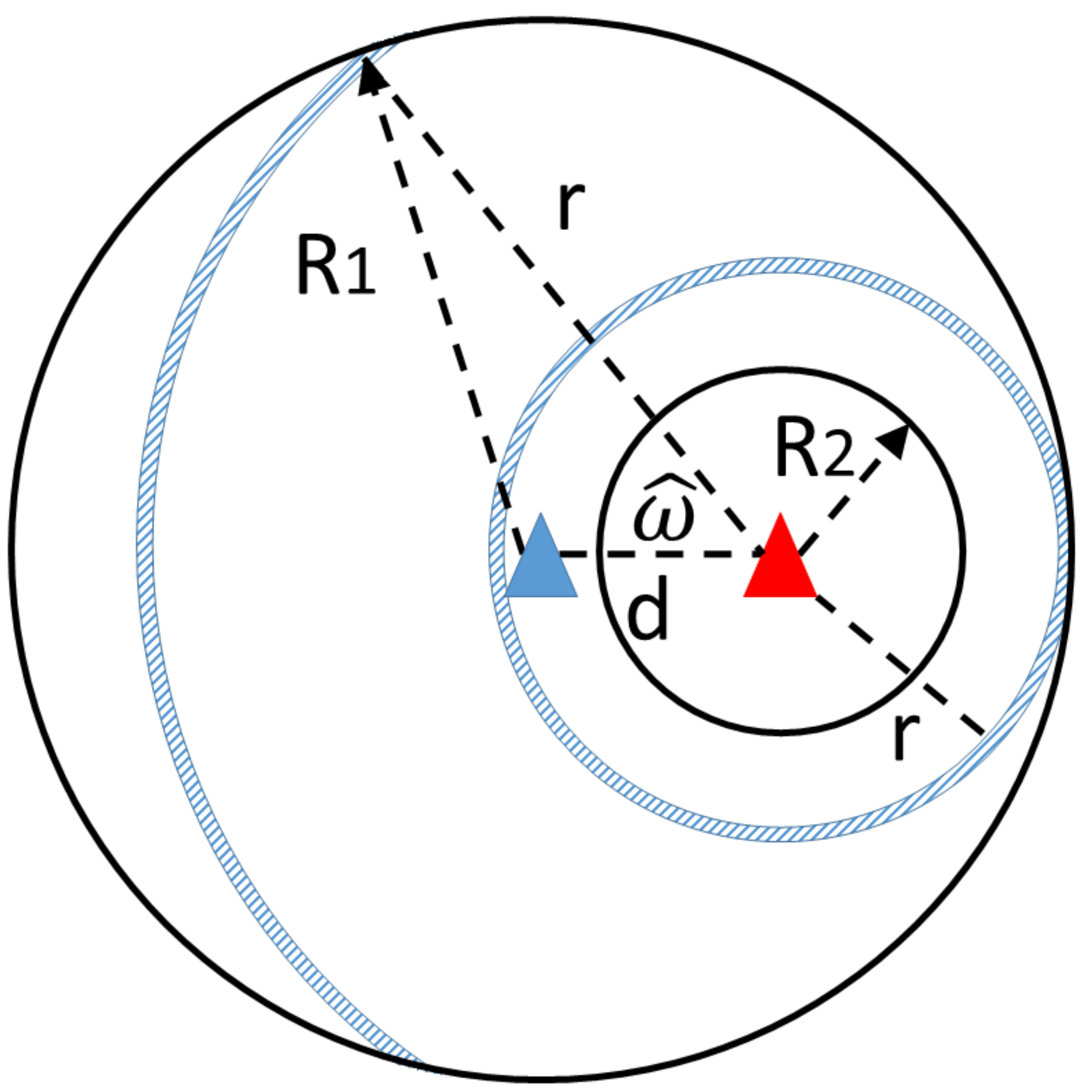}
\centering
\caption{Illustration of a disk region of radius $R_1$ with a circular hole with radius $R_2$. Their centers are $d$ apart.}
\label{fig:disCU}
\centering
\end{figure}
\fi

\section{Proof of Theorem~\ref{th:covD}}\label{ap:covD}
The uplink coverage probability of the ABS is given by
\begin{subequations}
\begin{align}
&\Pc^{u,\mathrm{A}}\!=\!\Pr(\textsf{SINR}^u_\mathrm{A}>\gammad^u)\nonumber\\
&\!=\!\Pr\left(G^u_A>\frac{\gammad^u}{P_\mathrm{AsD}PL_a(\Zd)}\left(P_\mathrm{TsUE} G^u_T PL_a(\Zc)+\sigma^2\right)\right)\nonumber\\
&\!=\!\mathbb{E}_{z}\!\!\!\left[\!\sum_{n=0}^{\mL-1}\!\!\!\frac{(\!-\!s_1)^n}{n!}\!\exp(\!-s_1\sigma^2)\!\sum_{k=0}^n\binom{n}{k}\frac{\pLOS}{(-\sigma^2)^{k-n}}\frac{\textup{d}^k}{\textup{d}s_1^k}\mathcal{L}_{\Id^u}(s_1)\right.\nonumber\\
&\left.\!\!+\!\!\!\!\sum_{n=0}^{\mN-1}\!\!\!\frac{(\!-\!s_2)^n}{n!}\!\exp(\!-s_2\sigma^2)\!\sum_{k=0}^n\!\!\binom{n}{k}\!\frac{\pNLOS}{(\!-\!\sigma^2\!)^{k\!-\!n}}\frac{\textup{d}^k}{\textup{d}s_2^k}\mathcal{L}_{\Id^u}(\!s_2\!)\!\right]\label{eq:prcovD1}\\
&\!=\!\mathbb{E}_{z}\left[\Pc^{u,\mathrm{L}}(P_\mathrm{AsD}|z)+\Pc^{u,\mathrm{N}}(P_\mathrm{AsD}|z)\right],\label{eq:prcovD2}
\end{align}
\end{subequations}
\noindent where \eqref{eq:prcovD1} comes from the fact that $G^u_A$ follows Gamma distribution with parameter $\mL$ and $\mN$ for LOS and NLOS aerial link respectively. $\Id^u=P_\mathrm{TsUE} G^u_T PL_a(\Zc)$, $s_1=\frac{\mL\gammad^u z^{\alphaL}}{P_\mathrm{AsD}\etaL}$ and $s_2=\frac{\mN\gammad^u z^{\alphaN}}{P_\mathrm{AsD}\etaN}$. Conditioned on the value of $h$, there are six possible cases for \eqref{eq:prcovD2}. Taking $\sqrt{(\frac{\Pmax\etaL}{\rhod})^{\frac{2}{\alphaL}}-R_2^2}<h<(\frac{\Pmax\etaN}{\rhod})^{\frac{1}{\alphaN}}$ as an example, the uplink coverage probability of the ABS equals to
\ifCLASSOPTIONpeerreview
\begin{align}
\Pc^{u,\mathrm{A}}\!&=\!\int_h^{(\frac{\Pmax\etaL}{\rhod})^{\frac{1}{\alphaL}}}\!\Pc^{u,\mathrm{L}}(\frac{\rhod}{\etaL} z^{\alphaL}|z)f_{\Zd}(z)\textup{d}z+\int_{(\frac{\Pmax}{\rhod})^{\frac{1}{\alphaL}}}^{\sqrt{h^2\!+R_2^2}}\!\Pc^{u,\mathrm{L}}(\Pmax|z)f_{\Zd}(z)\textup{d}z\nonumber\\
&+\int_h^{(\frac{\Pmax\etaN}{\rhod})^{\frac{1}{\alphaN}}}\!\Pc^{u,\mathrm{N}}(\frac{\rhod}{\etaN} z^{\alphaN}|z)f_{\Zd}(z)\textup{d}z+\int_{(\frac{\Pmax}{\rhod})^{\frac{1}{\alphaN}}}^{\sqrt{h^2\!+R_2^2}}\!\Pc^{u,\mathrm{N}}(\Pmax|z)f_{\Zd}(z)\textup{d}z.
\end{align}
\else
\begin{align}
\Pc^{u,\mathrm{A}}\!&=\!\int_h^{\left(\frac{\Pmax\etaL}{\rhod}\right)^{\frac{1}{\alphaL}}}\!\Pc^{u,\mathrm{L}}(\rhod\etaL^{-1} z^{\alphaL}|z)f_{\Zd}(z)\textup{d}z\nonumber\\
&+\int_{\left(\frac{\Pmax}{\rhod}\right)^{\frac{1}{\alphaL}}}^{\sqrt{h^2\!+R_2^2}}\!\Pc^{u,\mathrm{L}}(\Pmax|z)f_{\Zd}(z)\textup{d}z\nonumber\\
&+\int_h^{\left(\frac{\Pmax\etaN}{\rhod}\right)^{\frac{1}{\alphaN}}}\!\Pc^{u,\mathrm{N}}(\rhod\etaN^{-1}z^{\alphaN}|z)f_{\Zd}(z)\textup{d}z\nonumber\\
&+\int_{\left(\frac{\Pmax}{\rhod}\right)^{\frac{1}{\alphaN}}}^{\sqrt{h^2\!+R_2^2}}\!\Pc^{u,\mathrm{N}}(\Pmax|z)f_{\Zd}(z)\textup{d}z.
\end{align}
\fi

\begin{IEEEbiography}
[{\includegraphics[width=1in,height=1.25in,clip,keepaspectratio]{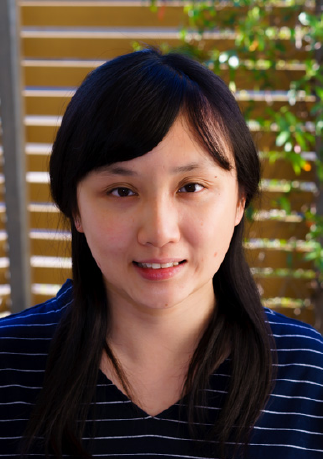}}]{Xiaohui Zhou} (S'15) received the B.Eng. degree (first class honours) from Australian National University, Canberra, Australia in 2014. She is currently working towards her Ph.D. degree at the Research School of Engineering, Australian National University, Canberra, Australia. She is a recipient of ANU University Research scholarship. Her research interests include drone communications, wireless information and power transfer, stochastic geometry and fifth-generation wireless networks.
\end{IEEEbiography}

\begin{IEEEbiography}
[{\includegraphics[width=1in,height=1.25in,clip,keepaspectratio]{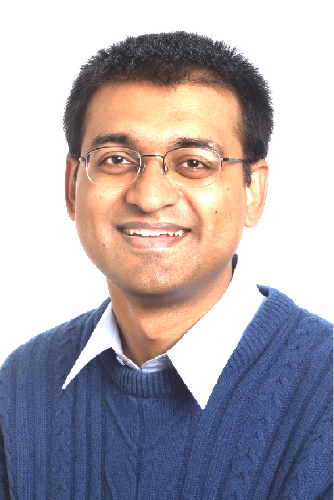}}]{Salman Durrani} (S'00-M'05-SM'10) received the B.Sc. (1st class honours) degree in Electrical Engineering from the University of Engineering \& Technology, Lahore, Pakistan in 2000. He received the PhD degree in Electrical Engineering from the University of Queensland, Brisbane, Australia in Dec. 2004. He has been with the Australian National University, Canberra, Australia, since 2005, where he is currently Associate Professor in the Research School of Engineering, College of Engineering \& Computer Science. His research interests include wireless information and power transfer, energy-harvesting-enabled wireless communications, drone communications, machine-to-machine and device-to-device communication, stochastic geometry modelling of finite area networks and synchronization in communication systems.

Dr. Durrani has co-authored more than 140 publications to date in refereed international journals and conferences. He was a recipient of the 2016 IEEE ComSoc Asia Pacific Outstanding Paper Award. He was the Chair of the ACT Chapter of the IEEE Signal Processing and Communications Societies from 2015 to 2016. He currently serves as an Editor of the IEEE TRANSACTIONS ON COMMUNICATIONS. He was awarded the 2018 ANU VC Award for Excellence in Supervision and the 2012 ANU VC Award for Excellence in Education. He is a Member of Engineers Australia, a Senior Fellow of IEEE, USA and a Senior Fellow of The Higher Education Academy, UK.
\end{IEEEbiography}

\begin{IEEEbiography}
[{\includegraphics[width=1in,height=1.25in,clip,keepaspectratio]{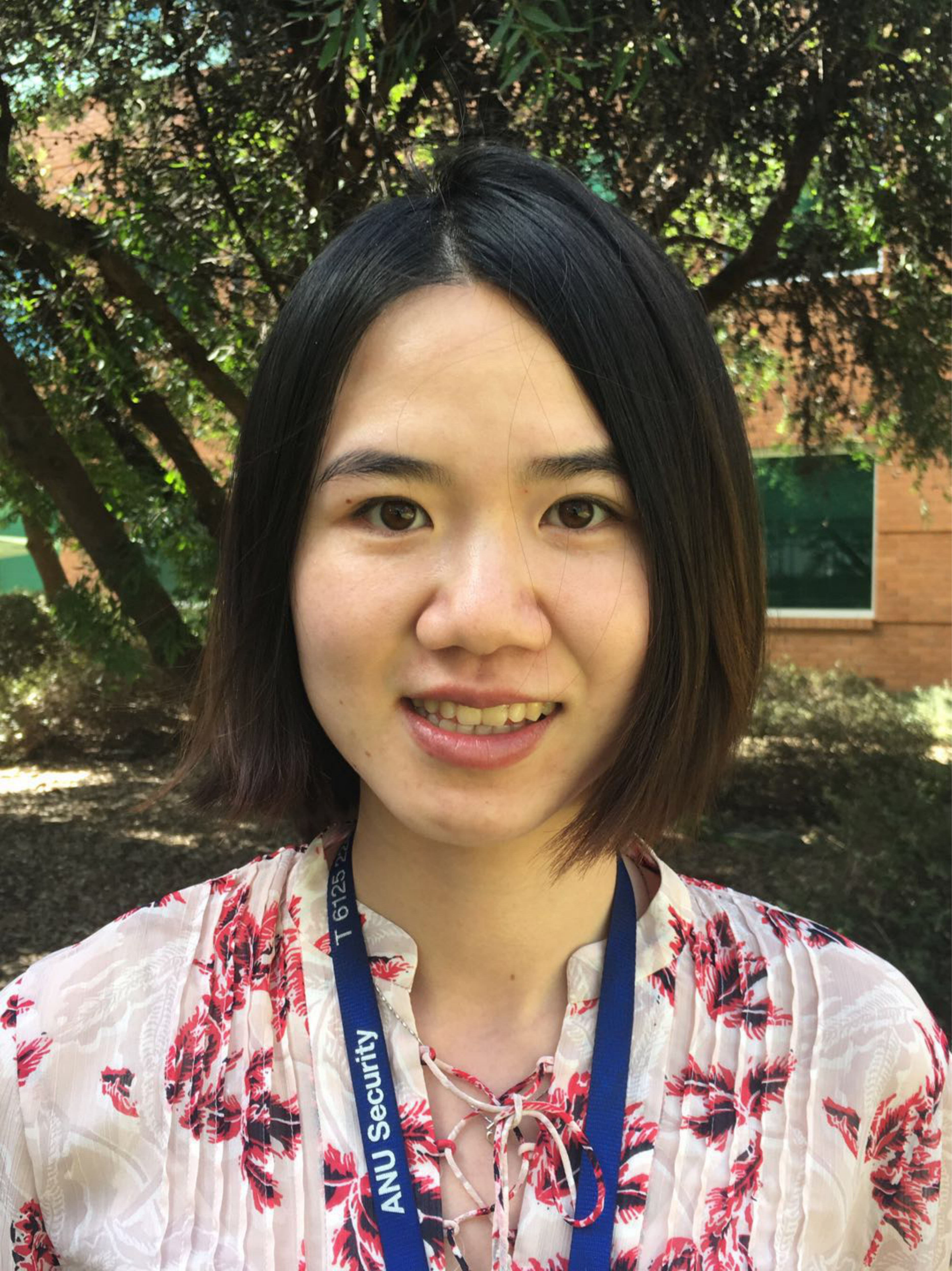}}]{Jing Guo} (S'12-M'17) received her BSc (first class honours) in electronics and telecommunications engineering from the Australian National University, Australia and the Beijing Institute of Technology, China in 2012, and PhD degree in telecommunications engineering from the Australian National University in 2016. She is currently employed as a Postdoctoral Research Fellow at the Research School of Engineering, Australian National University, Canberra, Australia. Her research interest lies in the field of wireless communications, including machine-to-machine communications and the application of stochastic geometry to wireless networks.
\end{IEEEbiography}

\begin{IEEEbiography}
[{\includegraphics[width=1in,height=1.25in,clip,keepaspectratio]{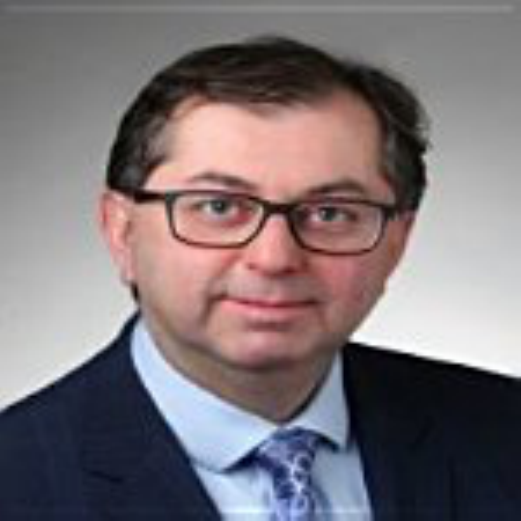}}]{Halim Yanikomeroglu} (F'17) was born in Giresun, Turkey, in 1968. He received the B.Sc. degree in electrical and electronics engineering from the Middle East Technical University, Ankara, Turkey, in 1990, and the M.A.Sc. degree in electrical engineering (now ECE) and the Ph.D. degree in electrical and computer engineering from the University of Toronto, Canada, in 1992 and 1998, respectively.

During 1993-1994, he was with the R\&D Group of Marconi Kominikasyon A.S., Ankara, Turkey. Since 1998 he has been with the Department of Systems and Computer Engineering at Carleton University, Ottawa, Canada, where he is now a Full Professor. His research interests cover many aspects of wireless technologies. Dr. Yanikomeroglu has supervised 20 PhD and 28 MASc students (all completed with theses); several of his PhD students received various medals. He has been one of the most frequent tutorial presenters in the leading international IEEE conferences. He has had extensive collaboration with large-scale (such as Huawei, Samsung, Blackberry, Telus, Nortel), medium-scale, and small-scale companies as well as the government labs. During 2012-2016, he led one of the largest academic-industrial collaborative research projects on pre-standards 5G wireless networks. Dr. Yanikomeroglu’s collaborations with industry have resulted in 25 granted patents (plus many applied).

Dr. Yanikomeroglu is a Fellow of the IEEE. He is a Distinguished Lecturer for the IEEE Communications Society and a Distinguished Speaker for the IEEE Vehicular Technology Society. He has been involved in the organization of the IEEE Wireless Communications and Networking Conference (WCNC) from its inception in 1998 in various capacities including serving as a Steering Committee member for a decade and the Technical Program Chair/Co-Chair of WCNC 2004 (Atlanta), WCNC 2008 (Las Vegas), and WCNC 2014 (Istanbul). He was the General Co-Chair of the IEEE 72nd Vehicular Technology Conference (VTC 2010-Fall) held in Ottawa, and the General Chair of the IEEE 86th Vehicular Technology Conference (VTC 2017-Fall) held in Toronto. He has served in the editorial boards of the IEEE Transactions on Communications, IEEE Transactions on Wireless Communications, and IEEE Communications Surveys \& Tutorials; he also served as a guest editor in various IEEE journal special issues. He was the Chair of one of the largest technical committees in IEEE, Technical Committee on Personal Communications (now called Wireless Communications Technical Committee and has 1,700+ members).

Dr. Yanikomeroglu is a recipient of the IEEE Ottawa Section Outstanding Educator Award in 2014, Carleton University Faculty Graduate Mentoring Award in 2010, the Carleton University Graduate Students Association Excellence Award in Graduate Teaching in 2010, and the Carleton University Research Achievement Award in 2009 and 2018. Dr. Yanikomeroglu spent the 2011-2012 academic year at TOBB U. of Economics and Technology, Ankara, Turkey, as a Visiting Professor. He is a registered Professional Engineer in the province of Ontario, Canada.
\end{IEEEbiography}


\begin{thebibliography}{10}
\providecommand{\url}[1]{#1}
\csname url@samestyle\endcsname
\providecommand{\newblock}{\relax}
\providecommand{\bibinfo}[2]{#2}
\providecommand{\BIBentrySTDinterwordspacing}{\spaceskip=0pt\relax}
\providecommand{\BIBentryALTinterwordstretchfactor}{4}
\providecommand{\BIBentryALTinterwordspacing}{\spaceskip=\fontdimen2\font plus
\BIBentryALTinterwordstretchfactor\fontdimen3\font minus
  \fontdimen4\font\relax}
\providecommand{\BIBforeignlanguage}[2]{{%
\expandafter\ifx\csname l@#1\endcsname\relax
\typeout{** WARNING: IEEEtran.bst: No hyphenation pattern has been}%
\typeout{** loaded for the language `#1'. Using the pattern for}%
\typeout{** the default language instead.}%
\else
\language=\csname l@#1\endcsname
\fi
#2}}
\providecommand{\BIBdecl}{\relax}
\BIBdecl

\bibitem{Zhou-2018}
X.~Zhou, J.~Guo, S.~Durrani, and H.~Yanikomeroglu, ``Uplink coverage
  performance of an underlay drone cell for temporary events,'' in \emph{Proc.
  IEEE ICC Workshop}, May 2018.

\bibitem{Yaliniz-2016}
I.~Bor-Yaliniz and H.~Yanikomeroglu, ``The new frontier in {RAN} heterogeneity:
  Multi-tier drone-cells,'' \emph{{IEEE} Commun. Mag.}, vol.~54, no.~11, pp.
  48--55, Nov. 2016.

\bibitem{Chandrasekharan-2016}
S.~Chandrasekharan, K.~Gomez, A.~Al-Hourani, S.~Kandeepan, T.~Rasheed,
  L.~Goratti, L.~Reynaud, D.~Grace, I.~Bucaille, T.~Wirth, and S.~Allsopp,
  ``Designing and implementing future aerial communication networks,''
  \emph{{IEEE} Commun. Mag.}, vol.~54, no.~5, pp. 26--34, May 2016.

\bibitem{Mozaffari-2018}
\BIBentryALTinterwordspacing
M.~Mozaffari, W.~Saad, M.~Bennis, Y.~N. Nam, and M.~Debbah. (2018) ``{A}
  tutorial on {UAV}s for wireless networks: Applications, challenges and open
  problems,'' submitted to \textit{{IEEE} Commun. Surveys Tuts.} [Online].
  Available: \url{https://arxiv.org/abs/1803.00680}
\BIBentrySTDinterwordspacing

\bibitem{Zeng-2018}
\BIBentryALTinterwordspacing
Y.~Zeng, J.~Lyu, and R.~Zhang. (2018) ``{C}ellular-connected {UAV}: Potentials,
  challenges and promising technologies,'' submitted to \textit{{IEEE} Wireless
  Commun.} [Online]. Available: \url{https://arxiv.org/abs/1804.02217}
\BIBentrySTDinterwordspacing

\bibitem{Mozaffari-2016c}
M.~Mozaffari, W.~Saad, M.~Bennis, and M.~Debbah, ``Unmanned aerial vehicle with
  underlaid device-to-device communications: Performance and tradeoffs,''
  \emph{{IEEE} Trans. Wireless Commun.}, vol.~15, no.~6, pp. 3949--3963, Jun.
  2016.

\bibitem{Chetlur-2017}
V.~V. Chetlur and H.~S. Dhillon, ``Downlink coverage analysis for a finite
  3-{D} wireless network of unmanned aerial vehicles,'' \emph{{IEEE} Trans.
  Commun.}, vol.~65, no.~10, pp. 4543--4558, Oct. 2017.

\bibitem{Azari-2017a}
M.~M. Azari, Y.~Murillo, O.~Amin, F.~Rosas, M.-S. Alouini, and S.~Pollin,
  ``Coverage maximization for a poisson field of drone cells,'' in \emph{Proc.
  IEEE PIMRC}, Oct. 2017.

\bibitem{Galkin-2017}
B.~Galkin, J.~Kibilda, and L.~A. DaSilva, ``Coverage analysis for low-altitude
  {UAV} networks in urban environments,'' in \emph{Proc. IEEE Globecom}, Dec.
  2017.

\bibitem{Azari-2017b}
M.~M. Azari, F.~Rosas, A.~Chiumento, and S.~Pollin, ``Coexistence of
  terrestrial and aerial users in cellular networks,'' in \emph{Proc. IEEE
  Globecom Workshops}, Dec. 2017.

\bibitem{Sekander-2018}
S.~Sekander, H.~Tabassum, and E.~Hossain, ``Multi-tier drone architecture for
  5{G}/{B}5{G} cellular networks: Challenges, trends, and prospects,''
  \emph{{IEEE} Commun. Mag.}, vol.~56, no.~3, pp. 96--103, Mar. 2018.

\bibitem{Khawaja-2018}
\BIBentryALTinterwordspacing
W.~Khawaja, I.~Guvenc, D.~W. Matolak, U.~C. Fiebig, and N.~Schneckenberger.
  (2018) ``{A} survey of air-to-ground propagation channel modeling for
  unmanned aerial vehicles''. [Online]. Available:
  \url{https://arxiv.org/abs/1801.01656}
\BIBentrySTDinterwordspacing

\bibitem{Amorim-2017a}
R.~Amorim, H.~Nguyen, P.~Mogensen, I.~Z. Kovács, J.~Wigard, and T.~B.
  Sørensen, ``Radio channel modeling for {UAV} communication over cellular
  networks,'' \emph{{IEEE} Wireless Commun. Lett.}, vol.~6, no.~4, pp.
  514--517, Aug. 2017.

\bibitem{Amorim-2017b}
R.~Amorim, P.~Mogensen, T.~Sorensen, I.~Z. Kovacs, and J.~Wigard, ``Pathloss
  measurements and modeling for {UAV}s connected to cellular networks,'' in
  \emph{Proc. IEEE VTC Spring}, Jun. 2017.

\bibitem{Hourani-2014b}
A.~Al-Hourani, S.~Kandeepan, and S.~Lardner, ``Optimal {LAP} altitude for
  maximum coverage,'' \emph{\it{{IEEE} Wireless Commun. Lett.}}, vol.~3, no.~6,
  pp. 569--572, Dec. 2014.

\bibitem{Hourani-2014a}
A.~Al-Hourani, S.~Kandeepan, and A.~Jamalipour, ``Modeling air-to-ground path
  loss for low altitude platforms in urban environments,'' in \emph{Proc. IEEE
  Globecom}, Dec. 2014.

\bibitem{Fotouhi-2017}
\BIBentryALTinterwordspacing
A.~Fotouhi, M.~Ding, and M.~Hassan. (2017) ``{D}rone{C}ells: Improving 5{G}
  spectral efficiency using drone-mounted flying base stations,'' submitted to
  \textit{{IEEE} Trans. Mobile Comput.} [Online]. Available:
  \url{https://arxiv.org/abs/1707.02041}
\BIBentrySTDinterwordspacing

\bibitem{Zeng-2016a}
Y.~Zeng, R.~Zhang, and T.~J. Lim, ``Throughput maximization for {UAV}-enabled
  mobile relaying systems,'' \emph{{IEEE} Trans. Commun.}, vol.~64, no.~12, pp.
  4983--4996, Dec. 2016.

\bibitem{Alzenad-2017a}
M.~Alzenad, A.~El-Keyi, F.~Lagum, and H.~Yanikomeroglu, ``3-{D} placement of an
  unmanned aerial vehicle base station ({UAV-BS}) for energy-efficient maximal
  coverage,'' \emph{{IEEE} Wireless Commun. Lett.}, vol.~6, no.~4, pp.
  434--437, Aug. 2017.

\bibitem{He-2018}
H.~He, S.~Zhang, Y.~Zeng, and R.~Zhang, ``Joint altitude and beamwidth
  optimization for {UAV}-enabled multiuser communications,'' \emph{\it{{IEEE}
  Commun. Lett.}}, vol.~22, no.~2, pp. 344--347, Feb. 2018.

\bibitem{Zhang-2017}
C.~Zhang and W.~Zhang, ``Spectrum sharing for drone networks,'' \emph{{IEEE} J.
  Sel. Areas Commun.}, vol.~35, no.~1, pp. 136--144, Jan. 2017.

\bibitem{Lagum-2017}
F.~Lagum, I.~Bor-Yaliniz, and H.~Yanikomeroglu, ``Strategic densification with
  {UAV-BS}s in cellular networks,'' \emph{{IEEE} Wireless Commun. Lett.},
  vol.~7, no.~3, pp. 384--387, Jun. 2018.

\bibitem{Wu-2017}
Q.~Wu, Y.~Zeng, and R.~Zhang, ``Joint trajectory and communication design for
  multi-{UAV} enabled wireless networks,'' \emph{{IEEE} Trans. Wireless
  Commun.}, vol.~17, no.~3, pp. 2109--2121, Mar. 2018.

\bibitem{Azari-2018}
M.~M. Azari, F.~Rosas, A.~Chiumento, A.~Ligata, and S.~Pollin, ``Uplink
  performance analysis of a drone cell in a random field of ground
  interferers,'' in \emph{Proc. IEEE WCNC}, Apr. 2018.

\bibitem{Alzenad-2018}
\BIBentryALTinterwordspacing
M.~Alzenad and H.~Yanikomeroglu. (2018) ``{C}overage and rate analysis for
  unmanned aerial vehicle base stations with {LoS/NLoS} propagation,'' accepted
  to appear in \textit{Proc. {IEEE} Globecom Workshop}. [Online]. Available:
  \url{https://www.researchgate.net/publication/324595044_Coverage_and_Rate_Analysis_for_Unmanned_Aerial_Vehicle_Base_Stations_with_LoSNLoS_Propagation}
\BIBentrySTDinterwordspacing

\bibitem{Absolute-2018}
\BIBentryALTinterwordspacing
{ABSOLUTE} ({A}erial {B}ase {S}tations with {O}pportunistic {L}inks for
  {U}nexpected and {T}emporary {E}vents). [Online]. Available:
  \url{https://cordis.europa.eu/project/rcn/106035_en.html}
\BIBentrySTDinterwordspacing

\bibitem{Morozs-2015}
N.~Morozs, T.~Clarke, and D.~Grace, ``Heuristically accelerated reinforcement
  learning for dynamic secondary spectrum sharing,'' \emph{{IEEE} Access},
  vol.~3, pp. 2771--2783, Dec. 2015.

\bibitem{Koulali-2016}
S.~Koulali, E.~Sabir, T.~Taleb, and M.~Azizi, ``A green strategic activity
  scheduling for {UAV} networks: {A} sub-modular game perspective,''
  \emph{{IEEE} Commun. Mag.}, vol.~54, no.~5, pp. 58--64, May 2016.

\bibitem{Yang-2017}
P.~Yang, X.~Cao, C.~Yin, Z.~Xiao, X.~Xi, and D.~Wu, ``Proactive drone-cell
  deployment: {O}verload relief for a cellular network under flash crowd
  traffic,'' \emph{{IEEE} Trans. Intell. Transp. Syst.}, vol.~18, no.~10, pp.
  2877--2892, May 2017.

\bibitem{Rupasinghe-2018}
N.~Rupasinghe, Y.~Yapici, I.~Guvenc, and Y.~Kakishima, ``Non-orthogonal
  multiple access for mm{W}ave drone networks with limited feedback,''
  \emph{\rm{to appear in} \it{{IEEE} {T}rans. {C}ommun.}}, 2018.

\bibitem{Lyu-2018}
J.~Lyu, Y.~Zeng, and R.~Zhang, ``{UAV}-aided offloading for cellular hotspot,''
  \emph{{IEEE} Trans. Wireless Commun.}, vol.~17, no.~6, pp. 3988--4001, Jun.
  2018.

\bibitem{Song-2014}
L.~Song, Z.~Han, and C.~Xu, \emph{Resource Management for Device-to-Device
  Underlay Communication}.\hskip 1em plus 0.5em minus 0.4em\relax New York, NY,
  USA: Springer, 2014.

\bibitem{Zhou-2017}
Z.~Zhou, K.~Ota, M.~Dong, and C.~Xu, ``Energy-efficient matching for resource
  allocation in {D2D} enable cellular networks,'' \emph{{IEEE} Trans. Veh.
  Technol.}, vol.~66, no.~6, pp. 5256--5268, Jun. 2017.

\bibitem{Novlan-2013}
T.~D. Novlan, H.~S. Dhillon, and J.~G. Andrews, ``Analytical modeling of uplink
  cellular networks,'' \emph{{IEEE} Trans. Wireless Commun.}, vol.~12, no.~6,
  pp. 2669--2679, Jun. 2013.

\bibitem{Yaliniz-2016b}
I.~Bor-Yaliniz, A.~El-Keyi, and H.~Yanikomeroglu, ``Efficient 3-{D} placement
  of an aerial base station in next generation cellular networks,'' in
  \emph{Proc. IEEE ICC}, May 2016.

\bibitem{3GPP-2017}
3GPP, ``Study on enhanced {LTE} support for aerial vehicles,'' \emph{TR
  36.777}, Dec. 2017.

\end{thebibliography}
\end{document}